\documentclass[traditabstract]{aa}
\usepackage{graphicx}
\usepackage{amsmath,amssymb}
\usepackage{inputenc}
\usepackage{xcolor}
\usepackage{txfonts}
\usepackage{natbib}
\usepackage{enumitem}
\usepackage{multirow}
\usepackage{booktabs}
\usepackage[modulo, switch]{lineno}
\bibpunct{(}{)}{;}{a}{}{,} 

\newcommand{\Kepler}{\textit{Kepler}}
\newcommand{\kepler}{\textit{Kepler} }

\newcommand{\Dnu}{\Delta \nu}

\newcommand{\numax}{\nu_\mathrm{max}}

\newcommand{\teff}{T_\mathrm{eff}}

\newcommand{\diamonds}{\textsc{D\large{iamonds}}}
\newcommand{\famed}{\textsc{FAMED}}
\newcommand{\glob}{\texttt{GLOBAL}}
\newcommand{\chu}{\texttt{CHUNK}}

\def\be{\begin{equation}}
\def\ee{\end{equation}}    
\def\ba{\begin{eqnarray}}
\def\ea{\end{eqnarray}}

\begin{document}

\title{Fast and Automated Peak Bagging with \textsc{Diamonds} (\textsc{FAMED})}
\author{E. Corsaro\inst{1},
J. M. McKeever\inst{2},
\and J. S. Kuszlewicz\inst{3,4}
          }
\offprints{Enrico Corsaro\\ \email{enrico.corsaro@inaf.it}}

\institute{INAF -- Osservatorio Astrofisico di Catania, via S. Sofia 78, 95123 Catania, Italy
\and Astronomy Department, Yale University, New Haven, CT 06511, USA
\and Max-Planck-Institut f\"{u}r Sonnensystemforschung, Justus-von-Liebig-Weg 3, D-37077 G\"{o}ttingen, Germany
\and Stellar Astrophysics Centre, Department of Physics and Astronomy, Aarhus University, Ny Munkegade 120, DK-8000 Aarhus C, Denmark
}

   \date{Received 12 March 2020; Accepted 10 June 2020}

\abstract
{Stars of low and intermediate mass that exhibit oscillations may show tens of detectable oscillation modes each. Oscillation modes are a powerful to constrain the internal structure and rotational dynamics of the star, hence tool allowing one to obtain an accurate stellar age. The tens of thousands of solar-like oscillators that have been discovered thus far are representative of the large diversity of fundamental stellar properties and evolutionary stages available. Because of the wide range of oscillation features that can be recognized in such stars, it is particularly challenging to properly characterize the oscillation modes in detail, especially in light of large stellar samples. Overcoming this issue requires an automated approach, which has to be fast, reliable, and flexible at the same time. In addition, this approach should not only be capable of extracting the oscillation mode properties of frequency, linewidth, and amplitude from stars in different evolutionary stages, but also able to assign a correct mode identification for each of the modes extracted. Here we present the new freely available pipeline \famed\,\,(Fast and AutoMated pEak bagging with \diamonds), which is capable of performing an automated and detailed asteroseismic analysis in stars ranging from the main sequence up to the core-Helium-burning phase of stellar evolution. This, therefore, includes subgiant stars, stars evolving along the red giant branch (RGB), and stars likely evolving toward the early asymptotic giant branch. In this paper, we additionally show how \famed\,\,can detect rotation from dipolar oscillation modes in main sequence, subgiant, low-luminosity RGB, and core-Helium-burning stars. \famed\,\,can be downloaded from its public GitHub repository (https://github.com/EnricoCorsaro/FAMED).} 

%
\titlerunning{Fast and Automated Peak Bagging with \textsc{Diamonds} (\texttt{FAMED})}
      \authorrunning{E. Corsaro et al.}
\maketitle

\section{Introduction}
Thanks to the photometric space missions CoRoT \citep{Baglin06}, NASA \kepler and K2 \citep{Borucki10,Koch10,Howell14K2}, and more recently NASA TESS \citep{Ricker14TESS}, the number of low- and intermediate-mass stars with detected oscillations now accounts for several tens of thousands of targets. The best characterized stars in terms of oscillations still remain those observed by the nominal \kepler mission (e.g., \citealt{Corsaro15cat}, hereafter C15, and \citealt{Lund17LEGACY}. This is because of its long observing time, which exceeds four years with a high duty cycle, and photometric precision. New high-quality observations are, however, being added thanks to TESS, especially in its continuous viewing zones (CVZs), where the observing length can be up to one full year \citep{Stassun19TIC}. In this scenario, we have tens of thousands of stars, mostly red giants, for which a detailed analysis of their oscillation properties, known as peak bagging, can be performed. The peak bagging analysis is an essential step to accomplish if one wants to extract the most possible from the asteroseismic observations. This is because it is with the individual oscillation mode properties that one is able to reconstruct the internal structure and rotational stratification of the star and, consequently, to obtain more accurate stellar ages \citep[e.g.,][]{Perez16,SilvaAguirre17}. However, the peak bagging analysis is significantly complicated by the large diversity of oscillation features that can be found in stars evolving from the main sequence (MS) to red giants (RGs). While analyzing a few targets, either manually or using semi-automated approaches, is still feasible, performing a peak bagging analysis on tens of thousands of stars may become an impractical path to follow. Although, efforts in this direction through the adoption of different peak bagging techniques are being done \citep[e.g.,][]{Handberg17,Garcia18pb,Themessl18,Kallinger19pb}. In addition to this, in this coming decade, we will face new observations from the all-sky mission ESA PLATO \citep{Rauer14PLATO}, which will increase the expected yield of stars for which a peak bagging analysis can be conducted by at least one order of magnitude.

In \cite{Corsaro14}, the authors introduced the so-called Bayesian multimodal fitting for peak bagging for the first time, showing that it can be successfully used on an MS solar-like oscillator, although the approach was still not competitive in terms of computational speed. In a recent work, \cite{Corsaro19} (hereafter C19) presents a significant improvement to the Bayesian multimodal approach and shows how it retains a great potential for overcoming the problem of extracting a large amount of asteroseismic information in a short time. In particular, C19 proves that this process can be automated by applying it to the case of an RGB star. However, at that time, the methodology was not extended and tested for stars in different evolutionary stages, nor was the actual mode identification discussed and included as an output of the computation. The mode identification typically requires a detailed knowledge of the possible oscillation features that can be observed from a given star and it is usually never returned as a direct output from the peak bagging analysis. The way the oscillation features change as a function of fundamental stellar properties, evolutionary stage, and rotation of the star, has been extensively discussed in the literature (see e.g., \citealt{Gizon03,Ballot08,Huber11,Mosser11universal,White11,Bedding11Nature,Mosser12probing,Beck12Nature,Corsaro12cluster}, C15, \citealt{Lund17LEGACY,Gehan18rotation}), and a detailed discussion about it is not the aim of this paper. Here we present a new pipeline, dubbed Fast and AutoMated pEak bagging with \diamonds (\famed), which exploits our knowledge of the stellar oscillation features, builds on the recent development in numerical sampling presented by C19, and is based on the public Bayesian inference tool \diamonds\,\,\citep{Corsaro14,Corsaro18corr,Corsaro18tutorial}\footnote{https://github.com/EnricoCorsaro/DIAMONDS.}. \famed\,\,is not only capable of quickly and efficiently extracting the asteroseismic properties of the individual oscillation modes, but it also incorporates a mode identification for each of the modes extracted up to the level of the rotationally split components. Finally, using both simulated and real datasets of well-characterized stars, we show how \famed\,\,can reliably perform the peak bagging analysis in stars ranging from the hot F-type MS to the core-Helium-burning phase of stellar evolution in an automated, fast, and flexible manner.

\section{Bayesian multimodal fitting}
\label{sec:multi_modal}
Bayesian approaches for detailed asteroseismic analysis have become more widely adopted mostly thanks to the larger availability of computational power and to the appealing possibility of incorporating a priori information on the free parameters to fit (e.g., \citealt{Gruberbauer09,Benomar09,Kallinger10Kepler,Handberg11,Corsaro14}, C15, \citealt{Davies16,Lund17LEGACY,Corsaro17spin,Handberg17,Vrard18pb}). These approaches are used to perform peak bagging by modeling the stellar power spectral density (hereafter PSD) with a mixture of Lorentzian profiles for as many oscillation modes as one intends to fit. Here each Lorentzian profile accounts for the free parameters of frequency, amplitude, and linewidth of a single oscillation mode to be modeled. This analysis, which we can refer to as a standard peak bagging, is intrinsically unimodal, meaning that the aim is to obtain a single-point solution that contains the estimates of all the free parameters of the fitting model. While this certainly allows for the highest precision on the parameter estimates, a unimodal approach has two main disadvantages: 1) it involves many free parameters to fit and therefore implies a slow computation; 2) it requires an input list of prior hyper-parameters that are often time consuming to retrieve and problematic to properly set up. These disadvantages cause standard peak bagging to be difficult to automate and to perform on a large sample of targets, hence severely limiting its wide applicability.

As originally anticipated by \cite{Corsaro14} and then more thoroughly discussed by C19, the \diamonds\,\,code offers the possibility to perform peak bagging in an innovative way, that is, using a Bayesian multimodal approach. This is possible thanks to the nested sampling Monte Carlo algorithm \citep{Skilling04}, well suited to efficiently sampling distributions that contain multiple local maxima. Here one can invert the condition presented in the unimodal case, and model the entire PSD with just one single Lorentzian profile. This in turn produces a degenerate solution (instead of a single-point solution), where the frequency centroid of the Lorentzian profile has many different possible outcomes. This multimodal approach can be easily set up because it only requires a few input prior hyper-parameters, which are straightforward to build up. Additionally the multimodal approach is extremely fast because of the low number of free parameters involved. 

\subsection{Islands peak bagging model}
\label{sec:isla}
In C19 it was shown that the so-called islands peak bagging model is a Lorentzian profile where the free parameters are its frequency centroid $\nu_0$ and amplitude $A$, with the FWHM of the profile, $\Gamma$, kept to a fixed value. The known correlation between oscillation mode height $H$ and $\Gamma$ is not a problem in this case because the FWHM is fixed, which means that $A^2$ and $H$ behave as the same quantity during the fitting process, except for a constant multiplication factor. A first improvement to the multimodal approach can thus be obtained by replacing $A$ with $H$. The advantage of having the height as a free parameter is that it is a direct observable in the PSD, meaning that it can be measured as the level of the PSD signal above the noise level. Conversely, the amplitude cannot be visualized in such a simple way (it corresponds to the frequency integral of the Lorentzian profile), and retrieving it from the PSD requires the height to be converted into amplitude according to the relation $A = \sqrt{\pi \Gamma H / 2}$ (for a single-sided PSD, e.g., \citealt{App14}). As a result, in order to have a reliable estimate of $A$, one should have at least a reasonable estimate of $\Gamma$. When $H$ is used instead, the corresponding prior hyper-parameters can be set up without any complication, and independently of the value chosen for $\Gamma$. If uniform prior probability density functions are adopted (as it is done for our applications), the prior hyper-parameters for $H$ can be set to zero for the lower limit, and to the maximum height measured from the region of the stellar PSD that contains the oscillation modes as the upper limit. The new islands peak bagging model therefore reads as
\begin{equation}
P_\mathrm{isla} \left( \nu, \Gamma; \nu_0, H \right) = \frac{H}{1 + 4 \left( \frac{\nu - \nu_0}{\Gamma} \right)^2}
\label{eq:isla}
\end{equation}
with $\nu_0$ and $H$ the free parameters specifying the frequency centroid and height, respectively, of the oscillation peak, and with $\Gamma$ the FWHM to be fixed to a specific value, which changes depending on the level of resolving power that is needed to fit the stellar PSD (see Sect.~\ref{sec:sampling_global} and \ref{sec:sampling_chunk} for more details). This is because fitting the islands peak bagging model to the PSD can be considered analogous to the effect obtained by smoothing the actual PSD signal at a resolution imposed by the FWHM of the Lorentzian profile.
 
 \begin{figure}
   \centering
  \includegraphics[width=8.5cm]{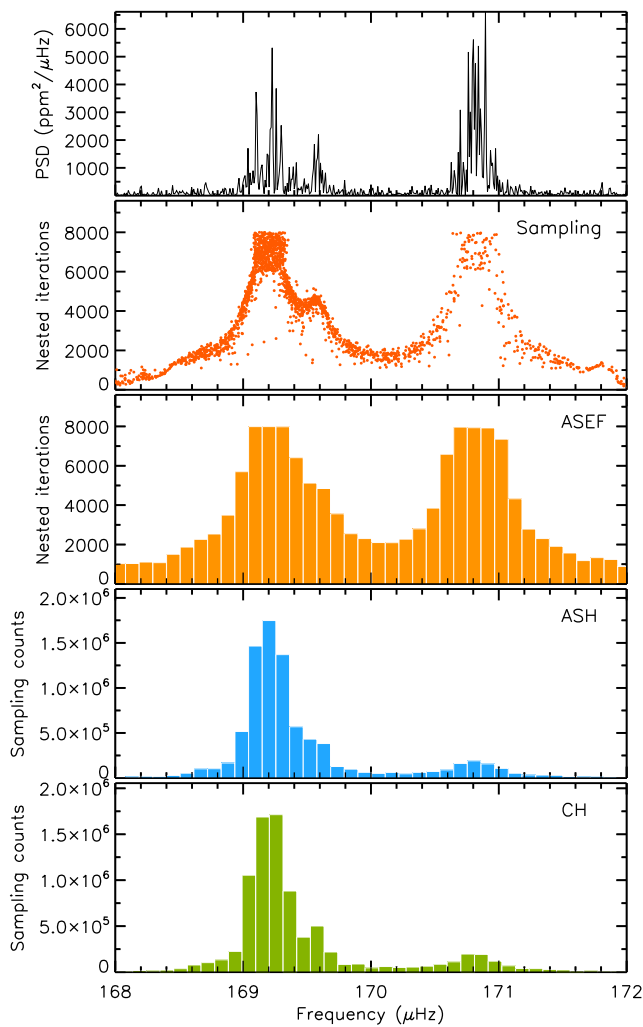}
    \caption{Different histograms discussed in Sect.~\ref{sec:multi_modal}, computed from the multimodal sampling obtained with \diamonds. \textit{Top panel}: a portion of PSD of the RGB star KIC~12008916, where the $\ell = 2, 0$ mode pair can be clearly observed. \textit{Second panel}: The sampling from \diamonds. \textit{Bottom three panels}: ASEF, ASH, and CH, respectively. For the ASH and CH, the right peak is almost entirely suppressed because of the low number of sampling points.}
    \label{fig:asef}
\end{figure}
 
 \subsection{Averaged shifted envelope function}
 \label{sec:asef}
A second important improvement to the multimodal approach concerns the counts histogram (CH) originally introduced by C19. The CH is the histogram built from the sampling obtained with \diamonds\,\,when applying the islands peak bagging model to the stellar PSD. This sampling is the set of points that was collected from the parameter space explored during the nested sampling process. The set of sampling points is sorted by increasing value of the likelihood function \citep{Skilling04,Sivia06}, meaning that as the sampling evolves, that is the nested iteration increases, we approach more closely to better solutions. In C19 it was shown that the CH can be used to automatically extract the individual oscillation mode frequencies because each oscillation mode will pop up as a peak in the CH. However, using a simple CH as shown in Fig. 1 of C19 presents a couple of drawbacks: 1) it requires that the sampling obtained during the first iterations be removed because it is contaminated by the noise structures in the PSD; 2) it may be subject to fluctuations depending on how the sampling is distributed with respect to the choice of histogram bins, hence with respect to the total number of bins, and to the actual density of the sampling for a given local maximum in the distribution. Here we propose a different tool that gets rid of the previous drawbacks, termed the averaged shifted envelope function (ASEF), which can be introduced with the following two steps: \textit{(i)} instead of computing a simple CH from the sampling, we obtain an averaged shifted histogram (ASH, \citealt{Hardle04}), which gives a more stable solution against fluctuations caused by the way the sampling points may fall into each bin. In the ASH, one considers the sum of the sampling point values (in nested iterations) falling in each bin, and shifts the bin position for a given number of realizations, thus averaging the different realizations into a single one at the end; \textit{(ii)} the ASH can still be affected by the noise structures that dominate the nested sampling at low values of the nested iterations and by the scarcity of the sampling in some regions of the parameter space. To overcome these problems, in the second step, instead of evaluating a standard ASH, the value of the histogram at each bin is now the maximum sampling value (in nested iteration) falling in the bin. The result is an averaged shifted histogram of the envelope function of the actual nested sampling obtained with \diamonds. 

The resulting ASEF does not require any cut-off to be applied to the sampling at low values of the nested iterations. The local maxima in the ASEF remain rather stable even if the fit is recomputed. This is adequate for our purpose, since we aim at extracting the local maxima without the need to perform multiple islands peak bagging fits to obtain a stable solution. Another important advantage of the ASEF is that a local maximum in the sampling can be well reproduced even if the density of its sampling points is scarce as compared to that of the other maxima. Finally, the ASEF preserves the same scale and ordinate value as the nested iteration values obtained in the sampling, thus offering the possibility to make a direct comparison with the sampling distribution. In Fig.~\ref{fig:asef} we can see a comparison of these three different metrics for a pair of quadrupole and radial modes observed in a RGB star. Despite the peak on the right side being under-sampled with respect to the peak on the left side (in this case because it falls closer to the edge of the frequency boundary), the final ASEF has the same amplitude for both peaks. Conversely, the ASH and CH yield a much lower amplitude for the peak on the right side because they are based on the sampling counts, thus limiting the possible detection of the peak. In addition, the CH is more affected by noise structures in the sampling, hence it turns out to be less stable than the corresponding ASH.

\section{\famed\,\,pipeline}
The~\famed\,\,pipeline is a parallelized multiplatform free software aimed at performing the peak bagging analysis in low- and intermediate-mass stars that exhibit solar-like oscillations. This includes stars that range from the MS to RGs. This means that stars in the subgiant (SG) phase of stellar evolution, as well as stars evolved toward the RGB tip, stars settled in the Helium-burning MS (or red clump, RC), and stars starting to evolve from the RC toward the asymptotic giant branch (AGB), can also be analyzed with this pipeline. \famed\,\,extracts the oscillation mode properties of frequency, amplitude, and linewidth, and provides a mode identification and detection probability for each mode by means of a fast and fully automated procedure. \famed\,\,exploits a combination of multimodal and unimodal fitting using \diamonds, and utilizes the free software \texttt{GNUparallel}\footnote{\texttt{GNUparallel} is a shell tool that can be downloaded at https://www.gnu.org/software/parallel/.} \citep{Tange2011GNUparallel} to run the fits with parallel. 

\begin{figure}
   \centering
  \includegraphics[width=9.0cm]{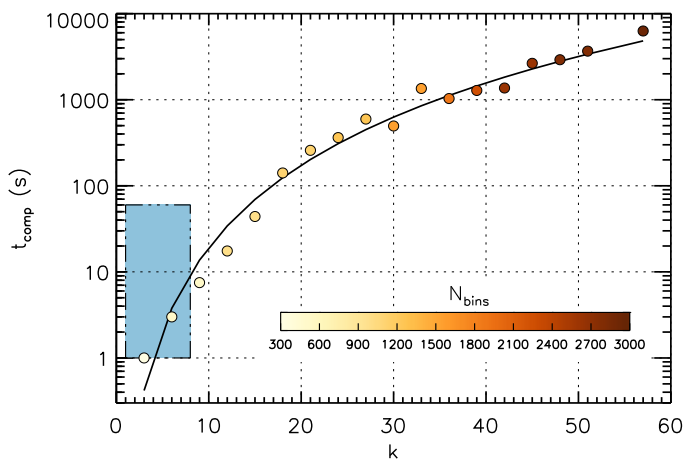}
    \caption{Computational time $t_\mathrm{comp}$ to run a fit with \diamonds\,\ as a function of the number of free parameters $k$ involved in the fit. The points correspond to the calibration performed by \cite{Corsaro18corr} on the star KIC~12008916 using a 2.7 GHz single-core CPU. Each point is a unimodal fit to the oscillation modes in the stellar PSD (see Sect.~\ref{sec:pb}). The color is the number of data bins involved in each computation. The blue box represents the region where \famed\,\,is operational when using \glob\,\,and \chu\,\,, exploiting a low-dimensional ($1 \le k \le 8$), hence fast computation ($1 \lesssim t_\mathrm{comp} \lesssim 60$\,s) with \diamonds.}
    \label{fig:time}
\end{figure}

\famed\,\,is constituted by modules executed in a sequential order from the most basic up to the most detailed in an onion-like structure. Although the pipeline currently accounts for four modules, the mandatory ones to produce usable outputs are the \glob\,\,and \chu\,\,modules (hereafter \glob\,\,and \chu\,\,for brevity). The \texttt{\'ECHELLE} and \texttt{COMPLETE} modules (hereafter \texttt{\'ECHELLE} and \texttt{COMPLETE} for brevity) are instead extra modules that the users may decide not to use, depending on their needs. Table~\ref{tab:famed} lists an estimate of the time required to accomplish each module and an overview of the outputs produced. In this work we focus on \glob\,\,and \chu\,\,only, while \texttt{\'ECHELLE} and \texttt{COMPLETE} will be presented in detail in a follow-up paper. Figure~\ref{fig:time} shows that the regime in which \glob\,\,and \chu\,\,are operating is low-dimensional through the use of multimodal fitting. \glob\,\,and \chu\,\,therefore constitute the major strength of the \famed\,\,pipeline in terms of processing speed\footnote{The \glob\,\,and \chu\,\,modules of \famed\,\,are currently available in IDL. A Python version is already under development. \famed\,\,can be downloaded from the public GitHub repository https://github.com/EnricoCorsaro/FAMED. Installing documentation and tutorials are provided.}.

\begin{table*}
\caption{Summary of the different modules of the \famed\,\,pipeline, with outputs and computational time indicated. The overall computational time is a reference time estimated by analyzing a four-years long \kepler dataset using \famed\,\,on a 2.6 GHz 6-core CPU. The overall computational time is subject to vary depending on the system configuration, the data quality, and the complexity of each individual application.}             
\centering                         
\begin{tabular}{l c c}       
\hline\hline
\\[-8pt]         
Module & Outputs & Overall computational time (min) \\ [1pt]
\hline
\\[-8pt]
\glob & \begin{tabular}{@{}c@{}} Low-precision and low-accuracy oscillation frequencies and 1-$\sigma$ uncertainties\\ $(n,\ell)$ mode identification, with $\ell = 0,1$\\Global $\Dnu_\mathrm{ACF}$, $\Dnu_0$, $\epsilon$, $\alpha_0$ \\ $N_\mathrm{chunks}$, frequency boundaries $s_n$ for each chunk \\ Flag for depressed dipole stars\end{tabular} &  $1.0$-$2.0$\\ [1pt] 
\hline
\\[-8pt]
\chu & \begin{tabular}{@{}c@{}} High-precision and high-accuracy oscillation frequencies and 1-$\sigma$ uncertainties\\ $(n,\ell)$ mode identification, with $\ell = 0,1,2,3$ \\  $m = 0, \pm1$ mode identification and $\cos i$ for peaks with detected rotation\\ Local $\epsilon$, $\delta\nu_\mathrm{02}$, $\Delta P_1$, $\Gamma_0$, $\Gamma_\mathrm{3}$ \\ Peak testing probabilities for detection, rotation, and duplicity  \\ Flags for peak blending and peak sinc$^2$ profile\\ Frequency ranges $r_{a,b}$ and divisions $d_{a,b}$ for each oscillation frequency \end{tabular}& $0.2$-$1.0$ per chunk\\ [1pt] 
\hline
\\[-8pt]
\texttt{\'ECHELLE} & \begin{tabular}{@{}c@{}}$\Dnu_{\ell = 0}$,  $\delta\nu_\mathrm{02}$,  $\delta\nu_\mathrm{01}$,  $\delta\nu_\mathrm{03}$, $\alpha_{\ell} $, $\beta_{0\ell}$, $\Delta \Pi_1$, $q$ \\  $m = 0,\pm1$ mode identification for all peaks in RGs and $\delta\nu_\mathrm{rot,core}$ for SG and RG \\ Evolutionary stage (RGB vs. RC) \\ Final validated list of oscillation modes \end{tabular}& $0.1$-$1.0$ \\ [1pt] 
\hline
\\[-8pt]
\texttt{COMPLETE} & \begin{tabular}{@{}c@{}}$\ell = 0, 1, 2, 3$ frequencies with highest precision and accuracy\\ Oscillation amplitudes and linewidths \\ $\cos i$ and $\delta\nu_\mathrm{rot}$ for MS only \\ Bayesian credible limits on all oscillation parameters\end{tabular} & $0.5$-$5.0$ per chunk \\ [1pt]
\hline
\end{tabular}
\label{tab:famed}
\end{table*}

\subsection{Estimating the background signal}
\label{sec:background}
\famed\,\,performs its analysis using only the original stellar PSD. While in some peak bagging applications one may decide to rely on a PSD normalized by a background level \citep[e.g.,][]{Buysschaert16}, \famed\,\,requires that the original signal, hence its statistics, are preserved. This is because \famed\,\,performs model hypothesis testing aimed at assessing the significance of the oscillation modes above the level of the background. For this procedure to be reliable one needs to incorporate the level of the background into the fitting models while keeping the statistics of the dataset unchanged (\citealt{Corsaro14}, C15). Therefore the user is required to estimate the background signal in the stellar PSD before proceeding with the actual peak bagging analysis with \famed.

As presented by C15; \cite{Corsaro17metallicity}, the background signal is typically modeled by a series of Harvey-like profiles and by the instrumental noise level (either white noise or both white and colored noise). To estimate the background, we recommend adopting the newest version of the Background code based on \diamonds\footnote{The latest version of the Background code extension for \diamonds\,\,is available at the public GitHub repository https://github.com/EnricoCorsaro/Background.}, which is a free software \citep[see also][]{Corsaro18tutorial}. In the Background code it is now possible to choose the background model  from a series of models that are commonly found in the literature \citep[e.g.,][]{Kallinger14}. Once the background fit is completed, the \famed\,\,pipeline will automatically read the resulting background outputs from \diamonds, and use the corresponding background level within its peak bagging analysis. It is also possible to feed in the background level estimated from a different fitting tool than the Background code. This will not affect the correct functioning of \famed, provided that the estimated background level is correctly computed (see the \famed\,\,GitHub repository for more details).

We caution that the background level adopted in the analysis should be quite accurate, especially in the region containing the oscillations. A poor background fit may hamper the reliability of the peak detection tests (Sect.~\ref{sec:peak_detection_test}) and the estimation of the peak amplitudes and linewidths (e.g., see the discussion in \citealt{Corsaro14}, Sect. 6.6).

\subsection{Asymptotic fitting code}
\label{sec:asymp}
Within the analysis done by \glob, \famed\,\,computes fits of the asymptotic relation of $p$ modes \citep{Tassoul80,Mosser11universal,Lund17LEGACY} by means of \diamonds. Following the recent developments by \cite{Lund17LEGACY}, we have implemented an asymptotic pattern for modes of angular degrees $\ell = 0,1,2,3$ that incorporates the curvature term on the large frequency separation $\Dnu$, which we refer to as $\alpha_\ell$, and on the small frequency spacings $\delta\nu_{0\ell}$, which we term $\beta_{0\ell}$. In particular, the generalized asymptotic relation adopted here takes the form 
\begin{equation}
\begin{split}
\nu_{n\ell} \simeq &\,\Dnu \left(n + \epsilon + \frac{\ell}{2} \right) + \\
& \frac{\alpha_{\ell} \Dnu}{2} \left( n - \frac{\numax}{\Dnu} \right)^2 - \\
& \beta_{0\ell} \Dnu \left( n - \frac{\numax}{\Dnu} \right) - \delta\nu_{0\ell} \, ,
\end{split}
\label{eq:asymp}
\end{equation}
where $\Dnu \equiv \Dnu_{\ell = 0}$, and we have explicitly removed the dependency of $\alpha_\ell$ and $\beta_{0\ell}$ on $\Dnu$, such that it is easier to set up their prior hyper-parameters for the fits. The asymptotic fits use a Gaussian likelihood that takes the radial orders $n$ as covariates, and as uncertainties those estimated on the observed frequencies, as obtained in Sect.~\ref{sec:freq_asef}. This is done by means of a new code extension of \diamonds, called Asymptotic, which has been made publicly available with the publication of this paper\footnote{The Asymptotic code extension for \diamonds\,\,is available at the public GitHub repository https://github.com/EnricoCorsaro/Asymptotic.}. The Asymptotic code allows $\epsilon$, $\alpha_{\ell}$, and $\beta_{0\ell}$ to be treated as either fixed or free parameters. For example, if the condition $\alpha_{\ell} = \beta_{0\ell} = 0$ is used, the curvature terms on $\Dnu$ and $\delta\nu_{0\ell}$ of the asymptotic relation are neglected during the fit. These features are useful within \glob\,\,and are discussed in Sect.~\ref{sec:global_modeid}. \texttt{\'ECHELLE} is instead heavily based on the use the Asymptotic code, as it will be presented in a follow-up paper. 

We note that the Asymptotic code does not incorporate the estimation of signal related to acoustic glitches because this is of small magnitude as compared to the level of uncertainties that are obtained in the  frequencies extracted from \glob\,\,and \chu. The glitch signal will be estimated from modules of the pipeline subsequent to \texttt{COMPLETE}.

\subsection{PeakBagging fitting code}
\label{sec:pb}
A large and fundamental part of the analysis done by \famed\,\,for \glob\,\,and \chu\,\,(and for \texttt{COMPLETE} too) is through the PeakBagging code extension of \diamonds. This code has already been made available and updated by \cite{Corsaro14}; C15; \cite{Corsaro18tutorial}, but here we release a new version of it that incorporates  additional features\footnote{The latest version of the PeakBagging code extension for \diamonds\,\,is available at the public GitHub repository https://github.com/EnricoCorsaro/PeakBagging.}. The new PeakBagging code can now be used in five configurations:
\begin{enumerate}
\item Multimodal fit: performs a multimodal fit using either one or two Lorentzian profiles;
\item Sliding-pattern fit: performs a multimodal fit using a mixture of Lorentzian profiles that are distributed in frequency according to the asymptotic pattern of $p$ modes;
\item Peak-testing fit: performs unimodal fits using a set of different peak testing models aimed at assessing the significance of the oscillation modes, the presence of blending between two adjacent modes, and the presence of rotational and duplicity effects;
\item Standard unimodal fit: performs the standard unimodal fit using a mixture of Lorentzian profiles;
\item Unimodal fit with rotation: performs the unimodal fit using a mixture of Lorentzian profiles with the rotational effect incorporated by using the additional free parameters of $\cos i$ and $\delta\nu_\mathrm{rot}$, for stellar spin-inclination angle and rotational splitting, respectively.
\end{enumerate}
\famed\,\,exploits a combination of all the five configurations provided by the new PeakBagging code. Configuration \#1 was already presented by C19, and further discussed in Sect.~\ref{sec:multi_modal} for the case of a single Lorentzian profile. Configuration \#1 with two Lorentzian profiles is presented in Sect.~\ref{sec:radial_quadrupole}. Configuration \#2 is described in Sect.~\ref{sec:sliding}, while configuration \#3 is addressed in Sects.~\ref{sec:peak_detection_test} and \ref{sec:peak_rotation_test}. Configurations \#4 and \#5 will be discussed in a follow-up paper as they are used in \texttt{COMPLETE}. A general description and application of configuration \#4 can already be found in \cite{Corsaro14} and in C15.

\section{The \glob\,\,module}
\label{sec:global}

\glob\,\,is the most basic but most important module of \famed. If \glob\,\,is not properly performed it will bias the entire analysis done by the pipeline. However, \glob\,\,is fast to be computed and the user can manually intervene to apply any relevant correction through the use of the input configuring parameters of \famed. This means that in the process of verifying the results, if needed, it is quite straightforward to repeat the analysis from the beginning.

The aim of \glob\,\,is to give an approximate frequency position of the oscillation modes by distinguishing between $\ell = 0$ and as $\ell = 1$ modes. This includes locating the frequency position of each chunk of PSD containing a pair of $\ell = 1,0$ modes. For evolved stars, showing $\ell = 1$ mixed modes, \glob\,\,focuses on $\ell = 0$ modes only. Executing \glob\,\,in a correct manner requires that the asymptotic terms $\Dnu$ and $\epsilon$ are properly estimated. To accomplish this task, \glob\,\,relies on two main parts, the multimodal sampling and its subsequent analysis.

\subsection{Global multimodal sampling}
\label{sec:sampling_global}
In the first part of \glob, the multimodal fit with \diamonds\,\,is performed using configuration \#1 of the PeakBagging code by means of a single Lorentzian profile (see Sect.~\ref{sec:pb}), over a frequency range that is centered around $\numax$, as obtained from the background fit. The frequency range covers about $4$-$4.5 \sigma_\mathrm{env}$ on each side of $\numax$ for MS, SG, and RG stars having $\Dnu > \Dnu_\mathrm{CL}$, $2.5 \sigma_\mathrm{env}$ for RGs having $\Dnu_\mathrm{tip} < \Dnu \leq \Dnu_\mathrm{CL}$, and $1.2 \sigma_\mathrm{env}$ for RGs having $\Dnu \leq \Dnu_\mathrm{tip}$. Here $\sigma_\mathrm{env}$ is the standard deviation of the Gaussian envelope of the oscillations that was obtained from the background fit, while $\Dnu_\mathrm{tip} = 3.2\,\mu$Hz and $\Dnu_\mathrm{CL} = 9\,\mu$Hz are set as configuring parameters of \famed\,\,and are based on the evolutionary stage classifications presented by \cite{Kallinger12} (see Table~\ref{tab:configuring}). This choice ensures that the covered frequency range is adequate to incorporate all the significant modes. In this initial part the large frequency separation $\Dnu$ is evaluated from the $\numax$-$\Dnu$ relation by \cite{Huber11} (see Sect.~\ref{sec:dnu_acf} for more details).

\begin{figure*}[t]
   \centering
  \includegraphics[width=18.5cm]{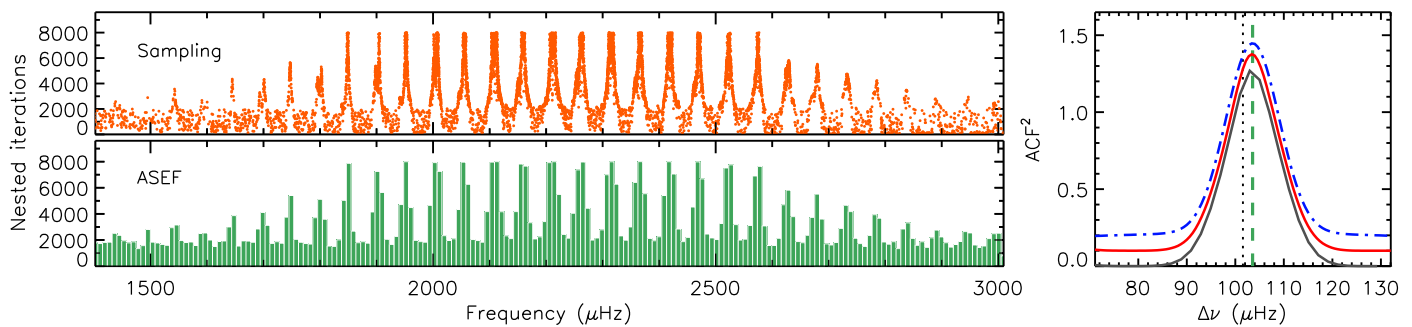}
    \caption{Estimating $\Dnu_\mathrm{ACF}$ in \glob\,\,for the G-type MS star KIC~12069424. \textit{Top-left panel}: global multimodal fit with \diamonds, with a regular structure in frequency resembling the asymptotic pattern of $p$ modes. \textit{Bottom-left panel}: the resulting ASEF, here in standard resolution for plotting purposes, containing $31$ local maxima. \textit{Right panel}: ACF$^2$ (computed at the frequency resolution of the PSD) of the high-resolution ASEF (solid black line), with the interpolated ACF$^2$ using a resolution higher than the PSD (about 100 bins in total, solid red line), and the final Gaussian fit to the interpolated ACF$^2$ (dot-dashed blue line). The red and blue curves have been offset by +0.1 and +0.2 in amplitude, respectively, for a clear visualization. The vertical dotted line is $\Dnu$ from the $\numax$-$\Dnu$ relation, while the vertical dashed green line is the final $\Dnu_\mathrm{ACF}$.}
    \label{fig:acf}
\end{figure*}

For performing the multimodal fit in \glob\,\,the FWHM of the Lorentzian profile from Eq.~(\ref{eq:isla}) is fixed to a value obtained using one of two regimes, both requiring an input value of $\teff$: \textit{(i)} for $\numax > \nu_\mathrm{max,thresh}$: $\Gamma$-$\numax$-$T_\mathrm{eff}$ relation calibrated by \cite{Ball18TESS}; \textit{(ii)} for $\numax \leq \nu_\mathrm{max,thresh}$: $\Gamma$-$\teff$ relations calibrated by C15.
with $\numax$ given as input from the background fit (see also Table~\ref{tab:configuring}). The adopted linewidth relations provide an estimate of $\Gamma$ for a radial mode. These relations will be updated from time to time in order to provide more accurate predictions of the oscillation linewidths for a wider range of input $\numax$ and $\teff$. We have tested that, in general, the entire analysis is not affected even if the input $\teff$ is off by up to about 200-300\,K. For stars evolved toward the tip of the RGB and toward the AGB, having $\Dnu \leq \Dnu_\mathrm{tip}$, the FWHM of the islands peak bagging model is however further reduced by a factor 10 because of the very narrow peaks that are found in these cool targets.

This choice of $\Gamma$ ensures that the oscillation envelope is sampled at a resolution that is sufficiently high to distinguish $\ell = 2,0$ mode pairs from $\ell = 1$ modes in stars from MS to RGs, but sufficiently low to avoid resolving the individual mixed modes in more evolved stars. This is done because here our purpose is mainly to distinguish between $\ell = 2,0$ pairs (with each pair approximated by a single $\ell = 0$ peak), and $\ell = 1$ peaks. In the case of stars with mixed modes, each $\ell = 1$ peak extracted from \glob\,\,is approximating the entire (or part of the) region of mixed modes between two adjacent $\ell = 0$ peaks. The probing power of the global multimodal sampling is already apparent from our application to a MS star (Fig.~\ref{fig:acf}). We refer to this first multimodal sampling as the global multimodal sampling because it is performed for \glob.

The strength in performing a multimodal sampling over the PSD is that, using a very simple and easy setup of the fitting model, it is able to recognize the presence of oscillation peaks without imposing any asymptotic pattern. This not only allows fitting stars in any evolutionary stage without the need to have a preliminary classification of their evolutionary stage, but to also automatically understand what kind of oscillation pattern we are dealing with based on how the multimodal sampling is distributed over the frequency range. 

\begin{table}
\caption{Main configuring parameters of \famed\,\,that are adopted for the peak bagging applications presented in this work. The left column shows the parameter name, while the right column lists its value. Each of these parameters can be tuned by the user from an input configuring parameter file.}             
\centering                         
\begin{tabular}{l c}
\hline\hline
\\[-8pt]         
Configuring parameter & Value\\ [1pt]
\hline
\\[-8pt]
$\nu_\mathrm{max,thresh}$ & $300\,\mu$Hz\\[1pt]
$\Dnu_\mathrm{thresh}$ & $30\,\mu$Hz\\[1pt]
$\Dnu_\mathrm{tip}$ & $3.2\,\mu$Hz \\[1pt]
$\Dnu_\mathrm{CL}$ & $9\,\mu$Hz \\[1pt]
$\Dnu_\mathrm{RG}$ & $15\,\mu$Hz \\[1pt]
$\Dnu_\mathrm{SG}$ & $90\,\mu$Hz \\[1pt]
$T_\mathrm{eff,SG}$ & $6350$\,K\\[1pt]
\hline
\end{tabular}
\label{tab:configuring}
\end{table}

\subsection{Large frequency separation $\Dnu_\mathrm{ACF}$}
\label{sec:dnu_acf}
The global multimodal sampling can now be used to measure $\Dnu$. This parameter is of critical to analyze the oscillation features of the star because it is directly related to the stellar mean density \citep{Ulrich86}, and because it defines the characteristic frequency spacing of $p$ modes in the asymptotic pattern \citep{Tassoul80}. The value of $\Dnu$ is initially estimated by means of the $\numax$-$\Dnu$ relation calibrated by \cite{Huber11} and it is used as a raw guess. The multimodal sampling is then converted into an ASEF, following the description presented in Sect.~\ref{sec:asef}. The ASEF is first computed at high resolution, having $N_\mathrm{bins} = 800$. A squared auto-correlation function (ACF$^2$) around the raw $\Dnu$ guess is computed from the high-resolution ASEF. The ACF$^2$ is then interpolated at a higher resolution and a Gaussian is fit to it\footnote{The Gaussian fit to the interpolated ACF$^2$ incorporates a background term which is modeled by a second-degree polynomial. This allows obtaining a more stable and reliable fit even if the ACF$^2$ signal is more contaminated by noise.}, with the centroid constituting our $\Dnu_\mathrm{ACF}$. The right panel of Fig.~\ref{fig:acf} depicts an example of what was just described for a MS star, where the ACF$^2$ of the ASEF shows a very clear and smooth peak. This can be explained by the highly regular structures that are already visible in the multimodal sampling obtained by \diamonds\,\,(Fig.~\ref{fig:acf} top-left panel) and in the obtained ASEF (visible in the bottom-left panel of the same figure). The advantage of computing $\Dnu_\mathrm{ACF}$ from the high-resolution ASEF is that it turns out to be a reliable estimate of $\Dnu$ of the star (often accurate to well below 1\,\%), and that it is measurable both for noisy datasets and for evolved stars that have a complicated oscillation mode pattern. This is because when computing the ASEF, one gets rid of the stochastic noise that is overlaid on the Lorentzian shape of each oscillation peak observed in the stellar PSD. As a result, using the ASEF yields a cleaner ACF$^2$ signal than in the case of using the PSD itself.

\subsection{Extracting oscillation frequencies and uncertainties}
\label{sec:freq_asef}
After measuring $\Dnu_\mathrm{ACF}$, the ASEF is recomputed at standard resolution, which here is typically on the order of $N_\mathrm{bins} \approx 200$ for MS stars, and of $N_\mathrm{bins} \approx 100$ for RGs due to the smaller frequency range of their oscillation envelope. 

\subsubsection{Hill-climbing algorithm}
With the standard-resolution ASEF in hand, following what was explained by C19, we adopt a hill-climbing algorithm to locate each local maximum in frequency. The hill-climbing algorithm proceeds from the left bound of the frequency range of the stellar PSD and moves gradually toward the right bound until a local maximum is found. This is done by confronting the value of the ASEF at each bin to those of the neighboring bins. A local maximum is thus characterized by having an ASEF value that is larger than that of its neighboring bins (both on the left and right side), with a total amount of increase in the ASEF during the rising phase of the local maximum that has to exceed a user-defined input threshold. The input threshold is provided in percentage of the global maximum ASEF value. We have found that an input threshold of 1\,\% is adequate for all applications involving \glob. Once a new local maximum is found, this process is repeated until the entire frequency range of the stellar PSD has been inspected.

\subsubsection{Frequency divisions and ranges}
\label{sec:ranges}
The extracted local maxima from the ASEF represent our raw set of frequencies that are used to obtain the final list of oscillation frequencies for \glob. The next step consists of obtaining the so-called frequency ranges, $r_{a,b}$, and divisions, $d_{a,b}$, for each local maximum found in the ASEF. Both $r_{a,b}$ and $d_{a,b}$ are pairs of values (in frequency units) that define two different kinds of boundary around each local maximum, with the subscript $a$ referring to the lower edge, and $b$ to the upper edge of the boundary. The divisions are the midpoints between two adjacent local maxima. The determination of the frequency ranges is instead more complicated, and it is initially done in two steps. Firstly, for a given local maximum, both left and right ranges are initially defined such that they correspond to the point on either side of the local maximum peak where the ASEF stops decreasing. The ranges so defined often provide a narrower boundary than that set by the divisions. In any case, the ranges cannot exceed the divisions.
Secondly, using the ranges so computed, the sampling from \diamonds\,\,falling within each $r_{a,b}$ boundary is used to compute a weighted mean frequency value and a corresponding weighted 1-$\sigma$ uncertainty. This is done using the squared value of the nested iteration of each sampling point as a weight (see the description by C19, Figure 2B). These uncertainties and frequencies are then used to re-adjust $r_{a,b}$ such that the range on each side of the frequency peak is falling between 1-$\sigma$ and 2-$\sigma$ uncertainty of the peak. The frequency and its uncertainty are then recomputed once again. This process is iterated twice to provide a range more closely centered around the peak, hence a more realistic uncertainty and accurate value of the final frequency. 

In this way, divisions, ranges, frequency estimates and their uncertainties are all extracted from the sampling and can be used for the subsequent analysis. The advantage of using frequencies and uncertainties estimated from the sampling is that they correspond to estimates computed from an actual fit to the PSD. For the remainder of this paper we refer to these frequencies and their uncertainties as $\nu_{f,i}$ and $\sigma_{f,i}$. The frequency uncertainties obtained here are used in the asymptotic fits done in \glob\,\,using the Asymptotic code extension of \diamonds\,\,(see Sect.~\ref{sec:skimming}).

\begin{figure}[h]
   \centering
  \includegraphics[width=8.8cm]{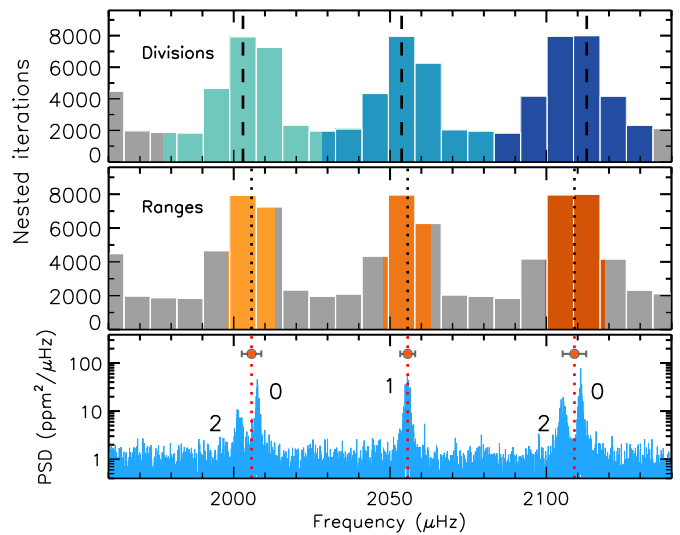}
    \caption{Evaluation of frequency divisions and ranges, and of frequency estimates and uncertainties, in a close-up view of the standard-resolution ASEF from \glob\,\,for the MS star KIC~12069424. \textit{Top panel}: frequency divisions $d_{a,b}$ (blue-tone shading) computed from three local maxima $\nu_i$ (vertical dashed lines). \textit{Middle panel}: frequency ranges $r_{a,b}$ (orange-tone shading) with overlaid the frequencies $\nu_{f,i}$ estimated from the sampling (vertical dotted lines). \textit{Bottom panel}: stellar PSD, with overlaid the frequencies $\nu_{f,i}$ (red bullets and vertical dotted lines) and corresponding 1-$\sigma$ uncertainties $\sigma_{f,i}$.}
    \label{fig:range}
\end{figure}

The difference between divisions and ranges is depicted in Fig.~\ref{fig:range}. As it is evident, the ranges are significantly smaller than the divisions but both ranges and divisions are useful for different applications within the analysis performed by \famed. This is especially true for \chu, as we discuss in Sect.~\ref{sec:chunk}. From the same figure, we can see that the local maxima $\nu_i$ and the estimated frequencies $\nu_{f,i}$ are in general different from one another, the latter being more accurate by definition. In addition, the local maxima do not come with any uncertainty, as this one is only a quantity estimated afterwards from the sampling itself once that the ranges are computed.

In the application of \glob\,\,the peaks corresponding to $\ell = 2$ modes are not resolved (see Sect.~\ref{sec:sampling_global}). As a result, the frequency $\nu_{f,i}$ covering an $\ell = 2,0$ mode doublet is biased because it falls roughly halfway between the two power peaks of the doublet, as visible in the bottom panel of Fig.~\ref{fig:range}. This bias is initially generated on purpose to obtain a general mode identification with high computational efficiency, but it is removed in a later stage, as explained in Sect.~\ref{sec:skimming}.

\subsection{Asymptotic phase term $\epsilon$}
\label{sec:epsilon}
At this stage, we have a list of frequencies $\nu_{f,i}$ comprising frequencies that correspond to $\ell = 2,0$ mode pairs, which for simplicity we refer to as $\ell  = 0$ or radial modes within \glob, and frequencies that correspond to $\ell = 1$ (or dipole) mode peaks.

To distinguish between $\ell = 0$ and $\ell = 1$ mode peaks in the stellar PSD it is necessary to have an estimate of the phase term $\epsilon$ of the asymptotic relation \citep{CD84}. As shown by \cite{White11}, in an \'echelle diagram $\epsilon$ sets the absolute position of each radial mode frequency. However, $\epsilon$ is not straightforward to obtain, especially during the MS and core-Helium-burning phase of stellar evolution, where it can be subject to large variation depending on the fundamental stellar properties \citep[e.g.,][]{White11}, as well as on the thermodynamic state of the stellar convective envelope \citep{Kallinger12,CD14}. For our purposes it is useful to consider that in the regime of hot stars (F-type) $\epsilon$ follows a relatively tight relation with $\teff$, as shown by \cite{White11} and by \cite{Lund17LEGACY}, while for RGB stars $\epsilon$ can be reliably predicted by the $\epsilon$-$\Dnu$ relation \citep{Mosser11universal,Kallinger12,Corsaro12cluster}. For solar-like oscillating stars in any evolutionary stage, one can compute $\epsilon$ by adopting the relation
\begin{equation}
\epsilon = \frac{\nu_0 \, \mbox{mod} \, \Dnu}{\Delta\nu}
\label{eq:epsi_sliding}
\end{equation}
which follows from evaluating the asymptotic pattern of radial modes in an \'echelle diagram. Here $\nu_0$ is the frequency of the central radial mode of the star (the closest to $\numax$). Given that $\Dnu$ is already known with good accuracy, computing $\epsilon$ requires that $\nu_0$ is properly evaluated. The phase term $\epsilon$ uniquely sets the mode identification for each oscillation mode.

\subsubsection{Sliding-pattern fit}
\label{sec:sliding}
\begin{figure}[tb]
   \centering
  \includegraphics[width=9.cm]{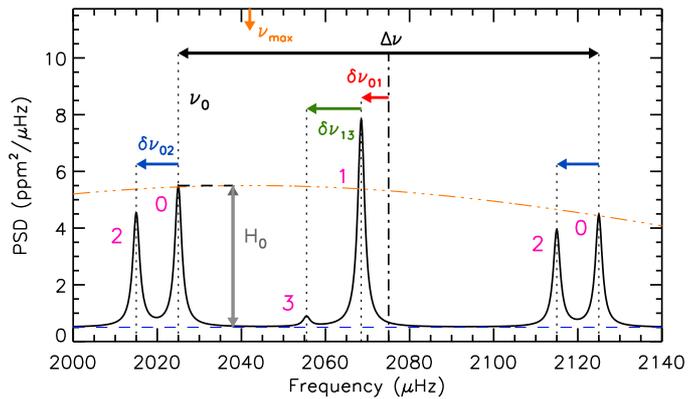}
    \caption{Sketch of a sliding-pattern model covering two consecutive radial modes, with angular degrees indicated. The parameter $\nu_0$ gives the location of the reference radial mode, while $\Dnu$, $\delta\nu_{02}$, $\delta\nu_{01}$, $\delta\nu_{13}$ are indicated by arrows. The vertical dotted lines mark the oscillation mode frequency, while the vertical dot-dashed line is the midpoint between the two radial mode frequencies of the pattern. The parameter $H_0$ is indicated by a vertical gray arrow. The Gaussian envelope $\mathcal{G} \left( \nu \right)$, rescaled by $H_0$, is shown by the double dot-dashed orange line, with its centroid $\numax$ indicated by the downward orange arrow. The horizontal dashed blue line is the background level estimated from the background fit. For illustration purposes, the rotational effect is not included (i.e., $\delta\nu_\mathrm{rot} = 0$ or $\cos i = 1$). }
    \label{fig:sliding}
\end{figure}

Locating the central radial mode of the star is in general not straightforward because its position is related to the phase term $\epsilon$, which has the effect of introducing a frequency offset in the asymptotic pattern that is given by the quantity $\epsilon \Dnu$. With this aim, we have developed the configuration \#2 of the PeakBagging code (see Sect.~\ref{sec:pb}), which allows fitting a so-called sliding pattern to the stellar PSD. The sliding pattern is a model defined as
\begin{equation}
\mathcal{P}_\mathrm{sliding} \left( \nu \right) = \mathcal{G} \left( \nu \right) R \left( \nu \right) \left[ \mathcal{P}_{03} \left( \nu \right) + \mathcal{P}_{12} \left( \nu \right) \right]  + \overline{B} \left( \nu \right)\, ,
\label{eq:sliding}
\end{equation}
where $R (\nu)$ is the squared response function given by the sampling of the dataset (see \citealt{Chaplin11}), and $\overline{B} \left( \nu \right)$ is the background level (fixed and without the Gaussian envelope) estimated from the background fit, and already incorporating the correction by the response function. The term $\mathcal{G}(\nu)$ is the unitary-height Gaussian used to modulate the peak heights in the sliding pattern according to $\numax$ and the Gaussian envelope standard deviation estimated from the background fit. The terms $\mathcal{P}_{03} \left( \nu \right)$ and $\mathcal{P}_{12} \left( \nu \right)$ are defined as
\begin{equation}
\mathcal{P}_\mathrm{03} \left( \nu \right) = \sum_{i=-(N_\mathrm{ord} - 1)/2}^{(N_\mathrm{ord} - 1)/2} \left[ \mathcal{L}^i_0 \left( \nu \right)  + \mathcal{L}^i_3 \left( \nu \right) \right] \, ,
\end{equation}
and
\begin{equation}
\mathcal{P}_\mathrm{12} \left( \nu \right) =  \sum_{i=-(N_\mathrm{ord} - 1)/2}^{(N_\mathrm{ord} - 1)/2} \sum_{m=-\ell}^{\ell} \left[ \mathcal{L}^i_{1m} \left( \nu \right) + \mathcal{L}^i_{2m} \left( \nu \right) \right] \, ,
\end{equation}
while
\begin{equation}
\mathcal{L}^i_0 \left( \nu \right) = \frac{H_0}{1 + \frac{4}{\Gamma^2_0} \left(\nu - \nu_0 - i \Dnu \right)^2} \quad ,
\end{equation}
\begin{equation}
\mathcal{L}^i_{1m} \left( \nu \right) = \frac{H_0 V^2_1 \xi_\mathrm{\ell m} (\cos i)}{1 + \frac{4}{\Gamma^2_1} \left( \nu - \nu_0 - i \Dnu - \Dnu/2 + \delta\nu_{01}  - m \delta\nu_\mathrm{rot} \right)^2} \quad ,
\end{equation}
\begin{equation}
\mathcal{L}^i_{2m} \left( \nu \right) = \frac{H_0 V^2_2 \xi_\mathrm{\ell m} (\cos i)}{1 + \frac{4}{\Gamma^2_0} \left( \nu - \nu_0 - i \Dnu + \delta\nu_{02} - m \delta\nu_\mathrm{rot} \right)^2} \quad ,
\end{equation}
\begin{equation}
\mathcal{L}^i_{3} \left( \nu \right) = \frac{H_0 V^2_3}{1 + \frac{4}{\Gamma^2_0} \left( \nu - \nu_0 - i \Dnu - \Dnu/2 + \delta\nu_{01} + \delta\nu_{13} \right)^2}  \quad ,
\end{equation}
are the Lorentzian profiles for $\ell = 0,1,2,3$ modes, respectively, with the additional rotational effect incorporated only for the modes $\ell = 1$ and 2 through the term $\xi_{\ell m} (\cos i)$ \citep{Gizon03}. This sliding pattern is tuned by up to eight free parameters, namely the already introduced $\nu_0$, the corresponding radial mode height $H_0$, $\Dnu$, the small frequency spacings $\delta\nu_{02}$, $\delta\nu_{01}$, $\delta\nu_{13}$, the rotational splitting $\delta\nu_\mathrm{rot}$, and the inclination angle in the form of $\cos i$. The linewidth of the central radial mode, $\Gamma_0$, is instead fixed to a value predicted by empirical relations (see Sect.~\ref{sec:sampling_global}). For dipole modes we adopt the linewidth $\Gamma_1 \equiv \Gamma_0 \eta_1$, which is thus related to $\Gamma_0$ through an input parameter $\eta_1$, termed dipole linewidth magnification factor. The term $\eta_1$, the number of radial orders to compute the sliding-pattern model, $N_\mathrm{ord}$, as well as the mode visibilities $V^2_{\ell}$, are all fixed numbers. An example of a sliding-pattern model is shown in Fig.~\ref{fig:sliding}.

The main free parameter is $\nu_0$, which controls the sliding of the entire pattern over the allowed frequency range. Adopting small frequency spacings to locate the other modes of the pattern is essential {{to make the central radial mode a reference mode}}. This in turn creates a multimodal fitting problem. Similarly to the multimodal fit performed in configuration \#1, the sliding-pattern fit utilizes a high threshold as a stopping condition for \diamonds\,\,(see also the discussion in C19).

The novelty of the sliding-pattern model is that it is highly configurable, where any of the free parameters ($\Dnu$, $\delta\nu_{02}$, $\delta\nu_{01}$, $\delta\nu_{13}$, $\delta\nu_\mathrm{rot}$, $\cos i$) can be arbitrarily fixed in order to either set any of them to a constant value or even completely remove one or more of them from the computation. For example, one could decide to fix the free parameter $\delta\nu_{01}$ to a constant value, meaning that the position of the dipole modes in the pattern is fixed with respect to that of the radial modes. Alternatively, it is possible to set $\delta\nu_{01}$ to a sufficiently large value (e.g., 99) in order to have the term $\mathcal{L}^i_{1m}$ completely removed from the pattern, which would therefore account for $\ell = 0,2,3$ modes only. We note that in this formulation the $\ell = 3$ modes do not incorporate the effect of rotation because of their low visibility.

Thanks to its flexibility, the sliding-pattern model can be adapted according to a preliminary identification of the evolutionary stage of the star. In particular: \textit{(i)} for MS and early SG stars ($\Dnu_\mathrm{ACF} > \Dnu_\mathrm{thresh}$), we adopt $V^2_1 = 1.5$, $V^2_2 = 0.62$, and $V^2_3 = 0.07$ obtained by \cite{Lund17LEGACY}, $\eta_1 = 1$, and $N_\mathrm{ord} = 7$ to adequately cover the large oscillation envelope (Sect.~\ref{sec:sliding_ms});
\textit{(ii)} for late SG and RG stars ($\Dnu_\mathrm{ACF} \leq \Dnu_\mathrm{thresh}$), we adopt $V^2_1 = 0.7$, $V^2_2 = 0.8$, and $N_\mathrm{ord} = 3$ because of the smaller frequency range of the oscillations as compared to less evolved stars, with $\eta_1 = 5$ in late RGB stars (Sect.~\ref{sec:sliding_rg}); \textit{(iii)} for depressed dipole stars, we adopt $V_1^2 = 0$, $V_2^2 = 0.62$ (Sect.~\ref{sec:sliding_depressed}).
We note that $\Dnu_\mathrm{thresh}$ has been tuned to $30\,\mu$Hz (Table~\ref{tab:configuring}). The mode visibilities adopted for evolved stars, $V_1^2$ and $V_2^2$, as well as the large value of $\eta_1$ for late RGB stars, are not to be intended as physically meaningful values but only as quantities that were calibrated for obtaining a more stable solution from the sliding-pattern fit. We remark that the evolutionary stages discussed in the following Sect.~\ref{sec:sliding_ms}, \ref{sec:sliding_rg}, and \ref{sec:sliding_depressed}, as identified by the pipeline, are not to be considered as either definitive or necessarily very accurate. These classifications have the main purpose of allowing the pipeline to choose a more optimized path for performing the subsequent analyses (see also the discussion in Sect.~\ref{sec:skimming}).

\begin{figure*}[t]
   \centering
  \includegraphics[width=18.5cm]{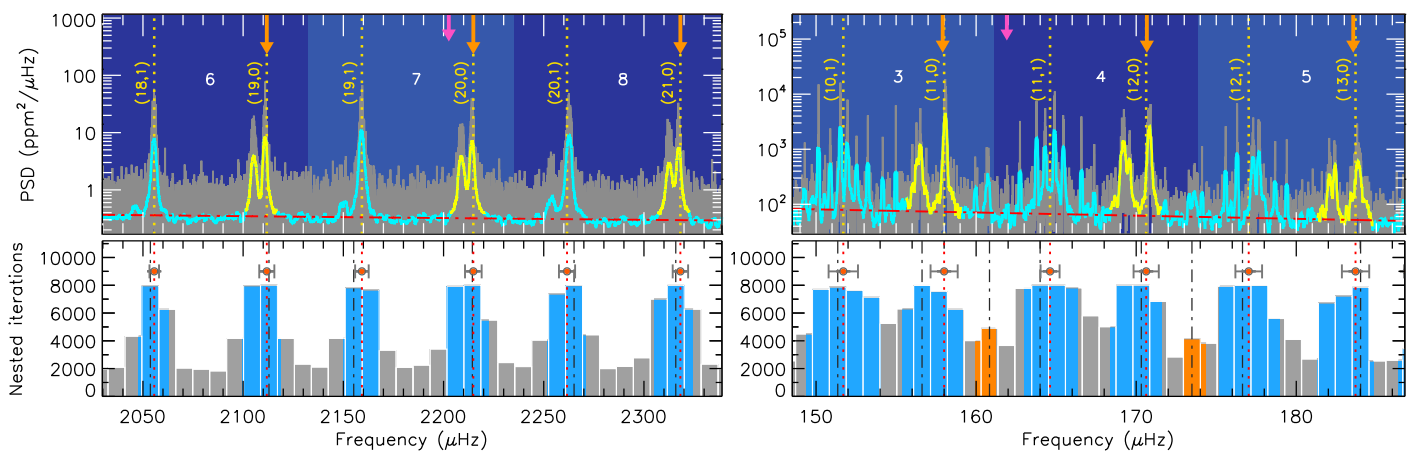}
    \caption{General mode identification, oscillation frequencies, and chunk division of \glob. Left plots are for the MS star KIC~12069424, while right plots are for the low-luminosity RGB star KIC~12008916. \textit{Top panels}: stellar PSD (dark gray) in logarithmic scale with a smoothing proportional to $\Gamma_\mathrm{global}$ overlaid (thick cyan curve, with yellow chunks representing $\ell = 2,0$ pairs). The level of the background estimated is shown by a horizontal dot-dashed red line. The vertical dotted yellow lines mark the frequencies $\nu_{f,i}$ resulting from Sect~\ref{sec:skimming}. The general mode identification $(n,\ell)$ (see Sect.~\ref{sec:global_modeid}) is indicated next to each frequency position. The orange downward-pointing arrows mark the asymptotic position of the radial modes, while the purple one shows the position of $\nu_\mathrm{max}$. The blue-shaded background (using alternate tonalities) indicates the different PSD chunks identified, with the individual chunk number reported. \textit{Bottom panels}: ASEF showing the position of the local maxima $\nu_i$ (vertical dashed black lines). The blue-shaded regions mark $r_{a,b}$ around each local maximum. The estimated frequencies $\nu_{f,i}$ shown in the top panels are indicated by vertical dotted red lines and red bullets, with their corresponding 1-$\sigma$ uncertainties $\sigma_{f,i}$. The orange-shaded regions mark $r_{a,b}$ for those local maxima corresponding to the frequencies $\nu_{f,i}$ that were discarded during the skimming process.}
    \label{fig:global}
\end{figure*}

\subsubsection{Main-sequence and early subgiant stars}
\label{sec:sliding_ms}
The oscillations in MS stars are characterized by a very regular pattern of $p$ modes. As the star evolves into an early SG, the $\ell = 2,0$ mode pairs remain regularly spaced, but some of the $\ell = 1$ modes start to undergo avoided crossings, thus becoming mixed modes \citep[with up to three mixed modes per radial order, in some cases, e.g.,][]{Benomar13}. For automatically distinguishing between the two regimes (MS vs. early SG), \famed\,\,analyzes the regularity of the frequencies $\nu_{f,i}$. The first step is to divide the frequencies falling within $\numax \pm 2 \Dnu_\mathrm{ACF}$ in two groups of odd and even frequencies. In the second step, for each frequency considered, the frequency modulo $\Dnu_\mathrm{ACF}$ is computed, and a median value for each group is obtained. In the third step, \famed\,\,finds the maximum deviation (in units of $\Dnu_\mathrm{ACF}$) between the frequency modulo $\Dnu_\mathrm{ACF}$ and the median value in each of the two groups. If this deviation is larger than an input threshold, which we set to 6\,\% of $\Dnu_\mathrm{ACF}$ for our applications, the star is assumed to be an early SG. This approach is quite reliable because in the central regions of the oscillation envelope we have the highest signal-to-noise, and because the frequencies $\nu_{f,i}$ in this region are typically marking alternate positions of $\ell = 0$ and $\ell = 1$ mode peaks. For improving the chances to correctly classify stars that may contain two or three mixed modes per radial order, the procedure described above is performed in an analogous manner by dividing the set of estimated frequencies in three groups instead of two. An example of the result obtained from \glob\,\,by applying such classification procedure is depicted in Fig.~\ref{fig:global} (left panels), where the pipeline was capable of correctly classifying the star as a MS by assessing the regularity of its $\ell = 1$ modes.

Once that the star has been assigned with a preliminary classification according to the scheme presented above, \famed\,\,proceeds with the setting up of the sliding-pattern model. For early SGs the sliding-pattern model: \textit{i)} utilizes a $\Dnu$ fixed to the value of $\Dnu_\mathrm{ACF}$; \textit{ii)} does not incorporate any $\ell = 1$ and $3$ peaks; \textit{iii)} does not account for rotation. These choices are motivated by the confusion created by the $\ell = 1$ mixed modes in the spectrum, having linewidths comparable to those of the radial modes and happening to fall close to, or even overlap with, the neighboring radial modes. Rotation and $\ell = 3$ peaks can be removed because they cannot be used to discriminate between $\ell = 1$ and $\ell = 0$ modes since the former ones are not included in the model. Moreover, it is necessary to fit the $\ell = 2$ modes through a varying $\delta\nu_{02}$ because this parameter can vary significantly from star to star in this regime \citep[see][]{White11}.

If the star is assumed to be a MS, despite the regular asymptotic pattern of $p$ modes, we can find complications caused by rotation (if resolved by the observations), and by the large mode linewidths. A potential source of bias in the sliding-pattern fit could be when the star has a high spin-inclination angle, $\cos i = 0$. Here the dipole modes are split into doublets by rotation ($m = \pm 1$ components), which could be confused with the adjacent $\ell = 2,0$ mode pairs. Including rotation on the $\ell = 1, 2$ modes, and an $\ell = 3$ peak too helps in getting rid of this potential degeneracy. On top of this, fitting a $\Dnu$ further improves the stability of the fit even if the signal-to-noise ratio (S/N) is low. This can be explained by the wide frequency range covered by the sliding pattern (seven radial orders), over which $\Dnu$ could be subject to an appreciable variation. The $\delta\nu_{02}$ parameter is also fit by allowing it to vary up to $\Dnu_\mathrm{ACF}/4$. In conclusion, the sliding-pattern for MS stars is fit using the largest set of free parameters available for this model. 

When MS stars are hot (F-type), with $\teff \geq T_\mathrm{eff,SG}\,$ (see Table~\ref{tab:configuring} for the adopted value), even the most flexible sliding pattern may still provide a wrong result if used with low S/N conditions. This is caused by the large linewidths of hot stars, which in turn generate important blending effects on the oscillation peaks. A strong peak blending can prevent us from unambiguously distinguishing between a $\ell = 2,0$ mode pair and a $\ell = 1$ mode peak (especially if split by rotation), which may look identical even to the most experienced eye (see Fig.~\ref{fig:hot_global}). To overcome this, \famed\,\,performs an additional check by comparing the $\epsilon$ value computed through Eq.~(\ref{eq:epsi_sliding}) using $\nu_0$ from the sliding-pattern fit, with $\epsilon$ obtained from the $\epsilon$-$\teff$ correlation \citep{Lund17LEGACY}. If the mode identification calculated from the two values of $\epsilon$ coincides, then the sliding-pattern fit is validated, otherwise the value of $\epsilon$ from the $\epsilon$-$\teff$ relation is taken as the true $\epsilon$ of the star.

\subsubsection{Late subgiant and red-giant stars}
\label{sec:sliding_rg}
When a star has evolved into a late SG and RG, its oscillation pattern still presents $\ell = 2,0$ modes regularly spaced by $\Dnu$, but at the same time it is populated by many dipolar mixed modes (with at least three or four of them in each radial order). Albeit the oscillation pattern is clearly more complicated than in MS and early SG stars, the advantage of having a high density of mixed modes is that one can still ignore the presence of $\ell = 1$ peaks in the pattern. This means that $\ell = 3$ modes and rotation are ignored as well. Another advantage of such evolved stars is that the frequency position of the $\ell = 2$ mode peaks follows a tight relation with $\Dnu$ \citep{Montalban10}, meaning that $\delta\nu_{02}$ can be fixed to the asymptotic value given by the $\delta\nu_{02}$-$\Dnu$ relation, for instance the one calibrated by \cite{Corsaro12cluster}. However, the curvature effects of the asymptotic pattern are in general more pronounced than in less evolved SGs, which calls for the need of fitting $\Dnu$. 

When the RG stars approach the RGB tip the oscillation pattern changes once again, a condition this one that we identify if $\Dnu_\mathrm{ACF} < \Dnu_\mathrm{tip}$. In this subclass of evolutionary stage, the number of dipole mixed modes that are observable in the PSD reduces to a single peak. The sliding-pattern model used to fit these stars therefore incorporates a $\ell = 1$ peak that, contrary to the case of MS stars, is set to have a large linewidth (with $\eta_1 = 5$ as already mentioned), but using a frequency centroid from the $\delta\nu_{01}$-$\Dnu$ relation \citep{Corsaro12cluster}. Any $\ell = 3$ peaks and the effect of rotation are here removed because useless.

For RGs evolving further into the Helium-burning MS, $\epsilon$ has the tendency to be lower than the value predicted from the $\epsilon$-$\Dnu$ relation for RGB stars in a way that is difficult to accurately predict \citep{Kallinger12,Corsaro12cluster}. This further motivates the choice of performing a sliding-pattern fit for evolved stars to measure their true $\epsilon$. If requested by the user, for low-luminosity RGB and late SG stars (which we have identified as fulfilling the condition $\Dnu_\mathrm{CL} \leq \Dnu_\mathrm{ACF} \leq \Dnu_\mathrm{thresh}$) \famed\,\,can perform an additional check against the $\epsilon$-$\Dnu$ relation calibrated by \cite{Corsaro12cluster}. This can be useful especially in low S/N or frequency resolution conditions.

We summarize the effect of performing a sliding-pattern fit on a low-luminosity RGB star in Fig.~\ref{fig:global} (right panels), where it is visible that the central $\ell = 0$ mode $(11,0)$, closest to $\numax$, has been properly located in the PSD.

\begin{figure}[t]
   \centering
  \includegraphics[width=9.0cm]{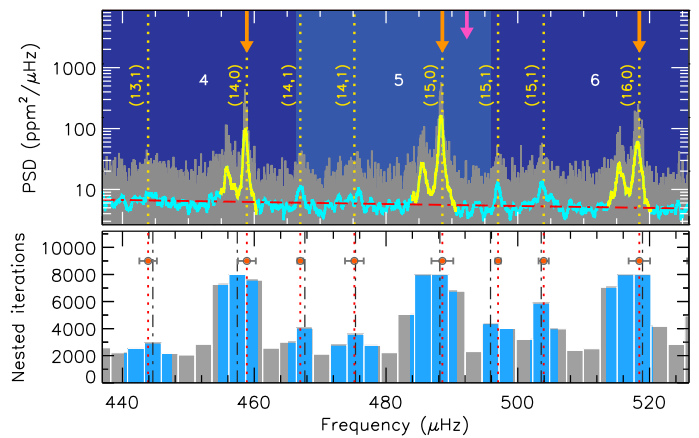}
    \caption{Similar description as for Fig.~\ref{fig:global} but for the depressed dipole SG star KIC~8561221.}
    \label{fig:global_depressed}
\end{figure}

\subsubsection{Depressed dipole stars}
\label{sec:sliding_depressed}
Stars having depressed dipole mode amplitudes \citep{Garcia14depressed,Stello16Nature,Mosser17depressed} are another class to be considered for investigation. These stars are found from the SG phase up to the RC. While the origin of this phenomenon is still under debate \citep[e.g.,][]{Fuller15Science,Mosser17depressed}, as shown by \cite{Stello16Nature} these stars are not rare and it is therefore useful to have the opportunity to perform a peak bagging analysis on them.

Given that the PSD of depressed dipole stars is mostly characterized by the presence of $\ell = 2,0$ modes, \famed\,\,checks whether there is a sufficiently large number ($50$\,\%) of estimated frequencies within $\numax \pm 2 \Dnu_\mathrm{ACF}$ for $\Dnu_\mathrm{ACF} > \Dnu_\mathrm{thresh}$ and within $\numax \pm \Dnu_\mathrm{ACF}$ for $\Dnu_\mathrm{ACF} \leq \Dnu_\mathrm{thresh}$, that have ASEF values well below the maximum nested iteration allowed in the global multimodal sampling ($< 3/4$). If this happens, then the star is flagged as a potential depressed dipole star. In this condition, if the star has $\Dnu_\mathrm{ACF} \leq \Dnu_\mathrm{thresh}$ the sliding-pattern model uses the same setup as for stars in the same $\Dnu$ range (Sect.~\ref{sec:sliding_rg}). Otherwise the sliding pattern is set up in a similar way as for the early SGs (Sect.~\ref{sec:sliding_ms}), with the difference that $\Dnu$ is fit instead of being fixed to the ACF value. This allows to better sample the position of the $\ell = 2,0$ mode regions because the $\ell = 1$ modes are not a source of power contamination for modes of different angular degree.

An example of the analysis of a depressed dipole star is illustrated in Fig.~\ref{fig:global_depressed}, where we can see that the ASEF peaks associated to potential $\ell = 1$ modes are significantly smaller in height with respect to those of the $\ell = 0$ modes, and always below 3/4 of the maximum nested iteration. In this case, \famed\,\,could successfully classify the star as a depressed dipole star just by using the frequencies estimated from the ASEF.

\subsection{General mode identification}
\label{sec:global_modeid}
Using $\epsilon$ from Eq.~(\ref{eq:epsi_sliding}) and $\Dnu_\mathrm{ACF}$, one can obtain an automated mode identification that provides the angular degrees $\ell = 1,0$ and the radial order $n$ for each frequency extracted from the ASEF. For this purpose, our starting point is to consider the asymptotic relations 
\begin{equation}
\nu_{n0} \simeq \Dnu \left( n + \epsilon \right) + \frac{\alpha_0 \Dnu}{2} \left( n - \frac{\numax}{\Dnu} \right)^2 \, ,
\label{eq:asymp_radial}
\end{equation}
for the radial modes and 
\begin{equation}
\nu_{n1} \simeq \Dnu \left( n + \frac{1}{2} + \epsilon \right) + \frac{\alpha_0 \Dnu}{2} \left( n - \frac{\numax}{\Dnu} \right)^2 - \delta\nu_{01} \, ,
\label{eq:asymp_dipole}
\end{equation}
for the dipole modes, where in the latter we have removed the curvature term on the small spacing, controlled by the parameter $\beta_{01}$ (see Eq.~\ref{eq:asymp}), and adopted the same $\alpha_0$ as for the radial modes. These choices are justified because the frequencies $\nu_{f,i}$ we are dealing with within \glob\,\,are less accurate and precise than those that are derived in \chu. By inverting Eq.~(\ref{eq:asymp_radial}) and Eq.~(\ref{eq:asymp_dipole}) for $\epsilon$, and by replacing the asymptotic frequency predictions $\nu_{n0}$ and $\nu_{n1}$ with a generic estimated frequency $\nu_{f,i}$, we can obtain two different estimates for $\epsilon$, namely
\begin{equation}
\epsilon_0 \simeq \frac{\nu_{f,i} - \Dnu \left[ n + \frac{\alpha_0}{2} \left( n - \frac{\numax}{\Dnu }\right)^2 \right]}{\Dnu}
\label{eq:epsi0}
\end{equation}
and
\begin{equation}
\epsilon_1 \simeq \frac{\nu_{f,i} - \Dnu \left[ n + \frac{1}{2} + \frac{\alpha_0}{2} \left( n - \frac{\numax}{\Dnu }\right)^2 \right] + \delta\nu_{01}}{\Dnu} \, .
\label{eq:epsi1}
\end{equation}
For a given input frequency $\nu_{f,i}$ (which could correspond to either a radial or a dipole mode), the associated radial order $n_i$ is found as the one minimizing the difference $\left| \epsilon_0 - \epsilon \right|$ (out of a set of possible $n$ values), with $\epsilon$ being the value obtained from the sliding-pattern fit (see Sect.~\ref{sec:sliding}) and $\epsilon_0$ given by Eq.~(\ref{eq:epsi0}). The radial order evaluated with this method is correct even if $\nu_{f,i}$ corresponds to a dipole mode. The angular degree $\ell$ is subsequently obtained as follows: \textit{i)} the radial order $n_i$ that was found is used to compute the reference values $\epsilon_0$ and $\epsilon_1$ through Eq.~(\ref{eq:epsi0}) and Eq.~(\ref{eq:epsi1}). The value of $\delta\nu_{01}$ used in Eq.~(\ref{eq:epsi1}) is set to the one obtained from the sliding-pattern fit in the case of MS stars as identified following Sect.~\ref{sec:sliding_ms}, to the asymptotic value from the $\delta\nu_{01}$-$\Dnu$ relation for stars having $\Dnu < \Dnu_\mathrm{tip}$, and to zero otherwise; \textit{ii)} the differences $\delta\epsilon_0 = \left| \epsilon_0 - \epsilon \right|$ and $\delta\epsilon_1 = \left| \epsilon_1 - \epsilon \right|$ are computed, with $\epsilon$ again the input value from the sliding-pattern fit; \textit{iii)} if $\delta\epsilon_0 < \delta\epsilon_1$ then the input frequency $\nu_{f,i}$ is better approximated by a radial mode frequency, hence it is flagged as a $\ell = 0$ mode, otherwise it is flagged as a $\ell = 1$ mode. This procedure is initially performed by setting $\alpha_0 = 0$ in both Eq.~(\ref{eq:epsi0}) and Eq.~(\ref{eq:epsi1}) because its value is not known at the beginning. 

The set of estimated frequencies $\{\nu_{f,i}\}$ is thus divided $\ell = 0$ and $\ell = 1$ peaks. In this subdivision, the radial modes are the most important ones for \glob. This is because the radial modes always fulfill the regular frequency pattern of $p$ modes (at least to first approximation) independent of the evolutionary stage of the star. The dipole modes are analyzed in detail in \chu\,\,(Sect.~\ref{sec:chunk}). An example of the general mode identification obtained by \famed\,\,is illustrated in Figs.~\ref{fig:global}, \ref{fig:global_depressed}. For the SG star showing depressed dipole modes, albeit some of the PSD peaks have been identified as $\ell = 1$ modes in \glob, they may be converted to $\ell = 3$ modes within \chu\,\,(see the application presented in Sect.~\ref{sec:depressed} for the same star).

\subsubsection{Improving the set of estimated frequencies}
\label{sec:skimming}
Although the general mode identification ($n$, $\ell$), with $\ell = 0,1$ can be applied to any of the frequencies in our set $\{\nu_{f,i}\}$, these frequencies may not always correspond to real oscillation modes. For example, there could be two frequencies $\nu_{f,i}$ flagged as radial modes within the same radial order, if for some reason (such as the presence of a neighboring dipole mixed mode) the ASEF has produced more than one local maximum in a region containing the real $\ell = 0$ mode. To overcome this we apply a skimming process on the set $\{\nu_{f,i}\}$ in a way that spurious frequencies can be identified and removed from the list. The entire process of the general mode identification is thus optimized as follows:
\begin{enumerate}
\item a correction to the frequencies $\nu_{f,i}$ identified as $\ell = 0$ modes is applied. These frequencies are initially biased to lower values because they correspond to $\ell = 2,0$ mode pairs in the PSD rather than to single $\ell = 0$ mode peaks (Sect.~\ref{sec:ranges}). Therefore the correction $\delta\nu_0$ on the radial mode frequencies is typically a positive term (see Fig.~\ref{fig:range}). The correction is $\delta\nu_0 \equiv \nu_0 - \nu_{0,f}$, where $\nu_0$ is the reference radial mode frequency obtained from the sliding-pattern fit, while $\nu_{0,f}$ is its closest radial mode frequency from the set $\{ \nu_{f,i} \}$. If no result from the sliding-pattern fit is used, then $\delta\nu_0 = 0$;
\item the asymptotic frequency predictions $\nu_{n0}$ for the estimated frequencies $\nu_{f,i}$ flagged as $\ell = 0$, are computed by means of the asymptotic pattern given by Eq.~(\ref{eq:asymp_radial}) by imposing $\alpha_0 = 0$ (in the first iteration);
\item the differences $\delta\nu_{0,i} = \vert \nu_{f,i} - \nu_{n0} \vert$ are evaluated for each radial mode frequency available;
\item if for a given $\nu_{f,i}$ flagged as $\ell = 0$, its $\delta\nu_{0,i}$ is below an input tolerance threshold, then the frequency is kept in the list of extracted frequencies, otherwise it is temporarily removed; The input tolerance threshold is given in units of $\Dnu$ and in our applications it has been set to 18, 20, 25\% for MS, SG and RG stars, respectively;
\item if more than one $\nu_{f,i}$ is flagged as a $\ell = 0$ mode with the same radial order $n$, then the one that is closer in frequency to the corresponding prediction from Eq.~(\ref{eq:asymp_radial}) is picked up. The other frequency is temporarily discarded from the $\{\nu_{f,i}\}$ set;
\item using the corrected and skimmed $\ell = 0$ frequencies obtained from the previous steps, as well as their corresponding radial orders $n_i$ and uncertainties $\sigma_{f,i}$, the Asymptotic code is adopted to fit the asymptotic pattern of radial modes. The asymptotic fit now incorporates the curvature term $\alpha_0$ as a free parameter, while $\epsilon$ is fixed to the value given by the sliding-pattern fit. In this way a more accurate estimate of $\Dnu$ than $\Dnu_\mathrm{ACF}$ is obtained, which we refer to as $\Dnu_0$;
\item the general mode identification is computed once again using the new set of estimated frequencies and by adopting the values of $\Dnu_0$ and $\alpha_0$ that were obtained from the asymptotic fit;
\item all the steps from \#1 to \#7 are repeated by re-starting each time from the original set of frequencies $\nu_{f,i}$ (uncorrected and unskimmed) in order to progressively improve the set, and for obtaining more reliable estimates of $\Dnu_0$ and $\alpha_0$. In particular, step \#2 is subsequently performed by imposing the value of $\alpha_0$ obtained from the asymptotic fit performed on step \#4 of the previous iteration.
\end{enumerate}

An example of the effect of the skimming process can be seen in the ASEF plotted in the bottom-right panel of Fig.~\ref{fig:global}. For the RGB star that is shown, two of the frequencies extracted from the ASEF in the selected frequency range -- indicated by the regions marked with orange shading -- are not retained in the final set of estimated frequencies. This is because these two frequencies, initially identified as radial modes because of their proximity to adjacent radial modes do not match the limiting threshold condition imposed at step \#4 of the skimming process.

We note that by increasing the tolerance threshold used to skim the set of frequencies with subsequent evolutionary stages (step \#4 of the skimming process), the overall result improves. This is mainly because for more evolved stars, the curvature effects of the asymptotic pattern becomes more important and the spectrum is more complex. Choosing an appropriate value of the tolerance threshold for the frequency skimming requires the star to be classified based on its evolutionary stage. Since the stellar evolutionary stage is not known a priori, in \glob\,\,and \chu\,\, \famed\,\,distinguishes between MS, SG, and RG stars as: \textit{(i)} MS stars by having either $\Dnu \geq \Dnu_\mathrm{SG}$ or at the same time $\Dnu_\mathrm{RG} \leq \Dnu \leq \Dnu_\mathrm{SG}$ and $\teff \geq T_\mathrm{eff,SG}$; \textit{(ii)} SG stars by having at the same time $\Dnu_\mathrm{RG} < \Dnu \leq \Dnu_\mathrm{SG}$ and $\teff < T_\mathrm{eff,SG}$; \textit{(iii)} RG stars by having $\Dnu \leq \Dnu_\mathrm{RG}$.
The thresholds used for MS and SG regimes are based on the $\Dnu$-$\teff$ diagram presented by \cite{App12freq} (see their Fig. 4), while that of RGs is approximated value of $\Dnu$ where the stars start to exhibit at least three or four mixed modes per radial order.
This classification is of course simple and schematic, and similarly to that adopted for setting up the sliding-pattern model, is not intended to provide the real evolutionary stage of the star. However, this scheme is used by \famed\,\,to improve the computational efficiency.

\subsection{Finding the chunks}
\label{sec:find_chunks}
The final task performed by \glob\,\,is to estimate the frequency separations $s_n$ that set the boundaries for each radial order $n$ in the PSD of the star. More precisely, these separations delimit chunks of PSD, each one containing a single pair of $\ell = 0,1$ modes. Each chunk is therefore approximately as wide as $\Dnu$. The number of chunks contained in the stellar PSD, $N_\mathrm{chunks}$, is uniquely estimated from the difference between the maximum and minimum radial order found from the set $\{\nu_{f,i}\}$.

The radial mode asymptotic frequencies are computed from Eq.~(\ref{eq:asymp_radial}) for each radial order covered by $\{\nu_{f,i}\}$. The upper frequency boundary (or upper separation) of each chunk is thus given by the asymptotic frequency of the radial mode increased by a fraction of $\Dnu$, which is set to 20\% for MS stars, and to 25\% for SG and RGs. The larger range in evolved stars is motivated by their more pronounced curvature effects in the asymptotic pattern of $p$ modes.

The upper separation of a given chunk is the lower separation of the subsequent chunk. In the way chunks are defined, the estimated frequencies contained inside a chunk corresponding to the radial order $n$ thus comprise one $\ell = 0$ frequency of radial order $n$, and one or more $\ell = 1$ mode frequencies of radial order $n-1$ (placed to the left side of the radial mode). Figure~\ref{fig:global} (top panels) shows an example of the chunk division obtained in both a MS and a RG star, while Fig.~\ref{fig:global_depressed} (top panels) shows the same process but for a SG star that has depressed dipole modes. 

Each chunk is identified by its frequency separations even if no $\ell = 0$ and/or $\ell = 1$ modes have been found in it. To understand this feature, let us suppose we have three radial orders from the general mode identification covered by a set of $\ell = 0$ frequencies $\nu_{f,0}$, $\nu_{f,1}$ and $\nu_{f,2}$, which have radial orders $n-1$, $n$, and $n+1$, respectively. If the frequency $\nu_{f,1}$, corresponding to the central radial order of the set, is removed during the skimming process, we end up with the frequencies $\nu_{f,0}$ and $\nu_{f,2}$. Nonetheless, when the chunks are identified, it is still possible to analyze the chunk corresponding to the central radial order of our example, and it is still possible to re-locate the $\ell = 0$ mode of the chunk during \chu. The minimum condition to analyze the chunk, however, is that it contains at least one frequency from \glob, that is at least one $\ell = 0$ or $\ell = 1$ mode.

\section{The \chu\,\,module}
\label{sec:chunk}
Although the information provided by \glob\,\,is already useful for understanding what kind of general oscillation features the star possesses and what is the mode identification scheme to follow, the output from \glob\,\,comes with some major limitations: \textit{(i)} the general mode identification is usually reliable for the radial mode frequencies, but the radial mode frequencies may not be very accurate because they approximate the $\ell = 2,0$ mode pairs instead of representing the single $\ell = 0$ mode peaks in the PSD; \textit{(ii)} the $\ell = 1$ modes identified are not thoroughly inspected, and may be unreliable; \textit{(iii)} the $\ell = 2$ modes are not extracted; \textit{(iv)} the $\ell = 3$ modes are not even contemplated because they are too small in amplitude with respect to $\ell = 0,1$ modes; \textit{(v)} the output frequency list is not yet validated in terms of detection against the noise level.
Overcoming these aspects requires investigating each chunk of the PSD identified by \glob\,\,by means of the module \chu. We note that the $\ell = 2$ modes extracted with \chu\,\,are not inspected neither from the point of view of rotation nor in terms of quadrupole mixed modes composition. This means that the $\ell = 2$ modes are treated as single peaks, as it is usually done in the standard peak bagging analysis, and that all {$\ell = 0$ and $\ell = 2$ modes extracted with \famed\,\,have a mode identification tag of the type $\left(n, \ell \right)$. In addition, for each chunk a local value of the $\epsilon$ parameter of the asymptotic pattern as evaluated from the chunk $\ell = 0$ mode, and of the frequency separation between the $\ell = 2,0$ modes, $\delta\nu_{02}$, are also provided.

In \chu\,\,the identification of $\ell = 1$ modes is done independently of the evolutionary stage of the star and of any theoretical asymptotic pattern by combining the frequencies extracted from the multimodal sampling (see Sect.~\ref{sec:sampling_chunk}) with the subsequent peak detection testing (Sect.~\ref{sec:test_1}). This data-driven approach is very efficient in obtaining a fast and automated extraction of $\ell = 1$. This is especially useful for evolved stars where the presence of mixed $\ell = 1$ modes can significantly complicate the oscillation pattern. The same data-driven approach allows handling without difficulties the analysis of RGs with depressed dipole mode power because each candidate dipole mode is tested separately from the others and independently of $\ell = 0$ and 2 modes.

\subsection{Chunk multimodal sampling}
\label{sec:sampling_chunk}
Similarly to \glob, \chu\,\,requires that an initial multimodal sampling is performed. This is done by using the PeakBagging code with configuration \#1 and one Lorentzian profile (Sect.~\ref{sec:pb}). The main differences with respect to the global multimodal sampling are that: 1) the prior for the frequency centroid of the island peak bagging model is now constrained by the frequency range of the chunk to analyze, as obtained from the frequency separations $s_n$ (see Sect.~\ref{sec:find_chunks}); 2) different values of the FWHM of the islands peak bagging model defined in Eq.~(\ref{eq:isla}) are used based on the evolutionary stage classification of the star. 

However, the frequency separations $s_n$ delimit the frequency region of each chunk without any overlap between consecutive chunks, which poses a risk in the detection of potential oscillation peaks that may unluckily fall at the edge of the chunk frequency range. To overcome this issue, \chu\,\,applies an extension to each frequency separation $s_n$, such that the lower frequency boundary of each chunk is shifted to smaller frequencies. The amount of chunk overlap adopted in this work has been set to be 15 \% of $\Dnu$ for MS stars, and to 25 \% of $\Dnu$ for SG and RG stars. A larger overlap in evolved stars is motivated by the presence of dipolar mixed modes. The chunk overlap is especially useful for chunks that are located toward the tails of the Gaussian envelope of the oscillations, where the global frequencies corresponding to $\ell = 0$ may be less accurate because of the low S/N of the dataset.

The FWHM of the islands peak bagging model for \chu, which we refer to as $\Gamma_\mathrm{chunk}$, is initially computed from the same relations presented in Sect.~\ref{sec:sampling_global}. By naming the FWHM used for the global multimodal sampling as $\Gamma_\mathrm{global}$, one has: \textit{(i)} for $\Dnu_\mathrm{RG} < \Dnu \leq \Dnu_\mathrm{SG}$ and $\teff \geq T_\mathrm{eff,SG}$, or for $\Dnu \geq \Dnu_\mathrm{SG}$: $\Gamma_\mathrm{chunk} = \Gamma_\mathrm{global} / 10$. This corresponds to MS stars; \textit{(ii)} for $\Dnu_\mathrm{RG} < \Dnu < \Dnu_\mathrm{SG}$ and $\teff < T_\mathrm{eff,SG}$: $\Gamma_\mathrm{chunk} = \Gamma_\mathrm{global} / 2$. This corresponds to SG and early RGB stars; \textit{(iii)} for $\Dnu_\mathrm{CL} < \Dnu \le \Dnu_\mathrm{RG}$: $\Gamma_\mathrm{chunk} = \Gamma_\mathrm{global} / 5$. This corresponds to low-luminosity RGB stars; \textit{(iv)} for $\Dnu_\mathrm{tip} < \Dnu \le \Dnu_\mathrm{CL}$: $\Gamma_\mathrm{chunk} = \Gamma_\mathrm{global} / 8$. This corresponds to RC stars (both 1$^\mathrm{st}$ and 2$^\mathrm{nd}$ RC) and to more evolved RGB stars; \textit{(v)} for $\Dnu \le \Dnu_\mathrm{tip}$: $\Gamma_\mathrm{chunk} \lesssim \Gamma_\mathrm{global}$. This corresponds to stars toward the RGB tip but can also include stars that have evolved off the RC toward the AGB. Here the parameters $\Dnu_\mathrm{SG}$, $\Dnu_\mathrm{RG}$, $\Dnu_\mathrm{CL}$, $\Dnu_\mathrm{tip}$, and $T_\mathrm{eff,SG}$ have already been defined in \glob\,\,(see Table~\ref{tab:configuring} for a summary). The reference value for $\Dnu$ used in \chu\,\,is $\Dnu_0$ estimated from \glob.

The general condition $\Gamma_\mathrm{chunk} < \Gamma_\mathrm{global}$ is necessary for obtaining a higher resolving power of the islands peak bagging model with respect to the case of \glob. The new multimodal sampling, henceforth termed chunk multimodal sampling, is performed over a frequency range that is about one order of magnitude smaller than that used in \glob, which implies that the frequency sampling is also much denser in terms of number of points. This allows maximizing the number of detections of potential oscillation peaks, especially for those peaks having the narrowest linewidths.

The large rescaling factor for computing $\Gamma_\mathrm{chunk}$ in MS stars stems from the large oscillation mode linewidths found in these stars. For stars evolved toward the tip of the RGB instead, the oscillation modes are the narrowest observed. Here the adopted $\Gamma_\mathrm{global}$ is already the smallest among the different evolutionary regimes, such that $\Gamma_\mathrm{chunk}$ can be chosen to be of the same order. The scaling values for SG and RG stars have been obtained using observations from the \kepler nominal mission, spanning up to more than four years.

The chunk multimodal sampling is obtained independently for each chunk, hence this process can be parallelized to improve the computational speed. In the automated analysis sequence, \famed\,\,first performs the multimodal sampling for all the chunks found in \glob, and then analyzes each chunk separately in progressively decreasing S/N order. In this way chunks having the best and most clear asteroseismic signal in the PSD are analyzed first. The chunk multimodal sampling is then exploited for computing a standard resolution ASEF ($N_\mathrm{bins} \approx 100$), with new chunk parameters $\nu_i$, $r_{a,b}$, $d_{a,b}$, $\nu_{f,i}$, and $\sigma_{f,i}$ all evaluated following the same description presented for \glob\,\,(Sect.~\ref{sec:freq_asef}). The threshold used for the hill-climbing algorithm to identify the local maxima of the ASEF now varies as a function of the evolutionary stage classification described in Sect.~\ref{sec:skimming}. In particular, our applications of \famed\,\,consider for the hill-climbing algorithm a 3\,\% threshold for MS stars, and a 5\,\% one for SG and RG stars. These thresholds help in minimizing the effect caused by the presence of noise peaks, thus their adoption significantly speeds up the overall computation.

The analysis of the oscillation peaks in \chu\,\,has now at its disposal the frequencies identified from \glob\,\,(henceforth referred to as global frequencies), as well as the asymptotic parameters $\Dnu_0$, $\epsilon$, $\alpha_0$. This information is essential for a detailed mode identification of the newly extracted frequency peaks.

\subsection{Radial and quadrupole modes}
\label{sec:radial_quadrupole}
An important task for a successful analysis in \chu\,\,is to obtain an accurate estimate of the frequency of the $\ell = 0$ and $\ell = 2$ modes contained in the chunk, if they are detectable.

\subsubsection{Finding the radial mode}
\label{sec:radial}
In Sect.~\ref{sec:find_chunks} we have seen that a single chunk can be analyzed within \chu\,\,only if at least one global frequency is found, which can be either a $\ell = 0$ or a $\ell = 1$ mode. From now on, we assume that this latter condition is always fulfilled. 

At the beginning, \chu\,\,attempts to either improve the global $\ell = 0$ mode frequency that is found in the chunk being analyzed, or to obtain one if it is not available. For this purpose \famed\,\,searches for a possible solution obtained with \chu\,\,on a neighboring chunk. For this solution to be available, the neighboring chunk must have been analyzed before the actual one. With the exception of the first chunk that is analyzed, that is the one with the highest S/N, all the remaining chunks can potentially exploit a solution from a neighboring chunk that was previously analyzed because having a higher S/N. For understanding how \famed\,\,is using this information, let us suppose we have a solution from \chu\,\,from the highest S/N chunk of the entire set, and we label this chunk as $c_i$. Very likely this chunk is centered around $\numax$ because this value corresponds by definition to the highest S/N of the PSD region containing the oscillations. We thus consider the chunk radial mode frequency $\nu_{n_i,0}$ of chunk $c_i$, which corresponds to the radial order $n_i$. If the new chunk that is analyzed after $c_i$ is located next to the right side, which we term chunk $c_{i+1}$, then the mean frequency of the chunk is higher than $\numax$. \famed\,\,can thus improve the value of the global radial mode frequency of the chunk $c_{i+1}$ by using the chunk radial mode frequency estimated for chunk $c_i$. The improved global radial mode frequency $\nu_{n_{i+1},0}$ of the chunk $c_{i+1}$, corresponding to the radial order $n_{i+1}$, can thus be written as
\begin{equation}
\nu_{n_{i+1},0} = \nu_{n_i,0} + \Dnu_0 \left[ 1 + \alpha_0 \left( n_{i + 1} - 0.5 - \frac{\numax}{\Dnu_0} \right) \right] \, ,                                                                                                                                                                                                                                                                                                                                                                                                                                                                                                                                           
\label{eq:curvature1}
\end{equation}
as follows by computing the difference $\nu_{n_{i+1},0} - \nu_{n_i,0}$ using Eq.~(\ref{eq:asymp_radial}, where $\Dnu_0$, $\alpha_0$ are our estimates of the asymptotic fit to the radial modes from \glob\,\,(see Sect.~\ref{sec:skimming}, step \#6), while the radial order $n_{i+1}$ is provided by the general mode identification (Sect.~\ref{sec:global_modeid}). In the unfortunate case that the previous chunk $c_i$ does not contain any $\ell = 0$ mode solution, \famed\,\,checks for the presence of a $\ell = 0$ mode solution in the chunk $c_{i-1}$, previous to $c_i$, and corresponding to radial order $n_{i-1}$. If a radial mode frequency $\nu_{n_{i-1},0}$ is thus obtained from the chunk $c_{i-1}$, the improved global radial mode frequency $\nu_{n_{i+1},0}$ of the chunk $c_{i+1}$ is calculated as
\begin{equation}
\nu_{n_{i+1},0} = \nu_{n_{i-1},0} + 2 \Dnu_0 \left[ 1 + \alpha_0 \left( \frac{2 n_{i + 1} - 1}{2} - 0.5 - \frac{\numax}{\Dnu_0} \right) \right] \, ,
\label{eq:curvature2}                                                                                                                                                                                                                                                                                                                                                                              
\end{equation}
again following from the difference $\nu_{n_{i+1},0} - \nu_{n_{i-1},0}$ using Eq.~(\ref{eq:asymp_radial}).
A similar argument can be applied for the specular case of a chunk $c_{i-1}$, falling to the left side of chunk $c_i$, hence having an average frequency lower than $\numax$. This approach is thus building upon those radial mode frequencies extracted from a higher S/N region of the stellar PSD, and that are therefore assumed to be more reliable in both frequency position and mode identification tag. The advantage of this approach is threefold: 1) it is able to maintain the accuracy of the global radial mode frequencies as we move away from $\numax$ toward lower S/N regions; 2) it allows avoiding misidentification of $\ell = 0$ modes into $\ell = 2$ (and vice versa) as we approach to the tails of the Gaussian envelope where the curvature effects of the asymptotic relation of $p$ modes are stronger; 3) it allows computing a reliable global radial mode frequency even if this is not available from \glob, thus opening the possibility to find quadrupole and radial modes contained in the chunk. If no global dipole mode frequency is found from \glob, this approach also provides a guess for a global dipole mode frequency based on the global radial mode frequency of the chunk.

If no global radial mode frequency can be obtained from the procedure explained above the pipeline does not proceed further in estimating the chunk radial mode frequency and in searching for a possible adjacent quadrupole mode. Dipole modes will be inspected and analyzed even if no radial mode is found.

Subsequently \famed\,\,is using the chunk using the ASEF computed out of the chunk multimodal sampling to find the frequency $\nu_{f,i}$ that is considered to be the most reliable candidate in representing the radial mode frequency of the chunk. The degree of reliability for the candidate chunk radial mode frequency includes the proximity to the global radial mode frequency, the amplitude of the corresponding ASEF local maximum, the number of sampling points falling within the ASEF local maximum, and the height of the peak from a smoothed PSD, previously computed using a smoothing factor given by $\Gamma_\mathrm{chunk}$. With the candidate chunk radial mode identified, it is then possible to establish limiting frequencies to ensure that the oscillation peaks to be identified inside the chunk are not placed beyond the actual chunk radial mode frequency and that at the same time they are above the previous radial mode. This prevents the pipeline from identifying the same peaks in two adjacent chunks.

\begin{figure*}[tb]
   \centering
  \includegraphics[width=17.5cm]{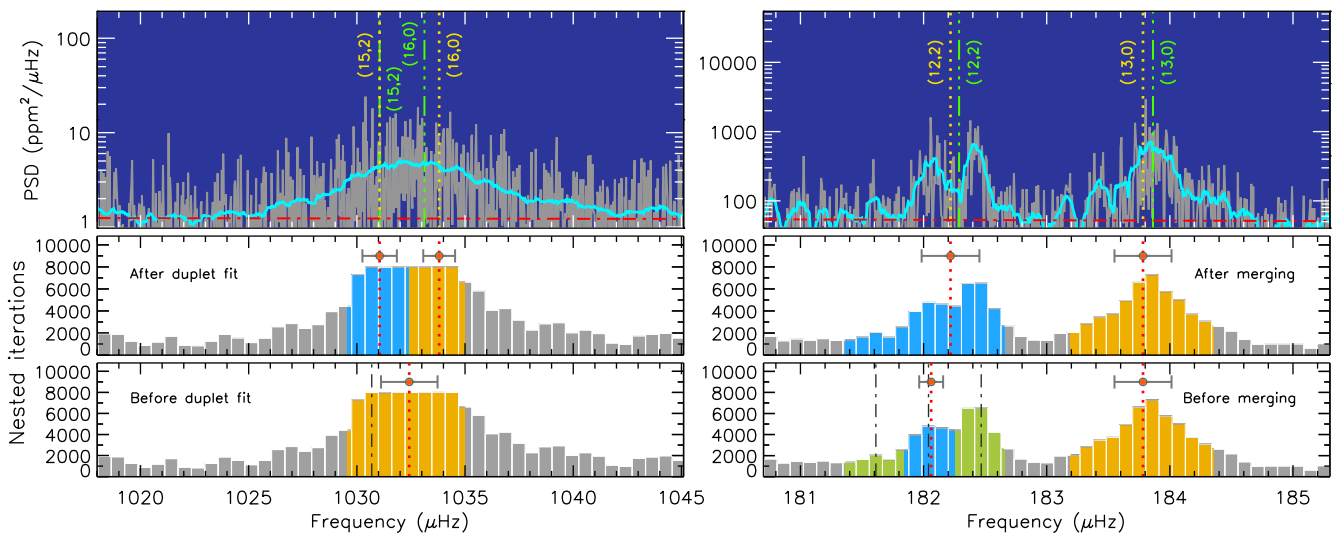}
    \caption{Identification of chunk quadrupole mode for the simulated F-type MS star 06550, having $\teff = 6416$\,K \citep{Ball18TESS} and corresponding to a 1-year length observation with TESS (left panels), and the RGB star KIC~12008916 observed with \kepler for more than four years (right panels). \textit{Top panels}: similar to Fig.~\ref{fig:global}, top panels, but now showing a PSD region containing a $\ell = 2,0$ mode pair, where the frequencies estimated from \chu\,\,are indicated by vertical dotted yellow lines, with included the corresponding ($n$,$\ell$) mode identification obtained by \famed. The cyan curve is the PSD smoothed by a factor proportional to $\Gamma_\mathrm{chunk}$. The vertical double-dot dashed green lines, and their mode identification, are from \cite{Ball18TESS} for the MS star, and from C15 for the RGB star. \textit{Middle and bottom panels}: chunk ASEF showing $r_{a,b}$ (blue shading for $\ell = 2$ and yellow for $\ell = 0$), $\nu_{f,i}$, $\sigma_{f,i}$, after and before applying a two-Lorentzian multimodal fit for the MS star, and a quadrupole mixed modes merging for the RGB star. The local maxima from the ASEF are marked by the vertical dot-dashed black lines, with green shading indicating the frequency ranges of those ASEF peaks that are merged.}
    \label{fig:doublet}
\end{figure*}

\subsubsection{Finding the quadrupole mode}
\label{sec:quadrupole}
In most stars that are less evolved, such as early SG and MS, the $\ell = 2,0$ oscillation mode pairs may often be partially or almost entirely blended, especially if the star has a high $\teff$ (hence large mode linewidths). Despite the chunk multimodal sampling is in principle able to distinguish peaks that are close to one another, it is usually not able to reliably separate an $\ell = 2$ from an $\ell = 0$ peak if a significant blending occurs. 
The small frequency separation $\delta\nu_{02}$ is not known a priori in such stars, and it is not straightforward to predict accurately \citep[e.g.,][]{White11}. As a consequence, it is not possible to understand whether the sampling will produce two separate maxima in the distribution or just a single broader one. To overcome this problem and solving any possible ambiguity, for all stars classified as SG and MS, having $\Dnu \geq \Dnu_\mathrm{RG}$, \famed\,\,performs an additional multimodal fitting on the sole region containing the $\ell = 2,0$ mode pair, located using the chunk radial mode frequency from the previous section. This time the new multimodal sampling, which we shall refer to as doublet multimodal sampling, is performed using configuration \#1 of the PeakBagging code presented in Sect.~\ref{sec:pb} but with two Lorentzian profiles. This configuration utilizes a two-Lorentzian islands peak bagging model, which in analogy to Eq.~(\ref{eq:isla}) can be defined as
\begin{equation}
\begin{split}
P_\mathrm{isla} \left( \nu, \Gamma; \nu_0, H, \delta\nu_{02} \right) &= \frac{H}{1 + 4 \left( \frac{\nu - \nu_0}{\Gamma} \right)^2} +
\frac{H V^2_2}{1 + 4 \left( \frac{\nu - \nu_0 + \delta\nu_{02}}{\Gamma} \right)^2}
\end{split}
\label{eq:isla2}
\end{equation}
where the first term on the right-hand side is the single-Lorentzian islands peak bagging model -- here used to model the $\ell = 0$ mode of the chunk -- while the second term represents the second Lorentzian profile that is modeling the $\ell = 2$ mode adjacent to the $\ell = 0$. The height of the $\ell = 2$ peak is modulated by the visibility of quadrupole modes $V_2^2$ (see Sect.~\ref{sec:sliding}). The linewidth $\Gamma$ is the same for both peaks, and it is fixed to the value predicted using the empirical relations from Sect.~\ref{sec:sampling_global} for the chunk radial mode frequency. Conversely to Eq.~(\ref{eq:isla}), Eq~(\ref{eq:isla2}) accounts for three free parameters, $\nu_0$, $H$, and $\delta\nu_{02}$, the latter allowing locating the position of the $\ell = 2$ mode with respect to its adjacent $\ell = 0$ mode. Figure~\ref{fig:doublet} (left panels) shows an example of the powerful capability of the doublet multimodal sampling in estimating an accurate position of quadrupole and radial modes even in the case of a MS star having a strong peak blending effect, where the mode linewidth is larger than $4\,\mu$Hz and with $\delta\nu_{02} \simeq 2\,\mu$Hz. The frequencies estimated by \famed\,\,have an excellent agreement with those obtained from the simulations \citep{Ball18TESS}, significantly below 0.1\,\%. In conclusion, for stars identified as SG and MS, the chunk radial mode frequency initially obtained as described in Sect.~\ref{sec:radial_quadrupole}, is here re-estimated using the solution from the doublet multimodal fit, which is used at the same time to estimate the chunk quadrupole mode frequency and to obtain a more accurate value of the lower limit frequency of the chunk.

For stars classified as RGs instead, having $\Dnu < \Dnu_\mathrm{RG}$, the two-Lorentzian islands peak bagging model is not used. This is because 1) the $\ell = 2$ and $\ell = 0$ modes are now well separated in frequency, given that their linewidth is smaller than $\delta\nu_{02}$, hence the chunk ASEF can properly distinguish them (see Fig.~\ref{fig:asef}); 2) the frequency separation $\delta\nu_{02}$ can be predicted with good accuracy from the $\Dnu$-$\delta\nu_{02}$ relation \citep{Huber11,Corsaro12cluster}. Consequently, the chunk quadrupole mode frequency corresponds to the frequency $\nu_{f,i}$ estimated from the chunk ASEF that is closest to the predicted quadrupole mode frequency $\nu_{n-1,2} = \nu_{n,0} - \delta\nu_{02}$, where $\nu_{n,0}$ is the chunk radial mode frequency and $\delta\nu_{02}$ is obtained from the $\Dnu$-$\delta\nu_{02}$ relation. If no $\nu_{f,i}$ frequency is found to be closer to the asymptotic position of the $\ell = 2$ mode than that of the chunk $\ell = 0$ mode, then the pipeline assumes that there is no $\ell = 2$ mode in the chunk.

Evolved stars can be characterized by the presence of $\ell = 2$ mixed modes, whose pattern in the PSD could be further complicated by the effect of rotation \citep[e.g., see][]{Deheuvels17}. The result is that the chunk ASEF could contain more than one local maximum in the region of the $\ell = 2$ mode (e.g., see Fig.~\ref{fig:doublet}, right panels). The current version of the pipeline does not deal with the detection and analysis of $\ell = 2$ mixed modes and therefore assumes that there is only one $\ell = 2$ peak, at most, in each chunk analyzed. For improving the estimation of the chunk quadrupole mode frequency, \famed\,\,checks whether there are frequencies estimated from the ASEF that fall inside the control range $\left[ \nu_{n-1,2} - \phi_2 \delta\nu_{02}/2, \nu_{n-1,2} + \phi_2 \delta\nu_{02}/2 \right]$, where $\phi_2$ allows to vary the extent of the control range, with $\phi_2 = 1$ for all our applications. If the condition is satisfied, the frequencies falling in the control range are discarded the corresponding ASEF peaks are merged to that of the identified $\ell = 2$ mode of the chunk. In this way the chunk quadrupole mode frequency and its associated 1-$\sigma$ uncertainty are recomputed using a larger $r_{a,b}$, which now incorporates the frequency ranges of the  frequencies falling inside the control range. This approach allows to take more accurately into account the $\ell = 2$ power barycenter in estimating the quadrupole mode frequency. A clear example of this procedure can be seen in the right panels of Fig.~\ref{fig:doublet}. Here the chunk quadrupole mode frequency identified in the bottom-right panel is clearly offset with respect to the one shown in the middle panel, the latter matching within 0.04\% to the value obtained by C15, who adopted a standard (nonautomated) peak bagging analysis. Nonetheless, we consider that the multimodal approach used by \famed\,\,retains potential for incorporating the analysis of quadrupole mixed modes in future developments of the pipeline.

Finally, as expected from the asymptotic theory of $p$ modes, the radial order of the identified chunk quadrupole mode frequency is obtained as that of the adjacent radial mode, already obtained in \glob, but decreased by one. 

\subsection{Peak detection and blending tests}
\label{sec:peak_detection_test}
With the identification of $\ell = 0,2$ modes, the remaining frequencies estimated from the chunk ASEF local maxima are initially flagged as $\ell = 1$ modes, independently of how many frequencies are found. Their corresponding radial order is given by that of the chunk radial mode frequency decreased by one. However, all the frequencies estimated and identified until this stage, including the chunk $\ell = 0,2$ modes, are still only candidate oscillation modes. This is because none of them has yet been validated against the level of noise in the PSD. As a result, a significant fraction of the  frequencies may turn out to be not significant, hence discarded from further analysis and from the final output of the pipeline. The following step, which is crucial for the reliability of the results produced by \chu, is therefore to test whether each frequency peak of the chunk, of any given $\ell$, is significant against the level of the background of the star. 

As shown by \cite{Corsaro14}; C15, testing the significance of a frequency peak in the stellar PSD can be done by performing a Bayesian model comparison that exploits the so-called Bayesian evidence. The advantage of the Bayesian evidence with respect to a standard $\chi^2$ minimization or a maximum likelihood estimation is, for example, that it takes into account both the model complexity and the fit quality such that the choice of the individual assumptions on the model free parameters can be tested in the light of the available data. The significance of a frequency peak that is assumed to be an oscillation mode has to be evaluated by confronting two fitting models, one incorporating the peak with a typical oscillation mode profile on top of a varying background level, and another one incorporating only the background level. This working scheme can be extended to test the significance of a pair of frequency peaks that are blended. If a peak blending occurs, by testing the peak significance for only one peak we would clearly ignore the hypothesis that the observed peak is in reality constituted by two different ones. Performing a Bayesian model comparison process therefore implies that: 1) the correct model at interpreting the data is among the fitting models that are considered in the comparison; 2) the assumptions on the model hypothesis, namely the prior probability distributions on the model free parameters, are part of the outcome, meaning that the obtained Bayesian evidence of each model also reflects the way we set up the prior hyper-parameters.

To manage this first phase of the peak testing, \famed\,\,makes use of a set of peak fitting models provided by the PeakBagging code, which are listed below:
\begin{itemize}
\item $\mathcal{M}_\mathrm{A}$ is a model to test the hypothesis that the frequency peak is consistent with noise. It has one free parameter, the background level amplitude $\sigma_\mathrm{noise}$, which is a dimensionless quantity changing the level of the background estimated from the background fit;
\item $\mathcal{M}_\mathrm{B}$ is a model to test that the frequency peak is a Lorentzian profile. It has four free parameters, namely $\sigma_\mathrm{noise}$, and frequency centroid, amplitude, and linewidth of the Lorentzian profile ($\nu_{i,0}, A_i, \Gamma_i$, respectively);
\item $\mathcal{M}_\mathrm{C}$ is a model to test that the frequency peak is a sinc$^2$ profile (C15). It has three free parameters, namely $\sigma_\mathrm{noise}$, and frequency centroid and height of the sinc$^2$ profile ($\nu_{i,0}, H_i$, respectively);
\item $\mathcal{M}_\mathrm{D}$ is a model to test the presence of a blended $\ell = 2,0$ mode pair, assuming that the two frequency peaks are Lorentzian profiles. It has seven free parameters, namely $\sigma_\mathrm{noise}$, and the frequency centroids, amplitudes, and linewidths of the two Lorentzian profiles ($\nu^{\ell = 0}_{i,0}, A^{\ell=0}_i, \Gamma^{\ell=0}_i$ for the radial mode, and $\nu^{\ell = 2}_{i,0}, A^{\ell=2}_i, \Gamma^{\ell=2}_i$ for the quadrupole mode, respectively).
\end{itemize}
The choice of fitting amplitudes instead of heights for models $\mathcal{M}_\mathrm{B}$ and $\mathcal{M}_\mathrm{D}$ is motivated by the known correlation between height and linewidth, which would increase the computational time needed to conduct each fit. The height is instead a free parameter for model $\mathcal{M}_\mathrm{C}$ because the sinc$^2$ profile does not require the fitting of a linewidth.

\famed\,\,therefore performs unimodal fits using the PeakBagging code by utilizing pairs of models from the above list and by comparing their Bayesian evidences to test whether the peak is significant above the level of noise (peak detection test), and whether the peak is actually a blended structure (peak blending test). The peak detection tests are divided in two groups, those for the candidate $\ell = 2,0$ modes, and those for the candidate $\ell = 1$ modes, while the peak blending test is only done for the candidate $\ell = 2,0$ mode pairs.

\subsubsection{FWHM of the chunk radial mode}
\label{sec:fwhm_radial}
The first frequency peaks to be assessed in terms of significance are the $\ell = 2,0$ modes of the chunk. In this section we shall assume that a candidate $\ell = 0$ mode has been identified. If this is not the case, then \famed\,\,goes directly to the peak testing of the $\ell = 1$ modes (Sect.~\ref{sec:test_1}). 

Before performing the actual peak testing phase, \famed\,\,evaluates the FWHM of the radial mode of the chunk by fitting the corresponding PSD peak by means of a Lorentzian profile. This is necessary to set up the peak testing that is discussed in Sects.~\ref{sec:test_20}, \ref{sec:test_1}, \ref{sec:peak_rotation_test}, \ref{sec:octupole_test}.

If the star has $\Dnu < \Dnu_\mathrm{thresh}$, the radial mode is treated as a single Lorentzian profile, namely by fitting frequency centroid, amplitude, and linewidth, with the background level fixed to the value of the background fit for obtaining a more stable solution of the FWHM (this is done using the fitting model $\mathcal{M}_\mathrm{E}$ presented in Sect.~\ref{sec:peak_rotation_test}). For stars having $\Dnu \ge \Dnu_\mathrm{thresh}$, the $\ell = 2,0$ peaks are fit simultaneously, meaning that the fitting model accounts for two Lorentzian profiles, again using a fixed background level (this is done using the fitting model $\mathcal{M}_\mathrm{G}$ presented in Sect.~\ref{sec:peak_rotation_test}). For performing the Bayesian inference with \diamonds, our pipeline adopts uniform prior distributions for all the free parameters. This choice is mainly motivated by the fast computation that is guaranteed with the adoption of this type of priors when using \diamonds\,\,as a fitting code (e.g., \citealt{Corsaro14}; C15, \citealt{Corsaro17spin}). The nested sampling algorithm implemented in \diamonds\,\,ensures that the adoption of uniform priors neither hampers the sampling quality nor the capability of \diamonds\,\,to obtain reliable estimates of the free parameters, especially in the low-dimensional regime adopted by \famed. The uniform prior hyper-parameters on the frequency centroid of the Lorentzian profiles are obtained from $r_{a,b}$ of the corresponding chunk frequencies, while those on the amplitude (and therefore on the linewidth) are obtained by computing an amplitude estimate using the peak heights measured from a PSD smoothed by $\Gamma_\mathrm{chunk}$, and a linewidth estimate from the empirical $\Gamma$-$\numax$-$\teff$ relations (see also Appendix~\ref{sec:prior} for more details). 

The peak fit utilizes a portion of the dataset that is centered around the peaks. If one Lorentzian profile is used, the dataset spans up to the boundaries set by the frequency divisions $d_{a,b}$ of the chunk radial mode frequency. If two Lorentzian profiles are used, the dataset spans from the frequency division $d_a$ of the quadrupole mode up to the frequency division $d_b$ of the radial mode. We also note that for obtaining a more reliable estimate of the peak linewidth, the peak fit can be performed multiple times in parallel, and the final FWHM estimate, henceforth $\Gamma_0$, is then taken as the median value of the set of FWHM estimates.

\subsubsection{Radial and quadrupole modes}
\label{sec:test_20}
\famed\,\,now proceeds with the peak detection and blending tests for the $\ell = 2,0$ peaks. According to C15, a peak is tested against the level of noise only if its height in the smoothed PSD is lower than 10 times the local level of the background, otherwise the peak is automatically considered as detected. This clearly speeds up the analysis. In the following we shall discuss the case of peaks that are not sufficiently prominent to be excluded from a peak detection and blending test.

For stars having $\Dnu < \Dnu_\mathrm{thresh}$, the peak blending test is not performed and the peak detection test on the $\ell = 2,0$ peaks is done separately for each peak. This is because for such evolved stars the two peaks are well separated from one another, their relative separation $\delta\nu_{02}$ can be predicted with good accuracy, and the chunk multimodal sampling is able to properly disentangle them. For testing the detection of each peak two fits are executed, one using model $\mathcal{M}_\mathrm{A}$, and the other using model $\mathcal{M}_\mathrm{B}$. Here the priors are again uniform and the hyper-parameters for the Lorentzian peak are set in a similar way as done for the fit to the FWHM of the chunk radial mode, but are this time based on the chunk radial mode linewidth $\Gamma_0$, which is expected to be more accurate than an empirical prediction. A detailed list of all the prior ranges is presented in Appendix~\ref{sec:prior}, Table~\ref{tab:priors_detection}. The adequateness of uniform prior distributions for conducting successful peak detection tests using \diamonds\,\,was already proved by \cite{Corsaro14}; C15. The resulting Bayesian evidences $\mathcal{E}_\mathrm{A}$ and $\mathcal{E}_\mathrm{B}$ for models $\mathcal{M}_\mathrm{A}$ and $\mathcal{M}_\mathrm{B}$, respectively, are then used to obtain the peak detection probability
\begin{equation}
p_\mathrm{BA} = \frac{\mathcal{E}_\mathrm{B}}{\mathcal{E}_\mathrm{A} + \mathcal{E}_\mathrm{B}} \, .
\label{eq:p_ba}
\end{equation}
The peak is thus considered to be significant if $p_\mathrm{BA} \geq 0.993$ \citep[see also][]{Trotta08}. This is equivalent to considering a strong evidence condition in the Jeffreys' scale of strength \citep{Jeffreys61} in favor of the model containing the peak, $\mathcal{M}_\mathrm{B}$. The adoption of a Monte Carlo fitting approach, such as that implemented in \diamonds, can produce variations on the output Bayesian evidence of the fit that translate into fluctuations up to the order of $10^{-3}$ in probability. For maximizing the yield of the peak detection test against these potential fluctuations the fits can be repeated multiple times in parallel. Multiple values of $p_\mathrm{BA}$ can therefore be obtained, with the final $p_\mathrm{BA}$ of the peak taken as the maximum of the entire set. If the peak is not deemed significant it is discarded from the set $\{ \nu_{f,i} \}$ of estimated frequencies of the chunk and it is not analyzed further.

For stars having $\Dnu \geq \Dnu_\mathrm{thresh}$, on top of the peak detection test that is carried out in a similar way as for evolved stars, \famed\,\,performs a peak blending test as well. For conducting this test, an additional fit using model $\mathcal{M}_\mathrm{D}$ is done. Here the two Lorentzian profiles of model $\mathcal{M}_\mathrm{D}$ combine together the same prior hyper-parameters adopted to fit each of the two individual profiles by means of model $\mathcal{M}_\mathrm{B}$. The corresponding probability of having two peaks detected at the same time is defined as
\begin{equation}
p_\mathrm{DA} = \frac{\mathcal{E}_\mathrm{D}}{\mathcal{E}_\mathrm{A} + \mathcal{E}_\mathrm{D}} \, ,
\label{eq:p_da}
\end{equation}
with $\mathcal{E}_\mathrm{D}$ the Bayesian evidence of model $\mathcal{M}_\mathrm{D}$. The probability of having only one peak detected is instead the same as Eq.~(\ref{eq:p_ba}). Finally, we can compute a peak blending flag based on
\begin{equation}
p_\mathrm{DB} = \frac{\mathcal{E}_\mathrm{D}}{\mathcal{E}_\mathrm{B} + \mathcal{E}_\mathrm{D}} \, ,
\label{eq:p_db}
\end{equation}
which only plays the role of a discriminant factor between model $\mathcal{M}_\mathrm{D}$ and $\mathcal{M}_\mathrm{B}$, and has not to be interpreted as a detection probability.
The probabilities $p_\mathrm{BA}$, $p_\mathrm{DA}$ are then used according to two different cases: \textit{(i)} for $p_\mathrm{DB} \geq 0.5$ the $\ell = 2,0$ peaks are treated as a blended structure. Here if $p_\mathrm{DA} \ge 0.993$, both the $\ell = 2,0$ modes are considered detected at the same time. If instead $p_\mathrm{DA} < 0.993$, then $p_\mathrm{BA} < 0.993$ and neither the $\ell = 0$ nor the $\ell = 2$ peaks are considered anymore in the subsequent analysis; \textit{(ii)} for $p_\mathrm{DB} < 0.5$, no blending is assumed and only one peak of the $\ell = 2,0$ pair can be considered in terms of significance. If $p_\mathrm{BA} \ge 0.993$, the peak having $\nu_{f,i}$ that is closest to the Lorentzian frequency centroid obtained from the fit of model $\mathcal{M}_\mathrm{B}$ is considered as the detected peak. This means that the detected peak can be either the chunk quadrupole or the radial mode, but not both. If instead $p_\mathrm{BA} < 0.993$ both the $\ell = 0$ and $\ell = 2$ peaks are considered as non detected and are not analyzed further.
 
\subsubsection{Dipole modes}
\label{sec:test_1}
\famed\,\,subsequently performs a peak detection test on the frequency peaks that were flagged as candidate $\ell = 1$ modes in order to be able to select only those that are statistically significant. 

If the star has $\Dnu \ge \Dnu_\mathrm{RG}$, the peak detection test is performed following the approach used for a single peak (Sect.~\ref{sec:test_20}). This means that for each candidate dipole peak, \famed\,\,performs two fits using models $\mathcal{M}_\mathrm{A}$ and $\mathcal{M}_\mathrm{B}$, hence computes the detection probability $p_\mathrm{BA}$ given by Eq.~(\ref{eq:p_ba}). 

For stars with $\Dnu < \Dnu_\mathrm{RG}$, \famed\,\,takes into account an additional complication to the peak detection testing. As shown by C15, stars that are already in the low-luminosity RGB exhibit a large fraction of dipole mixed modes that are unresolved even if observed for more than four years. This suggests that a sinc$^2$ could be a more adequate model than the Lorentzian profile in fitting these narrow peaks. For this purpose, additionally to fitting models $\mathcal{M}_\mathrm{A}$ and $\mathcal{M}_\mathrm{B}$, also model $\mathcal{M}_\mathrm{C}$ is fit (see Appendix~\ref{sec:prior}, Table~\ref{tab:priors_detection} for a list of priors). The detection probability of a sinc$^2$ peak is therefore given as
\begin{equation}
p_\mathrm{CA} = \frac{\mathcal{E}_\mathrm{C}}{\mathcal{E}_\mathrm{A} + \mathcal{E}_\mathrm{C}} \, ,
\end{equation}
where clearly $\mathcal{E}_\mathrm{C}$ is the Bayesian evidence of model $\mathcal{M}_\mathrm{C}$. For assessing whether the peak is better represented by a sinc$^2$ or a Lorentzian profile \famed\,\,computes the quantity
\begin{equation}
p_\mathrm{CB} = \frac{\mathcal{E}_\mathrm{C}}{\mathcal{E}_\mathrm{C} + \mathcal{E}_\mathrm{B}} \, ,
\end{equation}
which similarly to $p_\mathrm{DB}$ plays the role of a discriminant factor between models $\mathcal{M}_\mathrm{C}$ and $\mathcal{M}_\mathrm{B}$. Therefore \famed\,\,identifies two possibilities: \textit{(i)} if $p_\mathrm{CB} > 0.5$ the peak is flagged as a sinc$^2$ profile. Then if $p_\mathrm{CA} \ge 0.993$ the peak is deemed significant, otherwise it is discarded from any further analysis; \textit{(ii)} if $p_\mathrm{CB} \leq 0.5$ the peak is flagged as a Lorentzian profile. Then if $p_\mathrm{BA} \ge 0.993$ the peak is deemed significant, otherwise it is discarded from any further analysis.
Our sinc$^2$ profile test ensures that the statistical significance of dipole mode peaks in evolved stars is not underestimated by erroneously assuming that the oscillation lifetimes are temporally resolved. As a result the sinc$^2$ profile test becomes of increasing importance for decreasing durations of the observing length of the datasets in use (see our application in Sect.~\ref{sec:resolution}).

Nonetheless, the detected candidate dipole modes of the chunk are still considered as candidate dipole modes because one may want to look for additional fine-structure effects such as the presence of rotational multiplets or duplicity inside the peaks (Sect.~\ref{sec:peak_rotation_test}), and because one peak among them could be a $\ell = 3$ oscillation mode (Sect.~\ref{sec:octupole_test}).

\begin{figure*}[t]
   \centering
  \includegraphics[width=18.3cm]{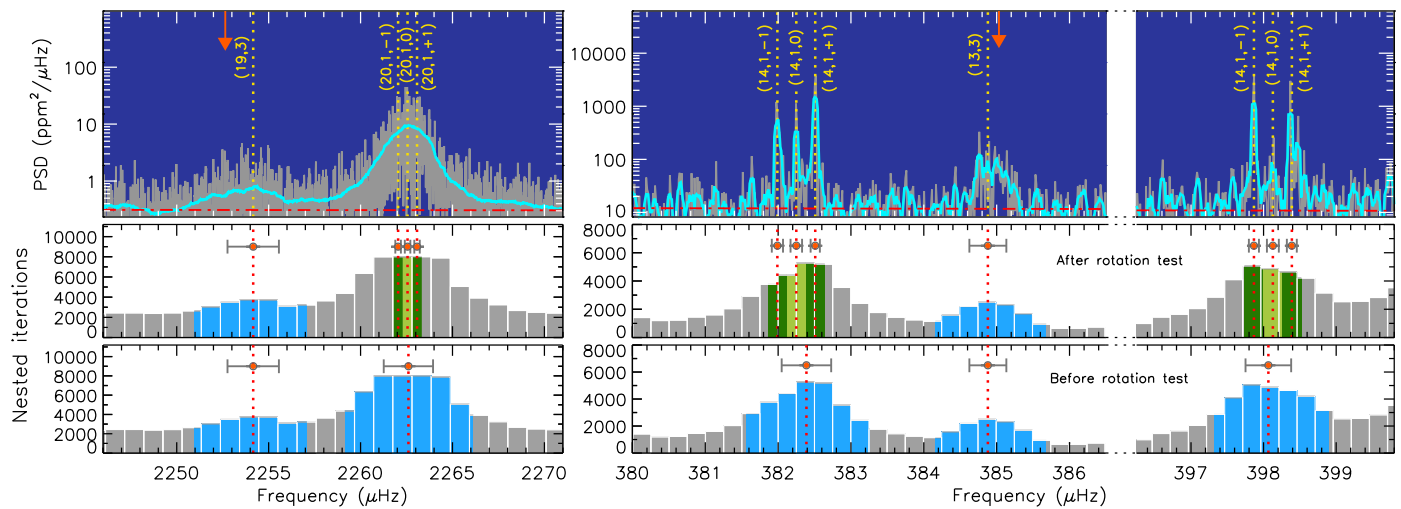}
    \caption{Detection of rotation and of octupole modes in the MS star KIC~12069424 (left panels) and in the late SG star KIC~4351319 (right panels) using a scheme similar to Fig.~\ref{fig:doublet}. \textit{Top panels}: a close-up to the chunk $\ell = 1$ rotational triplet and its adjacent $\ell = 3$ for the MS star, and to the $\ell = 1$ rotational triplets of two mixed modes of the same chunk for the late SG star, with the $\ell = 3$ mode included. The mode identification $\left(n,\ell,m\right)$ as obtained by \famed\,\,is indicated. The orange downward arrows mark the asymptotic position of the $\ell = 3$ modes. \textit{Middle and bottom panels}: ASEF showing $\nu_{f,i}$, $\sigma_{f,i}$, $r_{a,b}$, before and after the peak rotation tests.}
    \label{fig:rotation}
\end{figure*}

\subsection{Peak rotation and duplicity tests}
\label{sec:peak_rotation_test}
A second phase in the peak testing focuses on the detected candidate dipole modes. When analyzing these peaks, the resolving power of the chunk multimodal sampling may not be enough to recognize the presence of a possible fine structure within the peaks. This resolving power issue could show up in two different situations. Firstly, for SG and especially MS stars because the dipole mode linewidths are comparable with or even significantly larger than the frequency splitting among the rotationally split components. Secondly, for low-luminosity RGB and especially RC stars because of the low $\Dnu$ values and high dipole mixed mode density. Here the dipole mixed modes that exhibit a dominant $p$ character are significantly condensed in frequency range, exhibit larger linewidths than the g-dominated ones, hence are more difficult to resolve. Here a single chunk ASEF peak could be represented by: \textit{a)} a single mixed mode split by rotation; \textit{b)} two mixed modes not split by rotation but sufficiently close to one another to partially blend; \textit{c)} two neighboring rotational components, each one belonging to a different mixed mode; \textit{d)} two rotational components of the same dipole mixed mode that were extracted separately from a third component.

In summary, a candidate $\ell = 1$ mode could happen to be characterized by the effect of rotation or by the presence of duplicity. For exploring these different occurrences, \famed\,\,exploits another set of peak fitting models. Similarly to the set for peak detection and blending tests, these models are provided within the PeakBagging code and are:
\begin{itemize}
\item $\mathcal{M}_\mathrm{E}$ is a model to test the hypothesis that the frequency peak is a single Lorentzian profile, meaning that it does not contain any fine structure inside. This model is similar to model $\mathcal{M}_\mathrm{B}$ but has a fixed background level, that is it accounts for the three free parameters of frequency centroid, amplitude, and linewidth ($\nu_{i,0}$, $A_i$, $\Gamma_i$, respectively);
\item $\mathcal{M}_\mathrm{F}$ is a model to test that the frequency peak is a dipole oscillation peak split by rotation. It has five free parameters, namely the frequency centroid, amplitude, and linewidth of the Lorentzian profile, as well as the rotational splitting and stellar spin inclination angle of the rotational multiplet ($\nu_{i,0}, A_i, \Gamma_i, \delta\nu_\mathrm{rot}, \cos i$, respectively, where the spin inclination angle $i$ is in the form of $\cos i$ to flatten out the sin law of the isotropic distribution, see \citealt{Corsaro17spin}). The rotational multiplet accounts for $2 \ell + 1 = 3$ (azimuthal) $m$ components, and changes its shape following the formulation by \cite{Gizon03}. Similarly to model $\mathcal{M}_\mathrm{E}$, this model has a fixed background;
\item $\mathcal{M}_\mathrm{G}$ is a model to test that the frequency peak is constituted by two adjacent dipole oscillation peaks, each one represented by a single Lorentzian profile. It is analogous to model $\mathcal{M}_\mathrm{D}$ but has six free parameters, namely the frequency centroid of the left-side Lorentzian profile, the (positive) frequency separation between the two Lorentzian profiles, and the amplitudes and linewidths of the two Lorentzian profiles ($\nu_{i,0}^1,\delta\nu_\mathrm{split}, A_i^1,A_i^2,\Gamma_i^1,\Gamma_i^2$, respectively, with superscripts $1$ and $2$ indicating the left- and right-side peaks of the duplet, and with the centroid frequency of the right-side peak given as $\nu_{i,0}^1 + \delta\nu_\mathrm{split}$). Similarly to models $\mathcal{M}_\mathrm{E}$ and $\mathcal{M}_\mathrm{F}$, this model uses a fixed background level.
\end{itemize}
These models do not require that the background level is varying because this hypothesis was already tested during the peak detection test. The purpose of this second set of peak testing models is to assess whether a candidate dipole mode is either split by rotation (peak rotation test) or constituted by two different oscillation peaks (peak duplicity test). The prior distributions are uniform even for these fitting models, following the discussions in Sects.~\ref{sec:fwhm_radial} and \ref{sec:test_20}. The prior hyper-parameters adopted on the model free parameters are detailed in Appendix~\ref{sec:prior}, Table~\ref{tab:priors_rotation}. 
For each peak test, the portion of PSD used for the fit is always delimited by the frequency divisions $d_{a,b}$ of the given peak. If the frequency resolution of the dataset is comparable to the upper prior bound on the rotational splitting the rotation test is not performed because we cannot resolve the fine-structure of the peak. A similar discussion is applied to the peak duplicity test.

The fitting model $\mathcal{M}_\mathrm{F}$ does not incorporate any asymmetry in the splitting components \citep[e.g., see][]{Benomar18Science}, meaning that: 1) the rotational splitting $\delta\nu_\mathrm{rot}$ is the same on either side of the multiplet; 2) the linewidth is the same for all the rotational components of the multiplet. A more careful inspection of the rotational components aimed at reaching a higher level of accuracy on the estimated frequencies will be in charge of \texttt{COMPLETE}.

The peak rotation and duplicity tests performed by \famed\,\,represent an entirely novel application of the Bayesian model comparison process done with \diamonds. The peak rotation test, in particular, allows identifying in a complete automated manner the azimuthal components for those dipole modes that are split by rotation in stars from MS to late SGs and low-luminosity RGB, with some cases extending up to RC stars (see our application in Sect.~\ref{sec:rg}), hence to obtain an estimate of the spin inclination angle of the star directly from the parameter estimation of $\cos i$ in the fitting model $\mathcal{M}_\mathrm{F}$.

Following the approach used for the peak detection and blending tests, we can define the probability that a peak is split by rotation as
\begin{equation}
p_\mathrm{FE} = \frac{\mathcal{E}_\mathrm{F}}{\mathcal{E}_\mathrm{E} + \mathcal{E}_\mathrm{F}} \, ,
\label{eq:p_fe}
\end{equation}
where $\mathcal{E}_\mathrm{E}$ and $\mathcal{E}_\mathrm{F}$ are the Bayesian evidences of models $\mathcal{M}_\mathrm{E}$ and $\mathcal{M}_\mathrm{F}$, respectively. \famed\,\,therefore considers that the peak is better represented by a rotational multiplet if a weak evidence condition is met, which according to the Jeffreys' scale of strength translates into $p_\mathrm{FE} \ge 0.75$. The weak evidence condition is less stringent than that used for the peak detection test because we want to maximize the possibility to have the rotational effect detected in the peak. If however preferred by the user, one could strengthen the condition on the evidence by increasing the threshold on $p_\mathrm{FE}$, for example to that of a moderate evidence ($\ge 0.923$) or a strong evidence ($\ge 0.993$). As shown in our applications in Sect.~\ref{sec:applications}, a weak evidence threshold appears to represent a favorable condition for testing the presence of rotation.

For stars evolved into RG on top of the peak rotation test the candidate dipole peaks are checked against mode duplicity. The probability that a peak is a doublet is defined as
\begin{equation}
p_\mathrm{GE} = \frac{\mathcal{E}_\mathrm{G}}{\mathcal{E}_\mathrm{E} + \mathcal{E}_\mathrm{G}} \, ,
\end{equation}
with $\mathcal{E}_\mathrm{G}$ the Bayesian evidence of model $\mathcal{M}_\mathrm{G}$. By using a flag based on the quantity
\begin{equation}
p_\mathrm{GF} = \frac{\mathcal{E}_\mathrm{G}}{\mathcal{E}_\mathrm{F} + \mathcal{E}_\mathrm{G}} \, ,
\end{equation}
we can thus identify two conditions: \textit{(i)} if $p_\mathrm{GF} > 0.5$ the peak is considered as a potential doublet. In this condition the duplicity is confirmed if $p_\mathrm{GE} \ge 0.75$, using again a weak evidence condition as it is done for the peak rotation test; \textit{(ii)} if $p_\mathrm{GF} \leq 0.5$ the peak is considered as a potential dipole rotational triplet. Then, the peak rotation test is carried out using Eq.~(\ref{eq:p_fe}) by verifying the condition $p_\mathrm{FE} \ge 0.75$.
If the peak to be tested is significant but is flagged as a sinc$^2$ profile ($p_\mathrm{CB} > 0.5$ and $p_\mathrm{CA} \ge 0.993$) then it is excluded from the peak rotation and duplicity tests. This is because if the peak has a FWHM comparable to the frequency resolution of the dataset, there is no point to inspect its fine-structure. 

The left panels of Fig.~\ref{fig:rotation} show an example of the peak rotation test conducted on a chunk dipole mode of a MS star. Here the frequency corresponding to the central azimuthal component ($m=0$) extracted by \famed, has a remarkable agreement (on the order of $10^{-3}$\,\%) with the frequencies published by \cite{Davies15} and by \cite{Lund17LEGACY} for the same oscillation mode. \cite{Davies15} has also inferred a global spin inclination angle for 16 Cyg A of $i = 56^{+6}_{-5}$\,deg, which is in excellent agreement with the value of about $59$\,deg obtained by \famed\,\,from the peak rotation test. Another application of the peak rotation test, but this time for an evolved SG star observed by \kepler for more than four years and having $\Dnu \simeq 24\,\mu$Hz, is shown in the right panels of the same figure. Here we can see how the clear rotational triplets of two different mixed modes belonging to the same chunk have been correctly extracted and identified. The average stellar spin inclination angle obtained by \famed\,\,is of about 77\,deg, in line with an expected high-inclination configuration as already visible from the prominent $m=\pm 1$ components of the rotational multiplets (see our application in Sect.~\ref{sec:sg} and Fig.~\ref{fig:sg_application2} too).

We can instead see an example of the application of both the rotation and duplicity tests in Fig.~\ref{fig:duplicity} for a low-luminosity RGB star (see also Sect.~\ref{sec:rg}). On the PSD segment shown in the left panels, C15 has detected and identified two rotationally split dipole mixed modes (four peaks in total), corresponding to a configuration of high stellar spin inclination (hence only $m = \pm 1$ components). The ASEF obtained by \famed\,\,is not resolving the two close peaks in the central region, where each peak is the rotational component of a different mixed mode. By means of the peak duplicity test \famed\,\,is able to recover the actual frequency positions of the two peaks ($p_\mathrm{GF} > 0.5$ and $p_\mathrm{GE} \ge 0.75$), which show excellent agreement with those obtained in the literature. In the other PSD segment illustrated in the right side of Fig.~\ref{fig:duplicity} we can see the opposite situation, with a single ASEF peak decomposed after rotation has been detected ($p_\mathrm{GF} \le 0.5$ and $p_\mathrm{FE} \ge 0.75$). The mode identification obtained for this oscillation mode is in agreement with that found by C15 using a manual approach. In addition, C15 did not report the $m=0$ component of the rotational triplet because not well visible as they did not fit rotational multiplets but individual peaks only. This frequency component is instead delivered by \famed\,\,thanks to the peak rotation test. \texttt{\'ECHELLE} will however allow assigning an azimuthal number identification even to peaks that have been originated from peak doublets, namely not associated to any rotational multiplet within \chu.

The fits conducted for the peak rotation and duplicity tests are all unimodal and make use of the nested sampling configuration presented in Sect.~\ref{sec:peak_detection_test}. Nevertheless, if requested by the user the peak rotation (and as a consequence also the duplicity) test(s) can be deactivated, meaning that the output produced by \famed\,\,is  simplified by excluding any information on the azimuthal number mode identification, and on the stellar spin inclination angle. This condition could be useful if, for example, we know a priori that the datasets do not allow us to infer the rotation from the oscillation modes, and allows to significantly speed up the computation.

\begin{figure}[t]
   \centering
  \includegraphics[width=8.9cm]{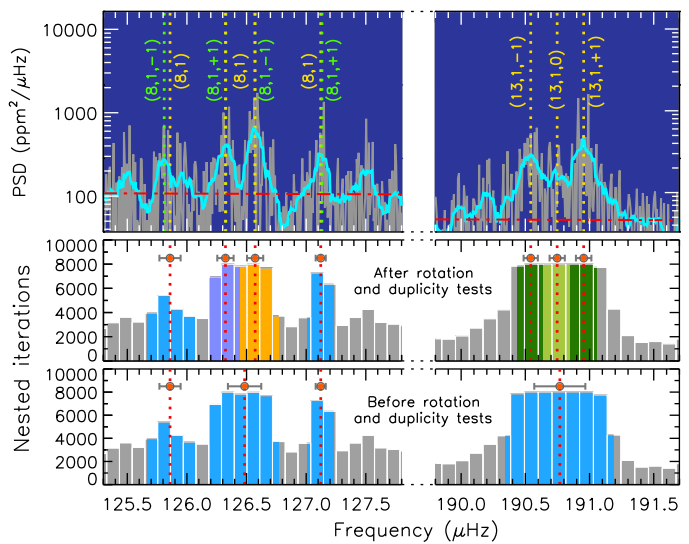}
    \caption{Detection of rotation and mode duplicity in the RG star KIC~12008916. Similar description to Fig.~\ref{fig:rotation} but showing the case of two different portions of the stellar PSD where both the peak rotation and duplicity tests have been performed. \textit{Left panels}: one of the ASEF local maxima is split in two peaks by duplicity. \textit{Right panels}: the ASEF local maximum is split in three peaks by rotation. The mode identification in green shows $m$ for the individual peaks as identified by C15.}
    \label{fig:duplicity}
\end{figure}

\subsection{Octupole modes}
\label{sec:octupole_test}

The last task in the analysis performed by \chu\,\,is that of searching for the presence of a possible $\ell = 3$ mode. 
This is accomplished independently of whether the peak rotation and duplicity tests are carried out, although we recommend having the latter tests performed for late SG and RG stars to further improve the reliability of their octupole mode detection.

If the star has $\Dnu \ge \Dnu_\mathrm{RG}$, the search for a $\ell = 3$ mode is conducted only if there are at least two significant candidate dipole modes. If only one candidate dipole mode is present, \famed\,\,privileges the mode identification of a $\ell = 1$ mode over that of a $\ell = 3$ mode given the higher mode visibility of the former. However, if the star has been flagged as a potential depressed dipole mode star (see Sect.~\ref{sec:sliding_depressed}), the octupole mode search is executed even if only one candidate dipole mode is detected. Here $\ell = 3$ modes may still be clearly visible because not suppressed (see our application in Sect.~\ref{sec:depressed}).

If the star has $\Dnu < \Dnu_\mathrm{RG}$, then the search for an octupole mode is done if at least one significant candidate dipole mode is obtained. Similarly to the case of less evolved stars, this guarantees that $\ell = 3$ modes in RGs can still be found even if no dipole modes are present, which clearly incorporates stars that have depressed dipole modes in this $\Dnu$ range.

In all cases introduced \famed\,\,computes a frequency range within which the potential octupole mode could be located. This frequency range is calculated from the asymptotic pattern of $p$ modes, which shows that in MS stars the $\ell = 3$ mode of a chunk is always placed below its neighboring $\ell = 1$ mode, and that its frequency is correlated with the value of $\delta\nu_{02}$ for all stars from MS to RGs. In particular, this is predicted by the small frequency separation $\delta\nu_{03}$ of the asymptotic relation presented in Eq.~(\ref{eq:asymp}). While in RGs the asymptotic position of the $\ell = 3$ modes is predictable with quite good accuracy, especially along the RGB, because it follows a tight $\Dnu$-$\delta\nu_{03}$ relation \citep{Huber11}, in MS and SG stars it is more uncertain because of the large variation of $\delta\nu_{02}$ found for these stars. In this latter case we can still exploit an useful approximation to express the small frequency spacings $\delta\nu_{0\ell}$ presented in Eq.~(\ref{eq:asymp}) as a function of $\delta\nu_{02}$ \citep[e.g., see][]{BK03}, namely 
\begin{equation}
\delta\nu_{0\ell} \simeq \frac{\ell \left( \ell + 1 \right)}{6} \delta\nu_{02} \, .
\end{equation}
From this relation it follows that $\delta\nu_{03} \simeq 2 \delta\nu_{02}$. By assessing $\delta\nu_{03}$ from mock samples prepared by \cite{Ball18TESS} we have experienced that $\delta\nu_{03}$ can vary between 1.5 and 4 times $\delta\nu_{02}$ in MS stars, and between 1.8 and 3 times $\delta\nu_{02}$ in SG stars. We have therefore implemented these limiting bounds to define a lower and upper frequency value to search for the octupole mode inside each chunk. For RG stars instead, we adopt as frequency range that obtained by varying $\delta\nu_{03}$ obtained from the $\Dnu$-$\delta\nu_{03}$ relation by \cite{Huber11} between $+12$ and $-16$\,\%, allowing for a wider lower range to account for the reduction in $\delta\nu_{02}$ found in RC stars \citep[e.g., see][]{Kallinger12}.

\famed\,\,therefore checks whether inside the search frequency range for octupole modes there is at least one $\nu_{f,i}$ fulfilling two constraints: 1) it corresponds to a detected peak according to the peak detection test (Sect.~\ref{sec:peak_detection_test}); 2) it is not flagged as a sinc$^2$ profile according to the sinc$^2$ profile test (Sect.~\ref{sec:test_1}). If a sinc$^2$ profile is better at reproducing the peak, assuming that the observation is sufficiently long to allow resolving the radial mode peaks of the star, then the FWHM of the mode is significantly smaller than that of an expected $\ell = 3$, which should instead be comparable to $\Gamma_0$, the linewidth of the chunk radial mode (see C15). It is then obvious that in such case there is no point of proceeding further. Additionally, the detection of octupole modes becomes less likely with the shortening of the observations because it is more likely that a peak is flagged as a sinc$^2$ profile (see the discussion in Sect.~\ref{sec:test_1}). 

If instead at least one frequency fulfilling the two constraints is found, \famed\,\,proceeds by evaluating the FWHM of the corresponding peaks by fitting model $\mathcal{M}_E$ to the peaks similarly to what has been done to evaluate $\Gamma_0$ (see Sect.~\ref{sec:fwhm_radial}). This procedure has the purpose of finding out the candidate octupole mode having the largest FWHM of the set, hereafter referred to as $\Gamma_\mathrm{3}$.

At this stage, the octupole mode test can be executed on the candidate octupole mode that has the largest FWHM, through the following two steps. In the first step, if the peak rotation and duplicity tests have been performed, \famed\,\,checks whether the peak is either a dipolar rotational triplet or a doublet, or none of the two. For RG stars: \textit{(a)} if the peak is a $\ell = 1$ rotational triplet ($p_\mathrm{GF} \leq 0.5$ and $p_\mathrm{FE} \ge 0.75$), then its mode identification remains that of a $\ell = 1$ mode by definition, and the peak is discarded from being a candidate $\ell = 3$ mode; \textit{(b)} if the peak is neither a rotational multiplet nor a doublet ($p_\mathrm{FE} < 0.75$ and $p_\mathrm{GE} < 0.75$), then \famed\,\,activates a further control check imposed on the second step; \textit{(c)} if the peak is a doublet ($p_\mathrm{GF} > 0.5$ and $p_\mathrm{GE} \ge 0.75$) the FWHM of each peak of the doublet, as obtained from fitting model $\mathcal{M}_\mathrm{G}$, is compared to $\Gamma_0$. If at least one of the two peaks of the doublet has a FWHM comparable to $\Gamma_0$, then similarly to point (b) \famed\,\,executes the second step. If none of the two peaks has a FWHM comparable to $\Gamma_0$, then the doublet is considered as a pair of $\ell = 1$ modes and the octupole mode test is stopped.
When the star is classified as either SG or MS, only (a) and (b) hold, with (b) discarding the information on $p_\mathrm{GE}$ because the peak duplicity test is not done for such stars. If instead the peak rotation test was not performed because either deactivated by the user, or because the frequency resolution did not allow to resolve any fine-structure inside the peaks, then \famed\,\,proceeds directly with the second step independently of the evolutionary stage classification of the star. In the second step, the ASEF value of the local maximum corresponding to the candidate octupole mode has to be smaller ($< 3/4$) than the limiting value imposed by the number of nested iterations. This is because of the low mode visibility of $\ell = 3$ modes, which prevents the chunk multimodal sampling from producing ASEF peaks that saturate to the limit. If $\Gamma_\mathrm{3}$ is comparable to $\Gamma_0$, and the above condition on the ASEF maximum is satisfied, the mode is finally flagged as a real $\ell = 3$ oscillation mode, having radial order given by that of the chunk radial mode minus two. Otherwise the octupole mode test is terminated without any $\ell = 3$ mode detection.

In case we have an octupole mode detection and the mode is a doublet peak as verified according to step \#1(c), before terminating \chu\,\,the peak is split up in its two components. Here the peak of the doublet that has the largest FWHM is kept as the real $\ell = 3$ mode of the chunk, while the other peak of the doublet is flagged as a $\ell = 1$ oscillation mode with radial order given by that of the chunk radial order minus one. In this regard step \#1(c) can be useful for the case of a RG star in which the $\ell = 3$ mode is falling very close to an adjacent dipole mixed mode, the latter being sensibly narrower than the $\ell = 3$ because of its partial $g$-mode character. The octupole mode test so introduced is therefore not only a method to detect $\ell = 3$ modes in the stellar PSD but to also disentangle them from possible $\ell = 1$ mixed modes that happen to be very close in frequency.

An example of the application of the octupole mode test for both a MS and an evolved SG star is shown in Fig.~\ref{fig:rotation}. The asymptotic prediction for the octupole mode in the MS star is accurate to within 0.1\,\%, and the octupole mode search range is sufficiently wide to locate the real $\ell =3$ peak in the PSD. The frequency estimated by \famed\,\,for the $\ell = 3$ oscillation mode of the MS star thus agrees on the order of $10^{-2}$\,\% with those extracted by \cite{Davies15} and by \cite{Lund17LEGACY} for the same oscillation mode, with also the radial order $n$ matching the literature value. For the evolved SG star presented in the same figure, the extracted octupole mode is found within 0.04\,\% from the predicted asymptotic value. Other examples of the effectiveness of the octupole mode test in different stellar evolutionary stages, as well as different S/N and frequency resolution conditions of the datasets, can be found in most of the applications shown in Sect.~\ref{sec:applications}.

\subsection{Detailed asteroseismic information}
\label{sec:detailed}
Thanks to the peak rotation test presented in Sect.~\ref{sec:peak_rotation_test}, \chu\,\,can even apply a mode identification to each $\ell = 1$ mode that is of the type $\left(n, \ell, m \right)$ in all stars ranging from MS up to late SGs and early RGB. For evolved RG stars instead, the dipole mixed modes usually have a mode identification that does not include the azimuthal component, hence of the type $\left(n, \ell \right)$, where the radial order of all the dipole mixed modes of the chunk is set to be equal to that of the chunk radial mode minus one. This is because in more evolved stars the rotational components of each dipole mixed mode may have frequency separations comparable to the spacing between adjacent mixed modes. This means that these rotational components are usually detected separately from one another from the chunk ASEF. In addition, the large rotational splitting can yield confusion when the rotational components of different dipole mixed modes cross to one another in frequency \citep[so-called crossing effects, e.g.,][]{Gehan18rotation}. Obtaining the azimuthal numbers for all the $\ell =1$ mixed modes detected in RG stars requires a dedicated analysis on the individual oscillation frequencies, which is in charge of \texttt{\'ECHELLE}, as it will be discussed in a forthcoming paper. Some exceptions to the mode identification scheme obtained by \chu\,\,for RGs may however occur when a rotational multiplet is contained within a single ASEF peak like for less evolved stars. This condition could be verified for p-dominated mixed modes, especially at high oscillation frequency, where the linewidth is larger, and in general for those RG stars where the linewidths are larger because of their higher $\teff$, such as the 2$^\mathrm{nd}$ RC stars (see Sect.~\ref{sec:rg} for an application). The rotational components of the $\ell = 1$ modes are computed at the end of \chu, and only after the octupole mode test has been performed.

Another quantity that is provided by \chu\,\,is the observed period spacing of the mixed dipole modes, $\Delta P_1$, which is computed locally for each chunk only if at least two $\ell = 1$ mixed modes are found within the same chunk. If more than two $\ell = 1$ modes are found, the local $\Delta P_1$ is computed as the mean value among the set of $\Delta P_1$ values that can be obtained, each one calculated using a pair of adjacent mixed modes. The value of $\Delta P_1$ can of course be used to understand whether a star is a RGB or a RC following the same approach presented by \cite{Bedding11Nature} using the $\Dnu$-$\Delta P_1$ diagram \citep[see also][]{Kallinger12,Corsaro12cluster,Stello13}.

The $\ell = 3$ modes extracted with \chu\,\,have a mode identification of the type $\left( n, \ell \right)$. The occurrence of the $\ell = 3$ modes identified by \famed\,\,can vary significantly from star to star, and it is to a large extent a strong function of the S/N of the oscillations in the stellar PSD (see the applications in Sects.~\ref{sec:noise} and ~\ref{sec:resolution}). Similarly for the case of $\ell = 1$ modes, the search for $\ell = 3$ modes is facilitated by the adoption of a data-driven approach based on the multimodal sampling.

Nevertheless, the oscillation mode frequencies extracted at the level of \chu, especially in the case of $\ell = 1$ mixed modes, are not yet tested in the light of the theoretical asymptotic patterns. This assessment is afterwards carried out by \texttt{\'ECHELLE}, and it is used by \famed\,\,to provide an additional level of validation to the extracted oscillation frequencies. In summary, \chu\,\,offers a detailed set of oscillation mode frequencies for each star, which thanks to their high level of precision and accuracy (often well below $0.1$\,\%) can already be suitable for subsequent stellar and asteroseismic modeling purposes. 

\section{Applications}
\label{sec:applications}
In this section we show different applications of the \famed\,\,pipeline that comprise stars spanning from F-type MS up to RC. These stars have already been investigated in the literature, so that a direct comparison with published results can be made. The configuring parameters of the pipeline described in this paper are the same for all the applications shown in this section, indicating that the current setup can already cover the analysis of a wide range of stellar fundamental properties and evolutionary stages. Where possible, the applications are illustrated in the form of \'echelle diagrams \citep{Grec83}, otherwise a direct visualization of the stellar PSD is given. For all the stars considered in this section, we have performed a preliminary background fit by means of the Background code (Sect.~\ref{sec:background}). All the datasets for the stars observed with NASA \kepler that are used in this paper were optimized for asteroseismic analysis following \cite{Garcia11,Garcia14gap} and \cite{Pires15}.

\begin{figure}[t]
   \centering
  \includegraphics[width=8.9cm]{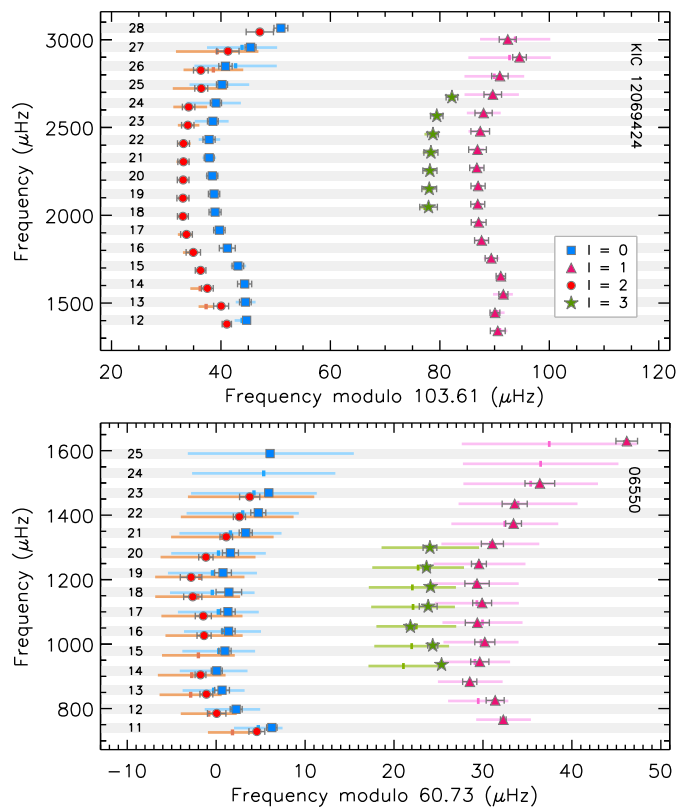}
    \caption{\'Echelle diagrams for the MS stars KIC~12069424 and the simulated one 06550. The colored symbols represent the frequencies extracted by \chu\,\,for angular degrees $\ell = 0,1,2,3$ (see legend), with their associated 1-$\sigma$ uncertainties, while the colored horizontal bars indicate the literature values varied by $\pm 1$\,FWHM of each peak, with the centroid marked by a vertical tick. For KIC~12069424, the adopted literature FWHM are the same for all the angular degrees within a single radial order, as provided by \cite{Lund17LEGACY}. The numbers on the left side of each panel indicate the radial order $n$ of the corresponding radial mode as obtained by \famed. The folding value of $\Dnu$ is $\Dnu_0$ obtained from \glob\,\,for the real star, and the one computed by \cite{Ball18TESS} for the simulated one. The x-axis range is adjusted to have the $\ell =2,0$ ridges appearing on the left side.}
    \label{fig:ms_application}
\end{figure}

\subsection{Main-sequence stars}
\label{sec:ms}
The first application that we present is that of the G-type MS star 16 Cyg A. This star, classified as a solar analog and having $\Dnu \simeq 104\,\mu$Hz and $\teff = 5825$\,K, is one of the best known and accurately analyzed dwarfs in terms of peak bagging analysis \citep{Metcalfe12,Lund17LEGACY}. This is because this star has been observed by \kepler for more than four years with high duty cycle, and exhibits excellent S/N oscillations in the stellar PSD that allow for the identification of a large number of oscillation modes. Most of the $\ell = 2,0$ modes are well disentangled from one another. Many $\ell = 3$ modes can be identified as well through the oscillation envelope. As shown in the top panel of Fig.~\ref{fig:ms_application} for this star, \famed\,\,can extract oscillation modes of different angular degrees throughout the frequency range investigated by \cite{Lund17LEGACY}, for a total of 58 oscillation mode frequencies correctly identified over 18 radial orders. In all cases, the estimated frequencies match within the literature FWHM, which is often smaller than the size of the symbol plotted in the figure. Additionally, \famed\,\,identifies and extracts two $\ell = 3$ modes for radial orders 18 and 19, one $\ell = 2$ mode for radial order $11$, and two more $\ell = 2,0$ modes for radial orders 27 and 28, respectively, that are not in the list published by \cite{Lund17LEGACY}. We note that the discrepancies between the \famed\,\,and literature set of frequencies may have the tendency to increase for very high and/or very low frequency modes. This can be explained by a combination of very large mode linewidths and low oscillation amplitudes for high frequency modes, and by the low amplitudes and high noise level for low frequency modes, which make it difficult to accurately locate the actual oscillation peak centroids independently of the approach that is used.

\begin{figure}[t]
   \centering
  \includegraphics[width=8.9cm]{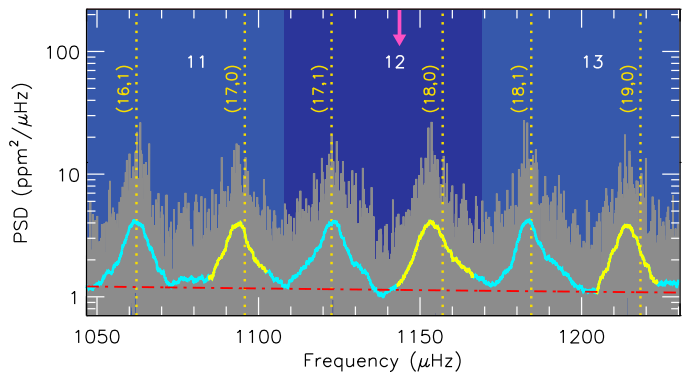}
    \caption{Three most central chunks of the stellar PSD of the MS star 06550 as seen from \glob, with similar description as in Fig.~\ref{fig:global}. The radial modes are slightly offset to the right side of the corresponding $\ell = 2,0$ peaks (yellow smoothing) as the result of the correction $\delta\nu_0$ (see Sect.~\ref{sec:skimming}). The frequencies of the radial modes obtained by \famed\,\,are more accurate than the corresponding peak maxima in the smoothed PSD because each of these peaks also contains the adjacent $\ell = 2$ modes in an almost entirely blended structure with the $\ell = 0$ modes.}
    \label{fig:hot_global}
\end{figure}

The second application is that of the F-type MS star 06550, which has been simulated by \cite{Ball18TESS} to reproduce a one year-length NASA TESS-like observation. The peculiarity of this target is that it has $\teff = 6416$\,K, hence it exhibits large oscillation mode linewidths throughout the oscillation envelope. The large mode linewidths coupled to $\Dnu \simeq 61\,\mu$Hz, hence a relatively small $\delta\nu_{02}$, cause a strong peak blending for both $\ell = 2,0$ and $\ell = 3,1$ modes in most cases. The mode identification is made challenging because of the confusion caused by the peak blending, which does not allow to understand whether any of the peak structures are either a $\ell = 2,0$ or a $\ell = 3,1$ mode pair. This effect is well observable in Fig.~\ref{fig:hot_global}, where the shape of the peaks associated to the two different mode pairs ($\ell = 2,0$ and $\ell = 3,1$) are very similar to one another and cannot be distinguished using only a visual inspection. Here the adoption of the sliding-pattern fit allows estimating the true reference radial mode frequency of the star. The final peak bagging result obtained with \chu\,\,is instead shown in the bottom panel of Fig.~\ref{fig:ms_application}. The comparison set used for this application is that of the theoretical frequencies and of their simulated linewidth. Despite the strong blending effects, as visible by the prominent superposition of mode linewidth for adjacent oscillation peaks, \famed\,\,correctly distinguishes the $\ell = 2,0$ mode pairs from those containing the $\ell = 3,1$ modes throughout the stellar PSD. The $\ell = 0$ and $\ell = 1$ modes with radial order 24, and the $\ell = 2$ mode with radial order 23, were not found because they did not pass neither the peak blending test nor the peak detection test, while the $\ell = 2$ mode of radial order 14 was not detected because it did not pass the peak blending test. Moreover, \famed\,\,can correctly locate and identify seven different $\ell = 3$ modes, thus providing for this simulated target a total of 47 oscillation mode frequencies covering 15 radial orders.

The value of the large frequency separation estimated from \glob, $\Dnu_0$ is accurate enough to vertically align the extracted mode frequencies in the \'echelle diagram. For KIC~12069424, $\Dnu_0$ agrees within 0.3\,\% with the one estimated by \cite{Lund17LEGACY}, while for the simulated star 06550 we find an agreement of about 0.8\,\%. This shows that \glob\,\,can be used to compute accurate values of $\Dnu$ in MS stars. In this way, \glob\,\,alone offers the opportunity to compute reliable estimates of stellar mass and radius for MS stars from scaling relations.

\subsection{Subgiant and early red-giant stars}
\label{sec:sg}
SGs and stars that have recently entered the RGB are challenging targets to analyze in terms of their oscillations. In these stars the period spacing of dipole $g$ modes, $\Delta \Pi_1$, can be significantly larger than that of stars that are more evolved along the RGB. In addition $\Delta\Pi_1$ can change sensibly from star to star depending on stellar mass and mean density, making it difficult to identify a simple $\Dnu$-$\Delta \Pi_1$ relation \citep{Benomar13}. This can in turn cause the dipolar mixed modes to be displaced all over the frequency range of a given radial order, and as a result these mixed modes can even be found inside the regions containing $\ell = 2,0$ modes, thus partially or entirely blending with them in some cases. The identification of $\ell = 2,0$ mode pairs is further complicated for early SGs because the linewidth of $\ell = 1$ mixed modes is still comparable to that of pure $p$ modes, thus yielding to particularly messy oscillation patterns in some cases.

\begin{figure}[t]
   \centering
  \includegraphics[width=8.9cm]{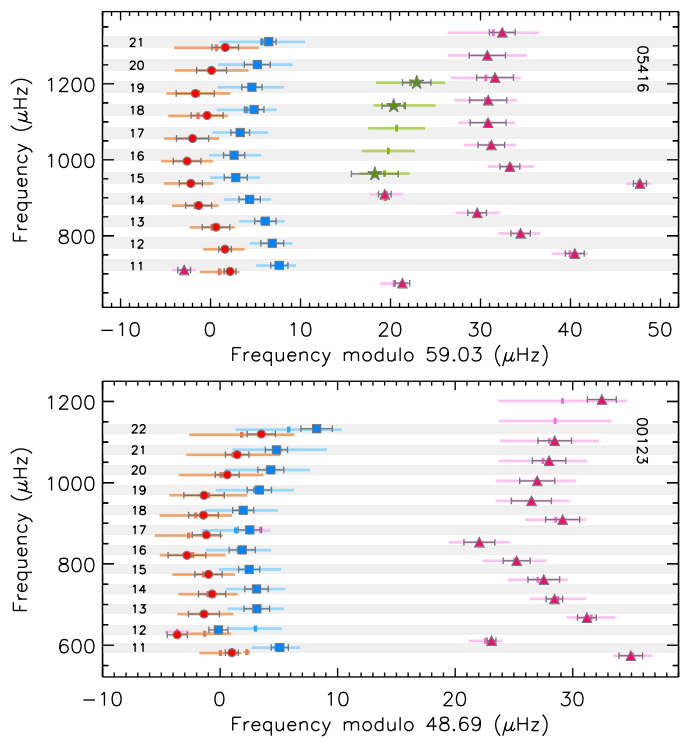}
    \caption{\'Echelle diagrams for the simulated SGs 05416 and 00123, with literature values from \cite{Ball18TESS}. Same description as for the bottom panel of Fig.~\ref{fig:ms_application}. The folding $\Dnu$ values are from the literature.}
    \label{fig:sg_application}
\end{figure}

\begin{figure}[t]
   \centering
  \includegraphics[width=8.9cm]{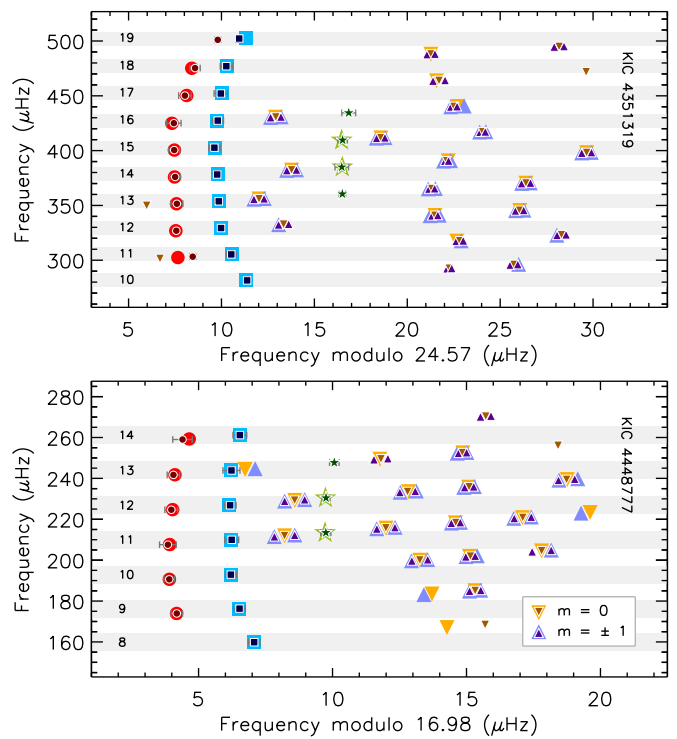}
    \caption{\'Echelle diagrams for the late SG star KIC~4351319 and for the early RGB star KIC~4448777. Similar symbol description as for the top panel of Fig.~\ref{fig:ms_application}. The frequency values obtained by C15 for KIC~4351319 (Di Mauro et al. in prep) and for KIC~4448777 \citep{DiMauro18} are shown using big symbols and light colors, with only detected oscillation modes considered. The values obtained by \famed\,\,are indicated by small symbols and dark colors. The azimuthal components of the $\ell = 1$ rotational triplets are distinguished with different colors and symbols (see legend). No uncertainties are shown for $\ell = 1$ modes because smaller than the symbol sizes in most cases.}
    \label{fig:sg_application2}
\end{figure}

For showing the capabilities of \famed\,\,in handling these targets, we have selected four different stars covering the regime of subgiants and early RGB. In the first case, we consider the simulated late F-type star 05416, which  corresponds to a one-year length NASA TESS-like observation \citep{Ball18TESS}. Given that this subgiant is not significantly evolved, with $\Dnu \simeq 58\,\mu$Hz, most of the chunks of its PSD contain only one $\ell = 1$ mixed mode each. However, the mixed modes are often displaced all over the range between two consecutive radial modes. The relatively high temperature of this target, $\teff = 6139$\,K, causes a partial peak blending for the $\ell = 2,0$ mode pairs and in one case with an adjacent $\ell = 1$ mixed mode as well. The top panel of Fig.~\ref{fig:sg_application} shows the comparison between the frequencies from \famed, for a total of 39 over 12 radial orders, and those by \cite{Ball18TESS}. We can see that there is a remarkable agreement between the two sets, including the $\ell =1$ mixed mode having radial order 10 identified next to the $\ell = 2$ mode of the same chunk. Three different $\ell = 3$ modes could be identified too, albeit the large mode linewidths and low octupole mode visibilities make them difficult to be clearly recognized in the PSD.

The second example is that of another subgiant simulated by \cite{Ball18TESS}, 00123, having $\Dnu \simeq 49\,\mu$Hz and $\teff = 6097\,$K. Although this star is slightly cooler than the previous target, here the peak blending for $\ell = 2,0$ and $\ell = 3,1$ mode pairs is more pronounced because of the smaller values of $\Dnu$ and $\delta\nu_{02}$. Nevertheless, \famed\,\,can perform the peak bagging analysis by providing correct oscillation frequency estimates and mode identification for all of the 37 peaks extracted over 14 radial orders. The $\ell = 2,0$ mode pair corresponding to radial order 12 is affected by the presence of a $\ell =1$ mixed mode next to it. The peak blending causes a bias in the estimation of these two frequencies (Fig.~\ref{fig:sg_application}). No $\ell = 3$ modes could be detected in this target, because either affected by peak blending or not enough significant with respect to the noise level.

In the third case we consider a late SG star observed with \kepler for more than four years, KIC~4351319, having $\Dnu \simeq 25\,\mu$Hz and $\teff = 4908$\,K \citep{Pinsonneault18}. The peak bagging analysis may result computationally demanding and challenging to perform on this target because it contains a very large number of $\ell = 1$ mixed modes with a clear rotational effect, and with mixed modes displaced all over the frequency range between consecutive radial orders. As a comparative set for our application we adopt the results from a standard peak bagging analysis performed with \diamonds\,\,following the method presented by C15, where the individual rotational components have been extracted and identified with the aim of investigating the internal angular momentum transport of the star (Di Mauro et al. in prep.). The automated approach performed by \famed\,\,yields a total of 86 oscillation mode frequencies covering 10 radial orders, with the azimuthal numbers of 20 dipole rotational multiplets correctly identified. The agreement with the comparative set is remarkable, although \famed\,\,appears to outperform the standard peak bagging approach by identifying all the rotational components of seven additional $\ell = 1$ mixed modes, one new $\ell = 2$ mode for radial order 18, and two more $\ell = 3$ modes for radial orders 12 and 15. The average value of $\Delta P_1$ estimated from \chu\,\,is $\sim45$\,s.

The last example of this section is the star KIC~4448777, observed by \kepler for more than four years. This star is an early RGB, with $\Dnu \simeq 17\,\mu$Hz and $\teff = 4750$\,K, that exhibits clear rotational effects on its $\ell = 1$ mixed modes and a significant core-to-envelope differential rotation. This star constitutes an ideal target for studies of internal angular momentum transport \citep{DiMauro16,DiMauro18}. From \cite{DiMauro18} a standard peak bagging analysis done with \diamonds is available for a direct comparison that includes the $\ell = 1$ rotationally split components. In the bottom panel of Fig.~\ref{fig:sg_application} we can see an excellent agreement between the individual frequencies and mode identification obtained with \famed\,\,and those from the literature. This is especially remarkable for all the $\ell = 1$ azimuthal components that were extracted by \famed\,\,using the peak rotation test. Only few differences arise in relation to those $\ell = 1$ mixed modes that were not found by our pipeline because not deemed significant, as well as for some other $\ell = 1$ mixed modes that were either not reported or reported but without detection by \cite{DiMauro18}, and which are instead detected with our analysis. \famed\,\,is also capable of detecting three $\ell = 3$ modes, one more with respect to \cite{DiMauro18}. The set of frequencies provided by the pipeline therefore accounts for 67 oscillation modes (including the $m$ components of 16 different dipole rotational triplets), spanning seven radial orders. From $\Delta P_1$ computed for the different chunks in  \chu, we find an average value of about 57\,s, which is also well compatible with what found by \cite{Bedding11Nature} for a star in this evolutionary stage.

The agreement between $\Dnu_0$ and those from the literature may vary depending on the frequency pattern of the oscillation modes. For the simulated stars 05461 and 00123 we find an agreement of about 1.2\,\% and 0.3\,\%, respectively, while for KIC~4351319 we have about 0.4\,\%, and a remarkable $\sim0.01$\,\% for KIC~4448777.

\subsection{Red giant branch and red clump stars}
\label{sec:rg}
The peak bagging analysis of RG stars faces a large number of $\ell = 1$ mixed modes. The applications that are presented in the following comprise a star in the low-luminosity RGB, one star evolving toward the tip of the RGB, and one star in the RC. It is worth mentioning that according to the evolutionary stage classification proposed by \cite{Kallinger12} \citep[but see also the peculiar cases discussed by][]{Corsaro12cluster}, high-luminosity stars having $\Dnu \lesssim \Dnu_\mathrm{tip}$ could comprise early AGB stars as well. These stars are analyzed by \famed\,\,in the same way as high-luminosity RGB stars by relying on the multimodal sampling conditions imposed for the regime $\Dnu \le \Dnu_\mathrm{tip}$.

\begin{figure}[t]
   \centering
  \includegraphics[width=8.9cm]{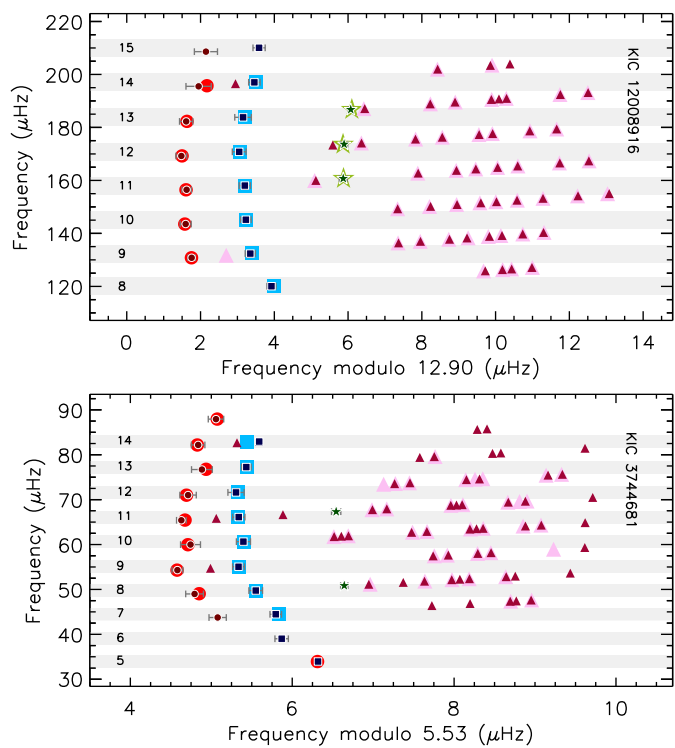}
    \caption{\'Echelle diagrams for the low-luminosity RGB KIC~12008916 and the 2$^\mathrm{nd}$ RC KIC~3744681. The literature values for the frequencies and mode identification are from C15 for the RGB star and from \cite{Kallinger19pb} for the RC. Same description as for the bottom panel of Fig.~\ref{fig:sg_application}. Only statistically significant frequencies are shown, as indicated from the literature. No uncertainties are shown for the $\ell = 1$ modes because smaller than the symbol sizes in most cases. For both stars, the folding $\Dnu$ is $\Dnu_0$ estimated by \famed.}
    \label{fig:rg_application}
\end{figure}

\begin{figure}[th]
   \centering
  \includegraphics[width=8.9cm]{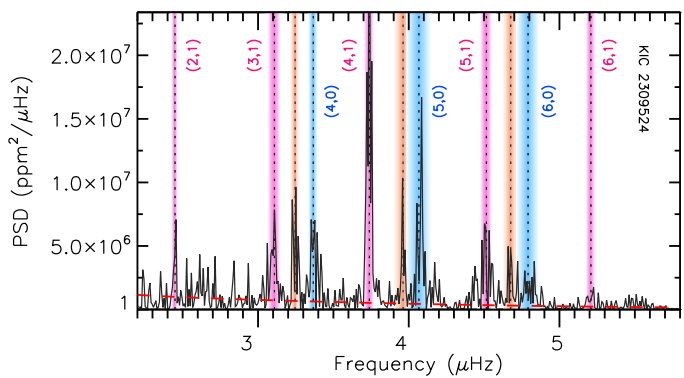}
    \caption{Peak bagging analysis of the high-luminosity RGB KIC~2309524. The vertical dotted lines mark the oscillation modes identified by \famed\,\,on the PSD of the star (black curve). The vertical colored bands, which are as wide as $3$-$\sigma$ uncertainty on each side of the estimated frequencies, indicate the mode angular degree (blue, red, and pink for $\ell = 0, 2, 1$, respectively). The mode identification $\left(n,\ell\right)$ is  overlaid for $\ell = 0,1$ modes and is omitted for $\ell = 2$ modes for visualization purposes only. The background level corresponds to the dashed red line.}
    \label{fig:rg_tip_application}
\end{figure}

Stars that are evolving along the RGB have a higher core-envelope density contrast as compared to RGs that are settled in the RC. As a consequence, the coupling between g- and p-mode cavities in the former stars is weaker than in the latter ones, meaning that the dipole mixed modes are more condensed in frequency during the RGB phase, and with a power barycenter that is regularly following that of pure $\ell = 1$ $p$ modes \citep{Montalban10,Corsaro12cluster}. In these stars the mode extraction and identification process is in principle more accessible.

RG stars placed in the lower part of the RGB are optimal targets to study oscillations because their dipole mixed modes, even if rotational effects are at play, are in general well separated in frequency from one another. This is because of their large values of $\Dnu$, on the order of $10$-$15\,\mu$Hz, and of their high $\numax$, around $100$-$200\,\mu$Hz, which often allows to analyze several consecutive radial orders. The low-luminosity RGB star considered for our application is KIC~12008916, which has been widely adopted in the literature as a benchmark RG thanks to its high S/N and clear oscillation pattern (C15}; \citealt{Davies16performance,Garcia18pb}). This star has $\Dnu \simeq 13\,\mu$Hz and $\teff = 5454$\,K, with a high spin-inclination angle that makes $\ell = 1$ mixed modes appear as rotational doublets in most cases. Although \chu\,\,is not identifying by default all the azimuthal numbers for each $\ell = 1$ rotational component if the star is classified as a RG, the oscillation frequencies are all extracted and identified using a pair $\left(n, \ell\right)$. In Fig.~\ref{fig:rg_application} we find a remarkable agreement between the oscillation frequencies detected and identified by \cite{Corsaro15cat} (where we have ignored their information on the azimuthal number) and those extracted by \famed, with a total of 67 oscillation frequencies covering eight radial orders. According to our analysis, \famed\,\,was capable of detecting a new $\ell = 0$ mode for radial order 15, a new $\ell = 2$ mode for radial order 14, as well as the three $\ell = 3$ modes previously reported. The correct extraction and identification of the $\ell = 3$ mode frequencies was possible even if for some of them $\ell = 1$ rotational components are found to be very close in frequency, thus potentially leading to confusion.

Stars that are evolving toward the tip of the RGB, also referred to as high-luminosity RGB areclassified as late K-type and early M-type giants \citep[e.g., see][]{Stello14}. For these stars only a handful of oscillation modes could be observed because of their very low $\numax$, on the order of a few $\mu$Hz, with no $\ell = 3$ modes expected to be found because of the prominent high-power noise at low frequency. In these stars the oscillation pattern is substantially simplified, with just one $\ell = 1$ mode peak per radial order as the result of the small $\Dnu$ and $\Delta \Pi_1$ values from the high core density and core-envelope density contrast. The difficulty in applying a Bayesian multimodal approach for performing a peak bagging analysis in these cool but luminous targets is represented by: 1) the lack of available radial mode linewidth estimates for stars of types later than mid-K; 2) the frequency resolution that can be achieved to resolve the individual oscillation peaks, which even in the best case of the full \kepler nominal mission often proves to be just enough to reach down to about $\numax \simeq 1\,\mu$Hz. In the current version of \famed, following the results by C15, \cite{Lund17LEGACY}, and \cite{Handberg17} on RGs, for stars having $\teff < 4900$\,K we have adopted a constant FWHM of the radial modes, which is set to $0.12\,\mu$Hz (new empirical relations connecting the mode linewidth to $\numax$, $\teff$, and metallicity in stars from F-type MS to M-type RGs will be presented in a separate paper, Corsaro et al. in prep.). The test target that was considered for our application is the late K-type star KIC~2309524, observed by \kepler for more than four years, and having $\Dnu \simeq 0.7\mu$Hz, and $\teff = 3938$\,K \citep{Pinsonneault18}. In Fig.~\ref{fig:rg_tip_application} we can clearly see that \famed\,\,was capable of correctly locating the position of the $\ell = 0$ modes from \glob, and to subsequently identify the $\ell = 1,2$ modes in \chu, for a total of 11 oscillation frequencies extracted. The $\epsilon$ value obtained from \glob\,\,is 0.57, in line with the $\epsilon$-$\Dnu$ trend reported by \cite{Kallinger12} for stars with similar $\Dnu$ (see their Fig. 4). This star was also analyzed by \cite{Kallinger19pb} with an automated peak bagging approach, therefore its oscillation mode frequencies are publicly available. The author detects the same three $\ell = 0$ modes found by \famed, with frequencies that agree on average around 0.3\,\%, and four $\ell = 2$ modes, three of which were also detected by our pipeline, with a very good average agreement for those in common, about 0.03\,\%. No $\ell = 1$ modes were however reported by \cite{Kallinger19pb} for this star so that a direct comparison cannot be done.

For RGs that are settled in the core-helium-burning phase of stellar evolution, the oscillation mode pattern is probably the most complex among all the evolutionary stages discussed until now. This is caused by four main aspects: 1) a value of $\epsilon$ of the asymptotic relation that does not follow the $\epsilon$-$\Dnu$ relation found for hydrogen-shell burning RGs, calling for the need of an independent evaluation on a star by star basis \citep{Kallinger12,Corsaro12cluster}; 2) larger oscillation mode linewidths with respect to their RGB counterparts because of their higher $\teff$ \citep{Corsaro12cluster}, which can yield important blending effects between neighboring oscillation peaks; 3) large values of $\Delta \Pi_1$ and strong coupling between the p- and g-mode cavities because of the lower core density and $\sim$10 times smaller core-envelope density contrast than in RGB stars of similar luminosity, resulting in $\ell = 1$ mixed modes that are significantly more displaced in frequency range; 4) rotational splittings comparable to the frequency separation between adjacent $\ell = 1$ mixed modes, which could often lead to strong crossing effects. These difficulties may prevent from performing a reliable peak bagging analysis in many cases, which also explains the lack of published oscillation frequencies for RC stars until only very recent times \citep{Kallinger19pb}. In this paper we show an application to the intermediate-mass 2$^\mathrm{nd}$ RC star originally analyzed by \cite{Deheuvels15}, KIC~3744681, and recently also peak bagged by \cite{Kallinger19pb}. This star is suited for the analysis of core-to-envelope differential rotation because it shows a wealth of $\ell = 1$ rotational multiplets throughout its PSD. This star has $\Dnu \simeq 5.5\,\mu$Hz, and $\teff = 5084$\,K, which produces relatively large mode linewidths and therefore partial peak blending effects in many cases. By means of the automated analysis done by \famed\,\,it was however possible to identify a total of 76 oscillation frequencies that cover 10 radial orders, as shown in the bottom panel of Fig.~\ref{fig:rg_application} in a comparison with the results by \cite{Kallinger19pb}, which do not have identification of the azimuthal number. The agreement for $\ell = 0$ and $\ell = 2$ modes is very good, with just one $\ell = 0$ mode likely misidentified as a $\ell = 2$ by \cite{Kallinger19pb}. \famed\,\,can even detect a $\ell = 2$ and a $\ell = 0$ mode, both corresponding to radial order 6, that were not reported in the literature set. We can see a remarkable agreement between the two sets of frequencies for those $\ell = 1$ mixed modes that are in common, although \famed\,\,delivers 19 more significant frequencies with respect to \cite{Kallinger19pb}. Furthermore, \famed\,\,reports the detection of two $\ell = 3$ modes, which are not listed in the literature. Another strength point of our approach is that despite most of the $\ell = 1$ mixed modes remain without an azimuthal number identification, it was still possible to detect rotation for three of these modes. This information extracted by \famed\,\,is already enough to compute an average inclination angle of the star, corresponding to about 70\,deg, well compatible with the value estimated by \cite{Deheuvels15} using a dedicated approach that focused on the $\ell = 1$ rotational multiplets of the star. The observed period spacing estimated by \famed\,\,has an average of about 146\,s, which is in agreement with the expected $\Delta P_1$ of a $2^\mathrm{nd}$ RC star \citep{Bedding11Nature}. The $\epsilon$ value obtained from \glob\,\,is 0.98, once again in line with the trend reported by \cite{Kallinger12}.

In conclusion, for the RG stars presented in this section, by estimating $\Dnu_0$ \famed\,\,reaches a level of agreement of about 0.3\,\% for KIC~3744681, and a remarkable one of $0.03$\,\% for KIC~12008916, as compared to the literature values. For KIC~2309524, our $\Dnu_0$ agrees around 0.1\,\% with that obtained by \cite{Kallinger19pb}, thus confirming that \glob\,\,can reliably estimate $\Dnu$ for evolved RGB stars as well.

\begin{figure}[t]
   \centering
  \includegraphics[width=8.9cm]{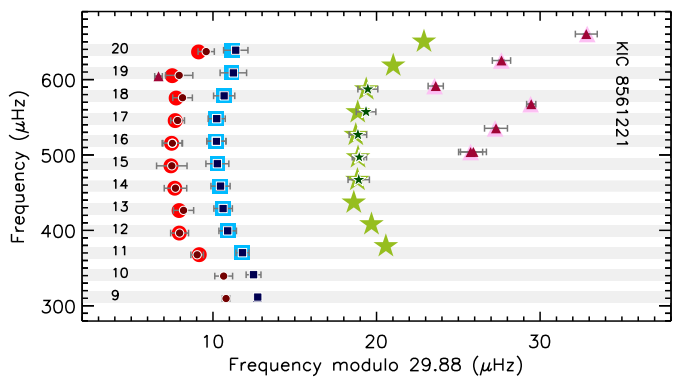}
    \caption{\'Echelle diagram for the depressed dipole subgiant star KIC~8561221. Same description as for Fig.~\ref{fig:rg_application}. The value of $\Dnu$ used for the \'echelle is that by \cite{Garcia14depressed}.}
    \label{fig:dep_application}
\end{figure}

\begin{figure*}[th]
   \centering
  \includegraphics[width=18.5cm]{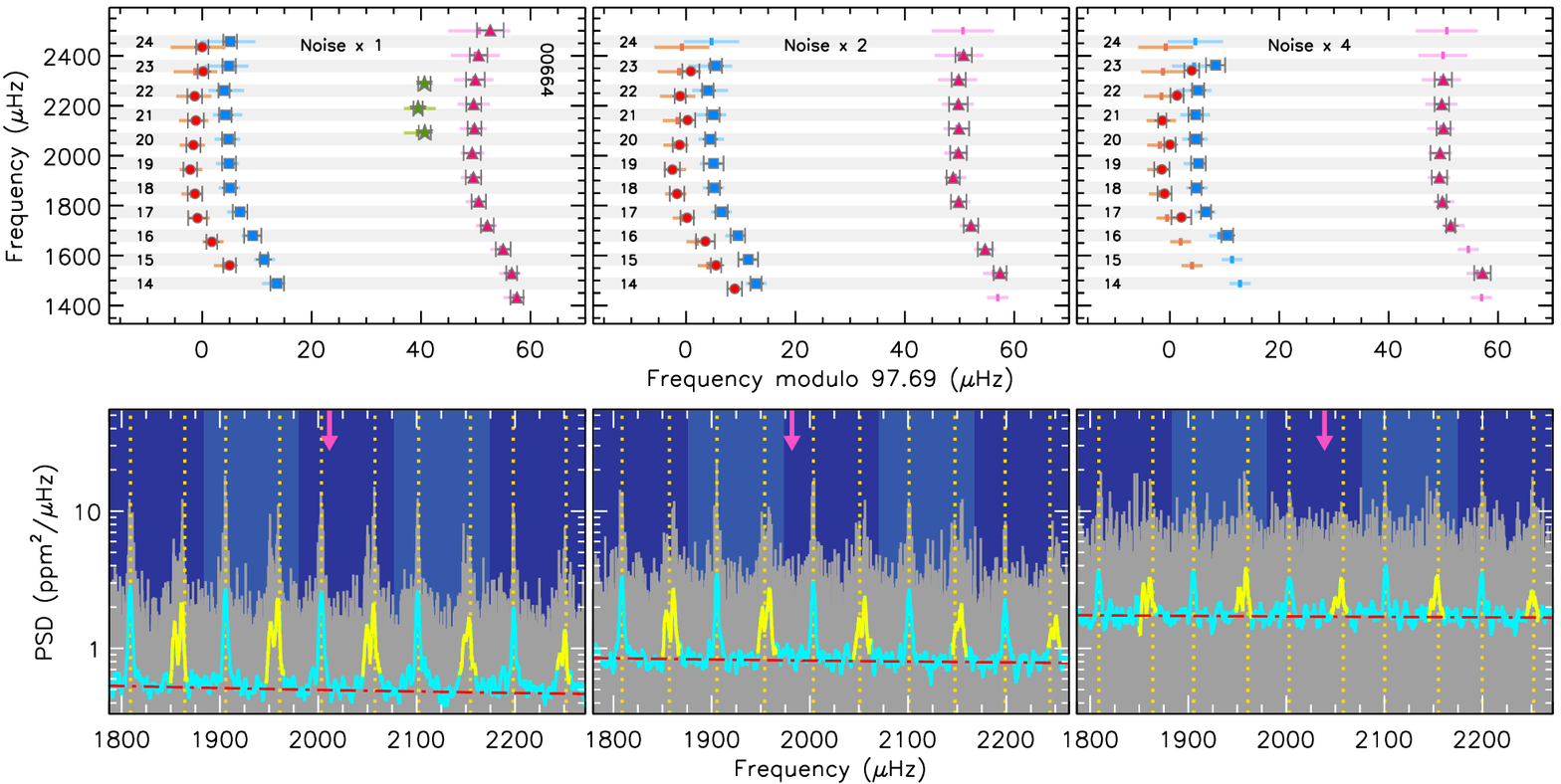}
    \caption{Peak bagging analysis of the simulated MS star 00664 by degrading the instrumental noise in the stellar PSD. \textit{Top panels}: \'echelle diagrams obtained for three different noise levels, with same description as in Fig.~\ref{fig:ms_application}. The folding $\Dnu$ value is that by \cite{Ball18TESS}. \textit{Bottom panels}: five most central chunks of the stellar PSD as seen from \glob, with similar description as in Fig.~\ref{fig:global}. The purple arrow marks the position of $\numax$ obtained from an independent background fit to each PSD.}
    \label{fig:noise_application}
\end{figure*}

\subsection{Depressed dipole stars}
\label{sec:depressed}
In Sects.~\ref{sec:sliding_depressed} and ~\ref{sec:detailed} we have discussed how \glob\,\,and \chu\,\,are suited for performing a peak bagging analysis even in RGs that show dipole mixed modes with depressed amplitudes. For ensuring that a peak bagging analysis is correctly performed in these targets, they have to be identified as potential depressed dipole mode stars during \glob. According to the present literature, these stars range from SGs to RGs, and are characterized by a dipole mode amplitude suppression affecting either a portion or the entire oscillation frequency range. For an application to this class of stars we have selected the late G-type star KIC~8561221, a SG having $\Dnu \simeq 30\,\mu$Hz and $\teff = 5290\,$K. This star was initially studied by \cite{Garcia14depressed}, who discovered the presence of a strong gradient in the amplitude of its $\ell = 1$ modes in moving from low to high oscillation frequencies. As shown in Fig.~\ref{fig:dep_application}, \famed\,\,is capable of detecting a large set of oscillation mode frequencies, 36 over 13 radial orders, including all the $\ell = 1$ modes previously identified by \cite{Garcia14depressed}, and five $\ell = 3$ modes, thus reaching a fine level of agreement with the literature values. Additionally, \famed\,\,detects a new $\ell = 1$ mode for radial order 18, which is located just next to a $\ell = 2$ mode. The presence of this dipole mixed mode could explain why the adjacent $\ell = 2$ mode reported in the literature appears suspiciously offset to lower frequencies with respect to the expected $\delta\nu_{02}$ of the star. Finally, our pipeline could detect two new $\ell = 2,0$ mode pairs in the two most low-frequency and low-S/N chunks. On top of the capabilities of \famed, we consider that the new modes identified by our pipeline in the low-S/N chunks are more likely to be found because of the better quality dataset used in this work, which corresponds to almost one year more observations with respect to that used by \cite{Garcia14depressed}. Nonetheless, all the frequencies reported by \cite{Garcia14depressed} do not come with detection probabilities, which may also explain the surprisingly large number of $\ell = 3$ modes reported by the authors, most of which located in low-S/N chunks where no detectable signal could be found by \famed. Finally, the agreement on $\Dnu$ obtained by \glob\,\,is of about 0.8\,\% only with respect to the literature value.

\subsection{Noise degradation}
\label{sec:noise}
Performing the peak bagging analysis in nonoptimal S/N conditions is an important requirement to maximize the yield of stars for which oscillation mode frequencies can be extracted. A larger ensemble of stars, including stars with low S/N data, can therefore account for both stars that are deemer and stars that are less evolved, hence showing smaller oscillation amplitudes \citep[e.g.,][]{Corsaro13}. This means that if a peak bagging analysis can be conducted in low S/N conditions, one is able to extend the asteroseismic probing potential to both stars that are placed more far away in our Galaxy and to stars that are younger, with direct implications on the study of galactic populations \citep[e.g.,][]{Miglio13,SilvaAguirre20,Chaplin20Nature} and of stellar dynamo and magnetic activity \citep[e.g.,][]{Garcia10Science,Bonanno14,Bonanno19EKEri}, respectively.

In the previous applications we have seen that the pipeline yields trustworthy results for a large variety of stars also for those PSD chunks having the lowest S/N within the oscillation envelope. This demonstrates that the multimodal sampling is sensitive to the presence of oscillation peaks even when these could be confused with noise from a simple visual inspection. While the best data quality reference currently available corresponds to that of the observations carried out by the nominal \kepler mission, other photometric missions, both current and future ones allowing to observe and detect stellar oscillations, may not be able to reach similar S/N conditions as \Kepler. In this section we therefore show how \famed\,\,can perform by varying the level of instrumental noise throughout the stellar PSD. For our application we select the simulated star 00664 from the mock sample of \cite{Ball18TESS}. This star is a late F-type MS, having $\Dnu \simeq 98\,\mu$Hz and $\teff = 6078$\,K. The outcomes from the pipeline for three different levels of the instrumental noise, namely x1, x2, and x4, are presented in Fig.~\ref{fig:noise_application}. \famed\,\,is capable of locating the correct position of the $\ell = 0$ modes even in the lowest S/N case. It is noticeable that the extracted oscillation mode frequencies remain in general rather accurate even if the noise level is increased by four times, going from a total of 36 oscillation frequencies extracted in the best S/N case (noise x1), where three $\ell = 3$ modes could be detected and properly identified too, to 23 oscillation frequencies in the worst S/N scenario (noise x4), where no $\ell = 3$ modes could be found because completely hidden by the noise. We thus have an overall decrease of about 36\,\% in the number of frequencies that could be extracted for this target when increasing the level of noise by four times.

The net result with decreasing S/N is, as expected, a progressive lowering of the number of oscillation modes detected by the pipeline, with the modes first disappearing from those radial orders with the lowest S/N, namely toward the tails of the oscillation envelope. This effect is explained by the drop of the detection probabilities obtained when performing the peak detection tests, because when the background noise increases it is more likely that the peak is confused with the background level in its proximity. Another effect produced by the noise degradation is a progressive lowering of the accuracy of $\ell = 2$ modes, which, after the $\ell = 3$ modes, are the most affected due to their lower visibility as compared to those of $\ell = 0$ and $\ell = 1$ modes. The estimation of $\Dnu$ from \glob\,\,reaches an accuracy level between 0.9 and 0.6\,\%, with no clear dependency on the actual noise level of the dataset. This means that as long as the oscillation modes can be sampled through the island peak bagging model, $\Dnu$ can be measured with high reliability, independently of the S/N of the stellar PSD.

\begin{figure*}[th]
   \centering
  \includegraphics[width=18.5cm]{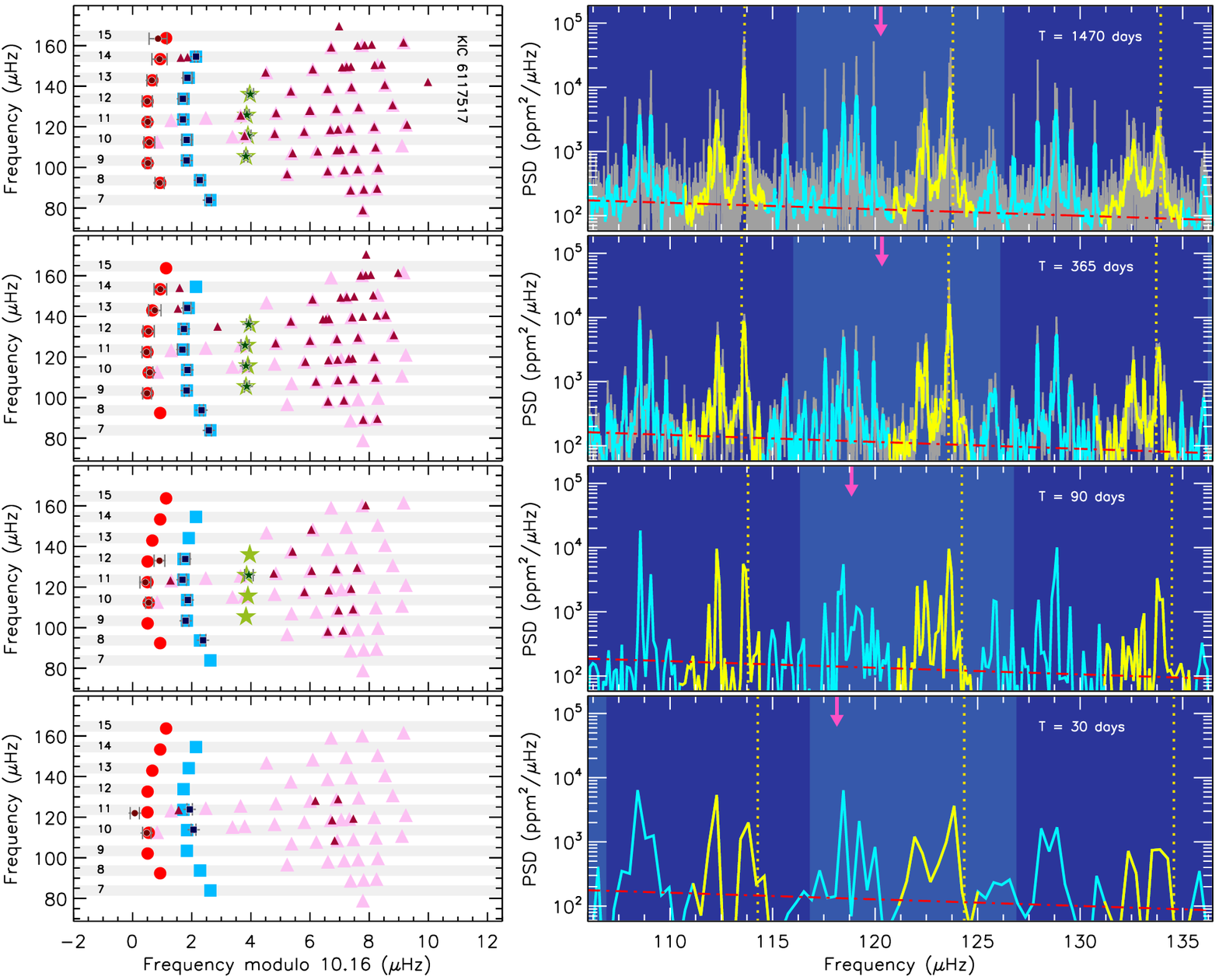}
    \caption{Peak bagging analysis of the low-luminosity RGB star KIC~6117517 by degrading the frequency resolution of the stellar PSD. \textit{Left panels}: \'echelle diagrams obtained for four different frequency resolutions, with same description as in Fig.~\ref{fig:rg_application}. The literature frequencies and the folding $\Dnu$ value are those by C15. \textit{Right panels}: three most central chunks of the stellar PSD (in gray) as seen from \glob\,\,for each test performed, with $\ell = 2,0$ regions in yellow. The purple arrows mark the position of $\numax$ obtained from a background fit to each test PSD. Only global radial mode frequencies are indicated (vertical dotted lines). The smoothing of the PSD is coinciding with the PSD itself for the two bottom panels because of the low frequency resolution.}
    \label{fig:resolution_application}
\end{figure*}

\subsection{Frequency resolution degradation}
\label{sec:resolution}
Another critical feature to consider is the impact of the frequency resolution on our capability to detect and properly identify the oscillation mode frequencies. Degrading the frequency resolution on the stellar PSD has three main effects: 1) the S/N decreases as the result of the decrease of the peak heights; 2) the observed linewidth of resolved oscillation modes is significantly affected by the frequency resolution of the dataset, such that as long as the frequency resolution is smaller than the true peak linewidth, the latter one decreases as the frequency resolution degrades because the mode becomes progressively less resolved, but as soon as the frequency resolution becomes larger than the true peak linewidth (in the worst scenario), meaning that the mode is completely unresolved, then the observed peak linewidth can become larger than the true value; 3) the number of frequency-resolved oscillation modes decreases because modes that are as close in frequency as $\sim$3-4 times the frequency resolution itself cannot be distinguished in general, although this limiting separation is subject to vary as a function of the S/N. Therefore by degrading the frequency resolution we expect that the overall number of modes that can be detected, and identified in the stellar PSD decreases, as well as the accuracy and precision obtained on the individual mode frequencies.

If the frequency resolution decreases it is more likely that the peak detection tests favor the scenario of a sinc$^2$ profile against that of a Lorentzian profile. This is more evident on evolved stars, where the densely populated $\ell = 1$ mixed modes forests are progressively simplified in structure as the frequency resolution decreases \citep[e.g., see the first results on red giants using CoRoT observations spanning about 150 days,][]{DeRidder09}. The $\ell = 1$ mixed modes having a stronger p-mode character are also more likely to remain observable thanks to their higher oscillation amplitudes as compared to those having a stronger g-mode character instead.

For testing the capabilities of \famed\,\,with a varying frequency resolution, we consider the low-luminosity RGB star KIC~6117517, observed by \kepler for more than four years, and with available oscillation mode frequencies from C15. This star has $\Dnu \simeq 10\,\mu$Hz, $\teff = 4687$\,K \citep{Pinsonneault18}, and exhibits a large number of oscillation modes, including some $\ell = 3$. Starting from the original KADACS light curve used in C15, on top of the reference PSD corresponding to an observing time of $T = 1470$\,days, we reproduce three additional datasets, for observing times $T = 365$\,days, $90$\,days, and $30$\,days. These observing times roughly reproduce the frequency resolution obtained from a 1-year observation with NASA TESS in the continuous viewing zones, a $\sim$3-months observation during a NASA K2 Campaign, and a typical $\sim$1-month observation with NASA TESS. The results are presented in Fig.~\ref{fig:resolution_application}. For $T = 1470$\,days (top panels), corresponding to a frequency resolution $\delta\nu_\mathrm{res} \simeq 0.008\,\mu$Hz, we find a very good agreement between the automated frequency extraction operated by \famed\,\,and the literature values by C15, for a total of 70 oscillation frequencies extracted by our pipeline and covering eight consecutive radial orders. Here \famed\,\,can detect and identify all the four $\ell = 3$ modes reported in the literature, as well as all the $\ell = 2,0$ modes and nearly all of the $\ell = 1$ mixed modes. The global radial mode frequencies obtained from \glob\,\,are quite accurate thanks to the high S/N and frequency resolution of the dataset. When degrading the frequency resolution to about four times the nominal value for $T = 365$\,days, the frequency set extracted by \famed\,\,reduces to 58 frequencies (hence by $\sim$17\,\%), with the four $\ell = 3$ modes reported in the literature still detected. The global radial mode frequencies from \glob\,\,are rather accurate once again. With one year of observation the situation has thus not changed dramatically, although most of the lost frequencies are $\ell = 1$ mixed modes that have smaller oscillation amplitudes, hence generally falling toward the wings of the $\ell = 1$ mixed mode region. But with an additional degradation of the frequency resolution, this time to about 16 times the nominal value, for $T = 90\,$days, the number of frequencies extracted by \famed\,\,reduces to 24 (meaning about 66\,\% less than in the nominal case), with one $\ell = 3$ mode still detected and correctly identified. Now it is evident that many of the modes with low amplitudes and/or corresponding to the chunks with the lowest S/N are being lost, while at the same time the detected $\ell = 1$ mixed modes are on average even more condensed to the $\ell = 1$ p-mode position than in the previous example. \glob\,\,now provides less accurate positions for the radial modes, although still good enough to allow \chu\,\,to properly locate and identify them in the PSD. Finally, we can see that the frequency position of some of the $\ell = 2,0$ modes as obtained from \chu\,\,starts to exhibit a lower level of accuracy with respect to the previous tests. In the last example, for $T = 30\,$days, we reach the worst scenario with a frequency resolution degraded to about 50 times the nominal value, thus corresponding to $\delta\nu_\mathrm{res} \simeq 0.40\,\mu$Hz, significantly larger than a typical radial mode linewidth measured in any RG star so far (e.g., C15); \citealt{Vrard18pb}). Here the worsening is evident, with only 10 oscillation frequencies extracted (i.e., about 86\,\% reduction with respect to the first example), covering three radial orders, that is the ones closest to $\numax$. The loss in the accuracy of the global radial mode frequencies from \glob\,\,is more pronounced than in the previous case, but it still suffices to allow the pipeline to properly identify the modes. The accuracy on the final frequencies is the lowest among the four examples provided, and this is now apparent on most of the $\ell = 2,0$ modes that are extracted. We stress that this loss in accuracy is not to be attributed to the extraction process made by \famed\,\,because the pipeline can adequately locate the position of each frequency peak in the PSD independently of the frequency resolution. This time the loss in accuracy is caused by the low frequency resolution that makes individual peaks appear as broad spike-like structures (see Fig.~\ref{fig:resolution_application}, bottom panel). 

The value of $\Dnu$ estimated by \glob\,\,for each example ranges between 0.6 and 0.1\,\%, showing that it is essentially independent of the frequency resolution that is adopted, even if only a few radial oscillation modes can be detected. In addition, the average $\Delta P_1$ estimated from the extracted dipole modes is of about 58\,s, 55\,s, 52\,s, and 49\,s, from the best to the worst frequency resolution test conducted. The estimate of $\Delta P_1$ is therefore not changing significantly as the frequency resolution degrades, even in the case of just one month of observations, and it is perfectly in line with the $\Delta P_1$ expected for a RGB star according to the literature \citep[e.g.,][]{Bedding11Nature}.

\section{Conclusions \& future work}
\label{sec:conclusions}
In this paper we have shown that the Bayesian multimodal approach is a very powerful technique for detailed asteroseismic analyses of solar-like oscillators. In particular, this is made possible by the relatively simple setting up of the approach, which allows us to obtain a large amount of information in a relatively short amount of time, even if using a single CPU system setup. The high computational speed that characterizes the \famed\,\,pipeline is largely supported by the adoption of \diamonds\,\,as a fitting code. \diamonds\,\,is used by \famed\,\,to perform up to hundreds of fits of different type within the peak bagging analysis of a single star, with most of these fits related to  peak testing. 
In Sect.~\ref{sec:applications} we have seen that \famed\,\,can be successfully exploited for a large diversity of stars of low and intermediate mass, ranging from F-type MS, characterized by strong peak blending effects between adjacent oscillation modes, to SGs in different stages of evolution, and RGs from the base of the RGB up to the RGB tip and finally to the RC, including peculiar stars that show depressed dipole mode amplitudes. The extraction of oscillation modes is adequately performed even for stars that exhibit a large number of dipolar rotational multiplets, which is an indication that rotation and stellar spin inclination are accessible quantities for our automated approach. We have seen that this is verified for stars from the MS up to even the challenging case of core-Helium-burning RGs. We have seen that the frequencies extracted with \famed\,\,by adopting the first two modules, \glob\,\,and \chu, are comparable in number and accuracy to those from the literature, with \famed\,\,being often able to slightly outperform standard approaches in detecting and identifying low-S/N oscillation modes (including $\ell = 3$ modes), modes with strong peak blending, and modes containing a fine-structure such as that of rotation. The adoption of the data-driven approach used by \famed\,\,for the automated extraction and identification of $\ell = 1$ mixed modes, including their rotational components when rotation is detected, yields robust results even without fitting an asymptotic pattern for mixed modes. 

We have also demonstrated that the pipeline has powerful abilities in analyzing datasets with low S/N and frequency resolution conditions, which make \famed\,\,well suited for studies involving a large number of targets, even if relatively far way and/or observed for just a few months. This is especially relevant in view of the observations that are being carried out by NASA TESS, where already $\sim$1 month of data can suffice to detect individual oscillation modes \citep[e.g.,][]{Huber19TOI,Campante19TESS,Chaplin20Nature,SilvaAguirre20,Jiang20TESS}, and for processing the huge amount of datasets that will be delivered by ESA PLATO in a near future. 
Furthermore, based on the results from our applications, we find that other outputs provided by \famed\,\,that can be of relevance in the context of an asteroseismic characterization are: 1) an accurate estimate of $\Dnu$, often well below 1\,\%, with an accuracy that is essentially not sensitive to the actual S/N and frequency resolution of the dataset; 2) the $\epsilon$ term of the asymptotic relation; 3) a value of the observed period spacing $\Delta P_1$ for each chunk analyzed; 4) the radial mode linewidths, $\Gamma_0$, for all the radial modes identified through the stellar PSD.

As already anticipated, the extra modules \texttt{\'ECHELLE} and \texttt{COMPLETE} will provide further validation on the oscillation modes extracted with \glob\,\,and \chu\,\,based on theoretical asymptotic relations, as well as additional parameters, mostly useful for analyzing evolved stars (see Table~\ref{tab:famed} for an overview). Future work will aim at incorporating new modules, for example for automatically extracting the acoustic glitches based on the methodology presented by \cite{Corsaro15glitch}, and a global stellar spin inclination angle following the approach presented by \cite{Kuszlewicz19angle}.

Interestingly, the multimodal approach used by \famed\,\,makes this pipeline suitable for applications that can differ from the peak bagging of solar-like oscillators. For example, \famed\,\,could be extended to another class of pulsators, provided that the analysis can still be performed in the Fourier domain (e.g., for DAVs type pulsators, see \citealt{Hermes17DAV}). Future developments of this kind could be included in the pipeline by exploiting its flexibility and modular structure.

In its current state, \famed\,\,by no means pretends to be a definitive and fully exhaustive approach to peak bagging. Our aim is, however, to provide a powerful, reliable, and at the same time easy-to-use tool to perform complicated and otherwise long-lasting analyses of stellar oscillations that can be at the disposal of the entire astrophysics community, including nonexperts in the asteroseismology field. For this reason \famed\,\,is subject to a continuous maintenance and development aimed at improving its overall functionality, reliability, computational stability, and efficiency. This will be possible by taking into account the testing and applications done by the future user community of the pipeline. In conclusion, we believe that \famed\,\,already represents a promising opportunity for overcoming the long-lasting problem of automatizing the complicated analysis of stellar oscillations. \famed, together with other possible approaches that may become available in the future, will definitely contribute to increase our capabilities to investigate large ensembles of oscillating stars in great detail.

\begin{acknowledgements}
E.C. is funded by the European Union's Horizon 2020 research and innovation program under the Marie Sklodowska-Curie grant agreement No. 664931. E.C. acknowledges support from PLATO ASI-INAF agreement n.2015-019-R.1-2018. E.C. acknowledges the participation in the PLATO PSM WP128 peak bagging exercises that helped in developing the \famed\,\,pipeline and for its calibration in the regime of MS and SG stars. We thank R. A. Garc\'{i}a for providing KADACS datasets of the \kepler targets used in the applications shown in this work and J. Tayar for providing NASA TESS datasets that contributed in improving the pipeline in the regime of SG and early RGB. We also thank M. B. Nielsen and G. R. Davies for useful discussions. 
\end{acknowledgements}

\bibliographystyle{aa} 


\appendix
\section{Prior hyper-parameters for peak tests}
\label{sec:prior}
Here we provide the uniform prior hyper-parameters that are adopted in the current version of \famed\,\,for conducting the peak testing of detection, blending, rotation, and duplicity. The priors currently used by \famed\,\,have been tuned to obtain an optimal reproducibility of the results found in the literature (Sect.~\ref{sec:applications}). The prior ranges are listed in Tables~\ref{tab:priors_detection} and \ref{tab:priors_rotation}. The quantity $S_{i,\mathrm{corr}}$ appearing in the upper prior bound for oscillation amplitudes is an estimate of the true peak height. This is obtained as the maximum value of the smoothed PSD within the frequency ranges of a given peak, which we refer to as $S_{i,\mathrm{max}}$, with the smoothing calculated using a factor $\Gamma_\mathrm{chunk}$, and is subsequently corrected by first subtracting the local level of the background and by then dividing by the response function of the sampling cadence such that the height estimate takes into account any degradation of the signal as we approach to the limiting Nyquist frequency.

In most cases, the prior ranges can be tuned from the input configuring parameter file of the pipeline through a specific parameter, whose value is also listed in Tables~\ref{tab:priors_detection} and \ref{tab:priors_rotation}. The user has the possibility to modify the lower bound of the prior on the linewidths, as well as any of the multiplication coefficients that appear for the upper bounds of the priors on linewidths and amplitudes. In the case of the amplitudes of model $\mathcal{M}_\mathrm{B}$ for less evolved stars, $\Dnu \ge \Dnu_\mathrm{thresh}$, the upper bound is selected as the maximum $S_{i,\mathrm{corr}}$ between that of the chunk radial mode and that of the chunk quadrupole mode. Following the results on standard peak bagging analysis of RGB stars by C15, we find that in evolved stars the upper prior bound on the linewidth is larger for $\ell = 2$ modes than for $\ell = 0$ ones to account for the presence of quadrupole mixed modes. Again for the case of evolved stars, the upper prior bound on the linewidth of $\ell = 1$ modes is smaller than for $\ell = 0$ modes because here the dipole modes are mixed modes. For model $\mathcal{M}_\mathrm{D}$ the prior ranges with subscripts indicating the angular degrees 0,2 imply that they are evaluated specifically for the chunk radial and quadrupole modes in relation to the free parameter they are associated with (e.g., $\ell = 0$ for $\nu_{i,0}^{\ell = 0}$ and so on). Finally, all the priors on linewidth and amplitudes incorporate the FWHM of the chunk radial mode, $\Gamma_0$, which is measured as explained in Sect.~\ref{sec:fwhm_radial}.

\begin{table*}
\caption{Prior hyper-parameters adopted for the peak testing performed within \chu. Each prior can be identified with its corresponding free parameter and fitting model, and may change as a function of the angular degree $\ell$ of the mode being tested and on the general evolutionary stage classification of the star (reported in terms of $\Dnu$ and $\teff$ thresholds), as presented in Sects.~\ref{sec:peak_detection_test}. The values of the configuring parameters used to set up each prior, if any, are listed in the last column. Some of the free parameters are the same for different models.}             
\centering             
\tiny
\begin{tabular}{l c c c c c}
\hline\hline
\\[-8pt]
Model & Free parameter (units) & $\ell$ & Evolutionary stage & Uniform Prior range & Configuring parameter value\\[1pt]
\hline   
\\[-6pt]
$\mathcal{M}_\mathrm{A}$, $\mathcal{M}_\mathrm{B}$ & $\sigma_\mathrm{noise}$ & 0,1,2,3 & Any & $\left[a_1,a_2\right]$ & $a_1 = 0.95$, $a_2 = 1.05$ \\ [1pt] 
\cmidrule{1-6}
\\[-8pt]
\multirow{3}[76]*{$\mathcal{M}_\mathrm{B}$} & \multirow{2}[8]*{$\nu_{i,0}$ ($\mu$Hz)} & \multirow{2}[3]*{0,2} & $\Dnu \geq \Dnu_\mathrm{thresh}$ &$\left[r_a (\ell = 2), r_b(\ell = 0) \right]$ & -- \\ [1pt]
\cmidrule{4-6}
\\[-8pt]
& & & $\Dnu < \Dnu_\mathrm{thresh}$ &$\left[\nu_{f,i} - \sigma_{f,i},\nu_{f,i} + \sigma_{f,i} \right]$ & --\\ [1pt]
\cmidrule{3-6}
\\[-8pt]
& & 1,3 & Any &$\left[\nu_{f,i} - \sigma_{f,i},\nu_{f,i} + \sigma_{f,i} \right]$ & --\\ [1pt]
\cmidrule{2-6}
\\[-8pt]
& \multirow{3}[28]*{$A_i$ (ppm)} & \multirow{2}[3]*{0} & $\Dnu \geq \Dnu_\mathrm{thresh}$ & $\left[0,\max_{\ell = 0,2} \left( \sqrt{\eta \Gamma_0 \pi S_{i,\mathrm{corr}}} \right) \right]$ & $\eta = 3$\\ [1pt]
\cmidrule{4-6}
\\[-8pt]
& & & $\Dnu < \Dnu_\mathrm{thresh}$ &$\left[0,\sqrt{\eta \Gamma_0 \pi S_{i,\mathrm{corr}}} \right]$ & $\eta = 3$ \\ [1pt]
\cmidrule{3-6}
\\[-8pt]
& & \multirow{2}[3]*{1} & $\Dnu \leq \Dnu_\mathrm{SG}$ and $\teff \geq T_\mathrm{eff,SG}$, or $\Dnu \geq \Dnu_\mathrm{SG}$ &$\left[0,\sqrt{\eta \Gamma_0 \pi S_{i,\mathrm{corr}} } \right]$ & $\eta = 6$ \\ [1pt]
\cmidrule{4-6}
\\[-8pt]
& & & $\Dnu < \Dnu_\mathrm{SG}$ and $\teff < T_\mathrm{eff,SG}$ & $\left[0,\sqrt{\eta \Gamma_0 \pi S_{i,\mathrm{corr}} } \right]$ & $\eta = 1.5$ \\ [1pt]
\cmidrule{3-6}
\\[-8pt]
& & \multirow{2}[3]*{2} & $\Dnu \geq \Dnu_\mathrm{thresh}$ &$\left[0,\max_{\ell = 0,2} \left(  \sqrt{\eta \Gamma_0 \pi S_{i,\mathrm{corr}} } \right) \right]$ & $\eta = 3$ \\ [1pt]
\cmidrule{4-6}
\\[-8pt]
& & & $\Dnu < \Dnu_\mathrm{thresh}$ &$\left[0,\sqrt{\eta \Gamma_0 \pi S_{i,\mathrm{corr}} } \right]$ & $\eta = 8$ \\ [1pt]
\cmidrule{3-6}
\\[-8pt]
& & 3 & Any & $\left[0,\sqrt{\eta \Gamma_0 \pi S_{i,\mathrm{corr}} } \right]$ & $\eta = 3$\\ [1pt]
\cmidrule{2-6}
\\[-8pt]
& \multirow{3}[23]*{$\Gamma_i$ ($\mu$Hz)} & 0 & Any &$\left[\Gamma_\mathrm{min},\eta \Gamma_0 \right]$ & $\Gamma_\mathrm{min} = 10^{-4}$, $\eta = 3$\\ [1pt]
\cmidrule{3-6}
\\[-8pt]
& & \multirow{2}[3]*{1} & $\Dnu \leq \Dnu_\mathrm{SG}$ and $\teff \geq T_\mathrm{eff,SG}$, or $\Dnu \geq \Dnu_\mathrm{SG}$ &$\left[\Gamma_\mathrm{min}, \eta \Gamma_0 \right]$ & $\Gamma_\mathrm{min} = 10^{-4}$, $\eta = 6$ \\ [1pt]
\cmidrule{4-6}
\\[-8pt]
& & & $\Dnu < \Dnu_\mathrm{SG}$ and $\teff < T_\mathrm{eff,SG}$ & $\left[\Gamma_\mathrm{min}, \eta \Gamma_0 \right]$ & $\Gamma_\mathrm{min} = 10^{-4}$, $\eta = 1.5$\\ [1pt]
\cmidrule{3-6}
\\[-8pt]
& & \multirow{2}[3]*{2} & $\Dnu \geq \Dnu_\mathrm{thresh}$ &$\left[\Gamma_\mathrm{min}, \eta \Gamma_0 \right]$ & $\Gamma_\mathrm{min} = 10^{-4}$, $\eta = 3$\\ [1pt]
\cmidrule{4-6}
\\[-8pt]
& & & $\Dnu < \Dnu_\mathrm{thresh}$ &$\left[\Gamma_\mathrm{min}, \eta \Gamma_0 \right]$ & $\Gamma_\mathrm{min} = 10^{-4}$, $\eta = 8$\\ [1pt]
\cmidrule{3-6}
\\[-8pt]
& & 3 & Any & $\left[\Gamma_\mathrm{min}, \eta \Gamma_0 \right]$ & $\Gamma_\mathrm{min} = 10^{-4}$, $\eta = 3$\\ [1pt]
\cmidrule{1-6}
\\[-8pt]
\multirow{3}[5]*{$\mathcal{M}_\mathrm{C}$} & $\sigma_\mathrm{noise}$ & 1 & $\Dnu < \Dnu_\mathrm{RG}$ &$\left[a_1,a_2\right]$ & $a_1 = 0.95$, $a_2 = 1.05$\\ [1pt] 
\cmidrule{2-6}
\\[-8pt]
& $\nu_{i,0}$ ($\mu$Hz) & 1 & $\Dnu < \Dnu_\mathrm{RG}$ & $\left[\nu_{f,i} - \sigma_{f,i},\nu_{f,i} + \sigma_{f,i} \right]$ & --\\ [1pt] 
\cmidrule{2-6}
\\[-8pt]
& $H_i$ (ppm$^2 / \mu$Hz) & 1 & $\Dnu < \Dnu_\mathrm{RG}$ & $\left[0,1.5 S_{i,\mathrm{max}}\right]$ & --\\ [1pt] 
\cmidrule{1-6}
\\[-8pt]
\multirow{4}[9]*{$\mathcal{M}_\mathrm{D}$} & $\sigma_\mathrm{noise}$ & 0,2 & $\Dnu \ge \Dnu_\mathrm{thresh}$ &$\left[a_1,a_2\right]$ & $a_1 = 0.95$, $a_2 = 1.05$\\ [1pt] 
\cmidrule{2-6}
\\[-8pt]
& $\nu_{i,0}^{\ell = 0}, \nu_{i,0}^{\ell = 2}$ ($\mu$Hz) & 0,2 & $\Dnu \ge \Dnu_\mathrm{thresh}$ & $\left[\nu_{f,i} - \sigma_{f,i},\nu_{f,i} + \sigma_{f,i} \right]_{\ell = 0,\ell =2}$ & --\\ [1pt] 
\cmidrule{2-6}
\\[-8pt]
& $A_i^{\ell = 0}, A_i^{\ell = 2}$ (ppm) & 0,2 & $\Dnu \ge \Dnu_\mathrm{thresh}$ & $\left[0,\sqrt{\eta \Gamma_0 \pi S_{i,\mathrm{corr}} } \right]_{\ell = 0,\ell =2}$ & $\eta = 3$\\ [1pt] 
\cmidrule{2-6}
\\[-8pt]
& $\Gamma_i^{\ell = 0},\Gamma_i^{\ell = 2}$ ($\mu$Hz) & 0,2 & $\Dnu \ge \Dnu_\mathrm{thresh}$ &$\left[\Gamma_\mathrm{min}, \eta \Gamma_0\right]$ & $\Gamma_\mathrm{min} = 10^{-4}$, $\eta = 3$\\ [1pt] 
\\[-8pt]
\hline
\end{tabular}
\label{tab:priors_detection}
\end{table*}

\begin{table*}
\caption{Prior hyper-parameters adopted for the peak rotation and duplicity testing performed within \chu. Similar description as for Table~\ref{tab:priors_detection}, following what presented in Sect.~\ref{sec:peak_rotation_test}. The $\ell = 3$ cases refer to those peaks flagged as candidate octupole modes according to Sect.~\ref{sec:octupole_test}, i.e., located inside the octupole mode search range, hence are not intended as yet confirmed $\ell = 3$ mode of the analysis.}             
\centering             
\tiny
\begin{tabular}{l c c c c c}
\hline\hline
\\[-8pt]
Model & Free parameter (units) & $\ell$ & Evolutionary stage & Uniform Prior range & Configuring parameter value \\[1pt]
\hline   
\\[-6pt]
\multirow{3}[46]*{$\mathcal{M}_\mathrm{E}$, $\mathcal{M}_\mathrm{F}$} & \multirow{2}[2]*{$\nu_{i,0}$ ($\mu$Hz)} & \multirow{2}[2]*{1,3} & $\Dnu \leq \Dnu_\mathrm{SG}$ and $\teff \geq T_\mathrm{eff,SG}$, or $\Dnu \geq \Dnu_\mathrm{SG}$ & $\left[\nu_{f,i} - \sigma_{f,i},\nu_{f,i} + \sigma_{f,i} \right]$ & --\\ [1pt]
\cmidrule{4-6}
\\[-8pt]
& & & $\Dnu < \Dnu_\mathrm{SG}$ and $\teff < T_\mathrm{eff,SG}$ & $\left[ r_a, r_b \right]_i$ & --\\ [1pt]
\cmidrule{2-6}
\\[-8pt]
& \multirow{3}[11]*{$A_i$ (ppm)} & \multirow{3}[5]*{1} & $\Dnu \leq \Dnu_\mathrm{SG}$ and $\teff \geq T_\mathrm{eff,SG}$, or $\Dnu \geq \Dnu_\mathrm{SG}$ &$\left[0, \sqrt{\eta \Gamma_0 \pi S_{i,\mathrm{corr}}} \right]$  & $\eta = 3$\\ [1pt]
\cmidrule{4-6}
\\[-8pt]
& & & $\Dnu_\mathrm{thresh} \leq \Dnu < \Dnu_\mathrm{SG}$ and $\teff < T_\mathrm{eff,SG}$ & $\left[0, \sqrt{\eta \Gamma_0 \pi S_{i,\mathrm{corr}}} \right]$  & $\eta = 1.5$  \\ [1pt]
\cmidrule{4-6}
\\[-8pt]
& & & $\Dnu < \Dnu_\mathrm{thresh}$ & $\left[0, \sqrt{\eta \Gamma_0 \pi S_{i,\mathrm{corr}}} \right]$  & $\eta = 1$  \\ [1pt]
\cmidrule{3-6}
\\[-8pt]
& & 3 & Any & $\left[0, \sqrt{\eta \Gamma_0 \pi S_{i,\mathrm{corr}}} \right]$  & $\eta = 3$ \\ [1pt]
\cmidrule{2-6}
\\[-8pt]
& \multirow{3}[11]*{$\Gamma_i$ ($\mu$Hz)} & \multirow{3}[5]*{1} & $\Dnu \leq \Dnu_\mathrm{SG}$ and $\teff \geq T_\mathrm{eff,SG}$, or $\Dnu \geq \Dnu_\mathrm{SG}$ &$\left[\Gamma_\mathrm{min},\eta \Gamma_0 \right]$  & $\Gamma_\mathrm{min} = 10^{-4}$, $\eta = 3$ \\ [1pt]
\cmidrule{4-6}
\\[-8pt]
& & & $\Dnu_\mathrm{thresh} \leq \Dnu < \Dnu_\mathrm{SG}$ and $\teff < T_\mathrm{eff,SG}$ & $\left[\Gamma_\mathrm{min}, \eta \Gamma_0 \right]$ & $\Gamma_\mathrm{min} = 10^{-4}$, $\eta = 1.5$ \\ [1pt]
\cmidrule{4-6}
\\[-8pt]
& & & $\Dnu < \Dnu_\mathrm{thresh}$ & $\left[\Gamma_\mathrm{min}, \eta \Gamma_0 \right]$ & $\Gamma_\mathrm{min} = 10^{-4}$, $\eta = 1$\\ [1pt]
\cmidrule{3-6}
\\[-8pt]
& & 3 & Any & $\left[\Gamma_\mathrm{min}, \eta \Gamma_0 \right]$ & $\Gamma_\mathrm{min} = 10^{-4}$, $\eta = 3$\\ [1pt]
\cmidrule{1-6}
\\[-8pt]
\multirow{2}[8]*{$\mathcal{M}_\mathrm{F}$} & \multirow{2}[3]*{$\delta\nu_\mathrm{rot}$ ($\mu$Hz)} & \multirow{2}[3]*{1,3} & $\Dnu \leq \Dnu_\mathrm{SG}$ and $\teff \geq T_\mathrm{eff,SG}$, or $\Dnu \geq \Dnu_\mathrm{SG}$ & $ \left[2\delta\nu_\mathrm{res}, \frac{2}{\theta}  \sigma_{f,i} \right] $ & $\theta = 2.8$ \\ [1pt] 
\cmidrule{4-6}
\\[-8pt]
 & & & $\Dnu < \Dnu_\mathrm{SG}$ and $\teff < T_\mathrm{eff,SG}$ & $ \left[2\delta\nu_\mathrm{res}, \frac{1}{\theta} \left(r_b - r_a\right) \right]_i$ & $\theta = 2.8$ \\ [1pt] 
\cmidrule{2-6}
\\[-8pt]
& $\cos i$ & 1,3 & Any & $ \left[0,1\right]$ & -- \\ [1pt] 
\cmidrule{1-6}
\\[-8pt]
\multirow{4}[8]*{$\mathcal{M}_\mathrm{G}$}  & $\nu_{i,0}^1$ ($\mu$Hz) & 1,3 & $\Dnu < \Dnu_\mathrm{RG}$ & $\left[ r_a, r_b\right]_i$& -- \\ [1pt] 
\cmidrule{2-6}
\\[-8pt]
& $\delta\nu_\mathrm{split}$ ($\mu$Hz) & 1,3 & $\Dnu < \Dnu_\mathrm{RG}$ & $\left[2\delta\nu_\mathrm{res}, r_b - r_a\right]_i$& --\\ [1pt] 
\cmidrule{2-6}
\\[-8pt]
& $A_i^1, A_i^2$ (ppm) & 1,3 & $\Dnu < \Dnu_\mathrm{RG}$ & $\left[0, \sqrt{\eta \Gamma_0 \pi S_{i,\mathrm{corr}}} \right]$ & $\eta = 1$\\ [1pt] 
\cmidrule{2-6}
\\[-8pt]
& $\Gamma_i^1, \Gamma_i^2$ ($\mu$Hz) & 1,3 & $\Dnu < \Dnu_\mathrm{RG}$ & $\left[\Gamma_\mathrm{min}, \eta \Gamma_0 \right]_i$ & $\Gamma_\mathrm{min} = 10^{-4}$, $\eta = 1$\\ [1pt]
\\[-8pt]
\hline
\end{tabular}
\label{tab:priors_rotation}
\end{table*}

\end{document}